\documentclass{article}

\usepackage{arxiv}

\usepackage[utf8]{inputenc} 
\usepackage[T1]{fontenc}    
\usepackage{hyperref}       
\usepackage{url}            
\usepackage{booktabs}       
\usepackage{amsfonts}       
\usepackage{nicefrac}       
\usepackage{microtype}      
\usepackage{lipsum}		
\usepackage{graphicx}
\usepackage{natbib}
\usepackage{doi}
\usepackage{rotating,enumerate}   
\usepackage{booktabs}
\usepackage{import}
\usepackage{bm,amsmath,multicol,multirow,amssymb,comment}
\usepackage{graphicx}
\usepackage{longtable,setspace}
\usepackage{caption,placeins,subcaption}
\usepackage{xcolor}
\usepackage{array}

\newtheorem{theorem}{Theorem}

\newtheorem{remark}{Remark}

\usepackage{xcolor} 
\usepackage{subcaption}
\usepackage[table]{xcolor}
\definecolor{lightgray}{gray}{0.9}

\newcolumntype{P}[1]{>{\raggedright\arraybackslash}p{#1}}

\title{While-alive regression analysis of composite survival endpoints}

\author{
 Xi Fang \\
    Department of Biostatistics \\
    Yale School of Public Health \\
    New Haven, CT, USA\\
   \And
 Hajime Uno \\
Department of Data Science \\
Dana-Farber Cancer Institute \\
Boston, MA, USA \\
  \And
 Fan Li \\
Department of Biostatistics \\
Yale School of Public Health \\
New Haven, CT, USA\\
  \texttt{fan.f.li@yale.edu} \\
}

\begin{document}
\maketitle

\begin{abstract}
Composite endpoints are frequently used in clinical trials to enhance the event rate and improve the statistical power. In the presence of a terminal event, the while-alive cumulative frequency measure offers a useful alternative to define composite survival outcomes, by relating the average event rate to the survival time. Although non-parametric methods have been proposed for two-sample comparisons, limited attention has been given to regression methods that directly address time-varying association effects in while-alive measures. We address this gap by developing a regression framework for exposure-weighted while-alive measures for composite survival outcomes that include a terminal component event. Our regression approach uses splines to model time-varying association between covariates and a generalized while-alive loss rate of all component events, and can be applied to both independent and clustered data. We derive the asymptotic properties of the regression estimator under both independent data and cluster-correlated data settings, and study the operating characteristics of our methods through simulations. Finally, we apply our regression method to analyze data two randomized clinical trials. The proposed methods are implemented in the \texttt{WAreg} R package.
\end{abstract}


\section{Introduction}
\label{sec:intro}

Composite endpoints are common in phase III clinical trials and consist of two or more component endpoints. 
For example, in cardiovascular disease research, a composite endpoint might combine time to death, myocardial infarction, stroke and hospitalization. The use of composite endpoints increases the efficiency of clinical trials by increasing the overall event rate, thereby improving the power to detect treatment effect \citep{freemantle2003composite}. 
Regulatory guidelines have underscored the importance of carefully constructing, analyzing, and reporting composite endpoints. For instance, the ICH-E9 guideline, \textit{Statistical Principles for Clinical Trials} \citep{european2020e9}, states: ``\emph{If a single primary variable cannot be selected from multiple measurements associated with the primary objective, another useful strategy is to integrate or combine the multiple measurements into a single or composite variable}.'' The U.S. Food and Drug Administration guideline, \textit{Multiple Endpoints in Clinical Trials} \citep{guidance2017multiple}, mentions that ``\emph{composite endpoints are often assessed as the time to first occurrence of any one of the components}'' while also noting that ``\emph{it may also be possible to analyze total endpoint events}.'' 

Traditional analyses of composite endpoints typically rely on time-to-first-event models, which implicitly assign equal clinical weight to fatal and non-fatal outcomes. When both recurrent and terminal events are present, various regression approaches have been proposed that differ in the primary target parameters and how death is handled. When estimating the treatment effect in a regression model, the strategies for handling death can be mapped to methods for handling intercurrent events under the ICH E9 estimands framework \citep{european2020e9}. A summary of representative regression models and their strategies for handling death is provided in Web Appendix Table 1. In particular, the while-alive approach has recently been highlighted in the ICH-E9(R1) Addendum as a viable strategy to handle death as an intercurrent event. Under this framework, the occurrence of events is evaluated relative to the duration of survival. By normalizing the event burden with respect to survival, while-alive estimands provide interpretable measures of the average event rate among patients while they are alive. Two particular forms of the estimand have been defined \citep{schmidli2023estimands}: the patient-weighted while-alive estimand, which is the expectation of the event-to-survival ratio across individuals \citep{ragni2024nonparametric}, and the exposure-weighted while-alive estimand, which is the ratio of the expected number of events to the restricted mean survival time \citep{mao2023nonparametric,wei2023properties}.

There are several previous efforts in studying while-alive measures. \citet{wei2023properties} established theoretical properties of exposure-weighted while-alive estimators under a gamma frailty model, \citet{mao2023nonparametric} developed a nonparametric estimator for the exposure-weighted while-alive estimand, and \citet{ragni2024nonparametric} derived the efficient influence function and constructed efficient nonparametric estimators for the patient-weighted while-alive estimand. While these contributions are useful for estimating treatment effects, they do not address association analysis under a regression framework. For regression analysis of while-alive measures, it is often necessary to address time-varying association effects. To accommodate time-varying effects, several strategies have been previously employed in survival analysis with a single endpoint. For example, kernel-based methods have been developed for Cox models \citep{cai2003local} to estimate time-varying hazard ratios. Spline-based methods approximate time-varying coefficients with flexible basis functions, and have been applied in the context of mean residual life regression \citep{sun2012mean} and restricted mean survival time regression \citep{zhong2022restricted}. Finally, landmark models \citep{van2008dynamic} provide a complementary approach by fitting survival models at prespecified landmark times, allowing dynamic prediction and assessment of time-varying effects in a practical fashion.

In this paper, we propose a new regression framework that incorporates time-varying coefficients into the while-alive measure for composite survival endpoints. We focus on the exposure-weighted while-alive estimand, operationalized through the generalized while-alive loss rate. We model the association between covariates and while-alive loss rate with spline functions, allowing effects to vary smoothly over follow-up time. 
We first address independent data, and then discuss cluster-correlated data under the working independence assumption. We establish the asymptotic properties of the proposed regression estimators and investigate their finite-sample performance using simulations. 
Finally, we illustrate the use of our methods by analyzing two clinical trials. To facilitate implementation, an \texttt{R} package \texttt{WAreg} (\url{https://github.com/fancy575/WAreg}) has been developed, along with a short tutorial in Web Appendix 15.

\section{Motivating data examples}\label{sec:data}

\begin{figure}[htbp!]
    \centering
    \includegraphics[width=0.7\linewidth]{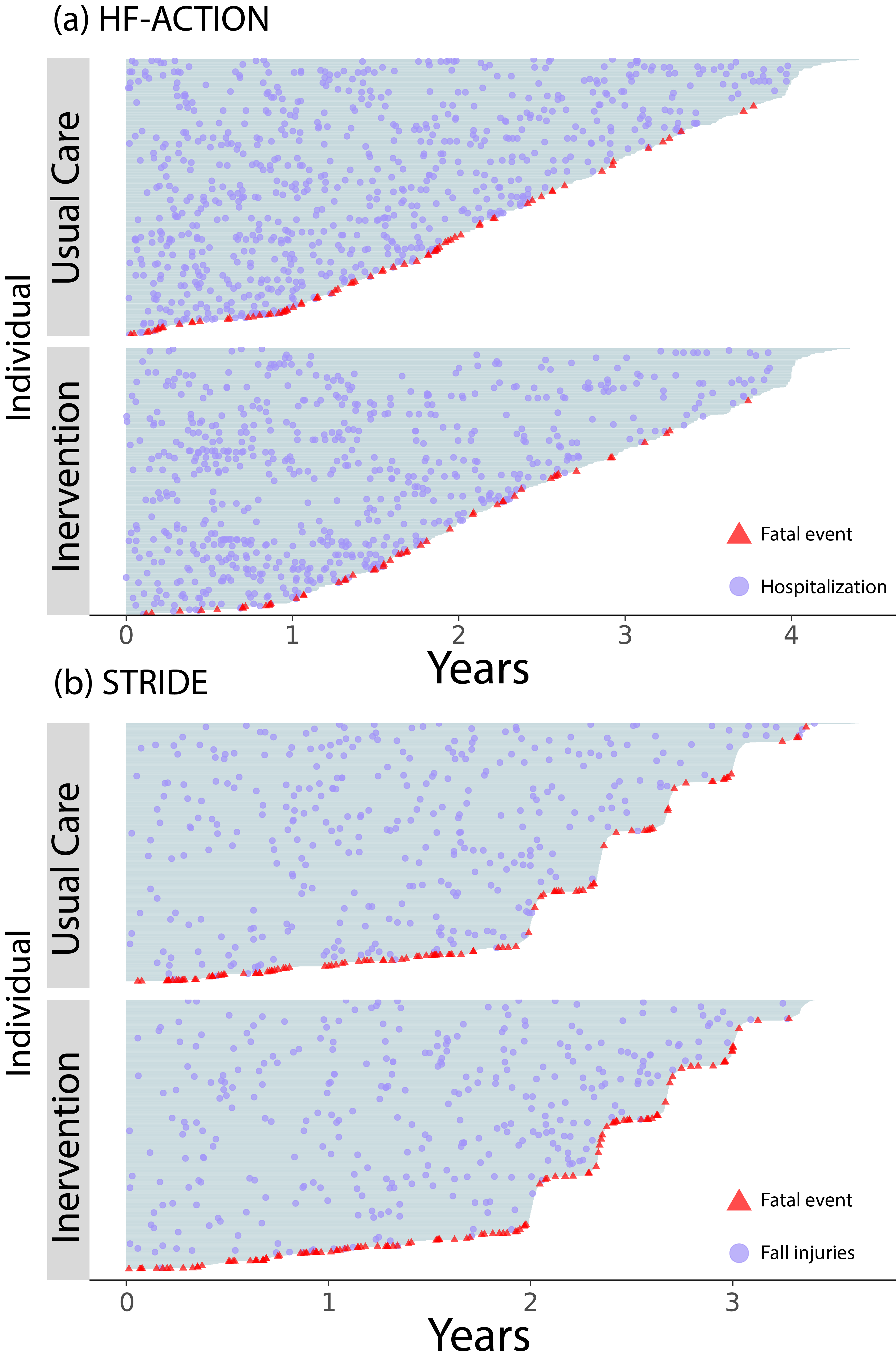}
    \caption{(a) Distribution of hospitalization and fatal events per subject by intervention group in the HF-ACTION trial. (b) Distribution of fall-related injuries and fatal events per subject by intervention group in the STRIDE trial. Each horizontal line shows one participant’s follow-up (blue ribbon) from randomization to the observed end of follow-up (death or censoring); purple circles mark recurrent hospitalizations/injuries and red triangles mark deaths. The staircase-like right edge occurs because many participants share common administrative or scheduled end times, yielding a visual alignment. Figures were generated using the \textit{reReg} R package \citep{chiou2023regression}.}
    \label{fig:event_plot}
\end{figure}

The first motivating example is the \textit{Heart Failure: A Controlled Trial Investigating Outcomes of Exercise Training} (HF-ACTION), a multicenter individually randomized trial. This study enrolled 2,331 patients with a median follow-up of 30 months to evaluate the effectiveness of exercise training in heart failure patients. Figure \ref{fig:event_plot}(a) shows the distribution of recurrent hospitalizations and mortality events for each subject, stratified by treatment group. The primary endpoint was a composite of all-cause mortality and all-cause hospitalization. Initial analysis using a Cox proportional hazards model for the time to the first event, adjusted for etiology, found statistically insignificant reductions in the exercise training group compared to usual care (hazard ratio = \(0.92\), with \(95\%\) Confidence interval  \(= (0.83, 1.03) \); p-value $= 0.14$) \citep{o2009efficacy}. However, this approach only considered the first occurrence of mortality or hospitalization, ignoring subsequent events and treating hospitalizations and deaths as equally important. To address these limitations, more advanced methods were applied to the HF-ACTION data. For example, \citet{mao2016semiparametric} considered a semiparametric regression model under the proportional mean framework to analyze weighted endpoints, but did not directly address the issue of differential follow-up time across patients. In their analysis, there was no statistically significant reduction in the mean frequencies of the weighted composite events at the 0.05 level, regardless of weights assigned to hospitalization and death. However, the p-values are smaller compared to that under the time to first event analysis.

A second example is the \textit{Strategies to Reduce Injuries and Develop Confidence in Elders} (STRIDE) trial, a pragmatic cluster randomized study evaluating the effectiveness of a multifactorial intervention for preventing fall-related injuries. A total of 86 primary care practices were randomly assigned to either the intervention or enhanced usual care (control) group, with 43 practices in each arm. There were 2,802 participants in the intervention group and 2,649 in the control group, with a maximum follow-up of 44 months. In the original analysis, a multistate survival model accounting for competing risks of death and clustering was used to evaluate time to first fall-related injury. The estimated hazard ratio was \(0.90\) (95\% Confidence Interval: \([0.83 - 0.99]\); p-value \(= 0.004\)). Figure \ref{fig:event_plot}(b) shows the distribution of recurrent fall injuries and mortality events for each subject, stratified by intervention group. This study requires consideration of both recurrent fall injuries and fatal events \citep{bhasin2020randomized} in clustered settings. In both examples, there is interest in exploring the impact of treatment and other risk factors on the composite of recurrent and terminal events through a single regression model, to provide useful summary measures. This motivates the while-alive regression methods, which are specifically designed to provide a framework for regression-based analysis of composite survival endpoints in both independent and cluster-correlated data settings.

\section{Methods}
\label{sec:method}

\subsection{While-alive measure}
Let \( D \) denote the time to the terminal event (death), and define \( N_D(t) = \mathbb{I}(D \leq t) \), where \( \mathbb{I}(\cdot) \) is the indicator function. With recurrent events, let \( N_k(t) \), \( k = 1, \dots, K \), denote the counting processes for \( K \) (\( K \geq 1 \)) distinct types of recurrent events. Since death is a terminal event, the counting processes for recurrent events are constrained by the survival time \( D \), and are given by $N_k(t) = N_k(D \wedge t) = \sum_{q=1}^{\infty} \mathbb{I}(T_{k,q} \leq D \wedge t)$, where \( T_{k,q} \) denotes the \( q\)th recurrence time of the \( k\)th event type, and \( D \wedge t = \min(D, t) \). This ensures that no recurrent events are observed after the terminal event. 
The complete event history up to time \( t \) is \( \mathcal{H}(t) \), which includes the times and types of all recurrent events up to \( t \), as well as whether the terminal event has occurred.

To characterize the cumulative events over time, we define \( \mathcal{L}(\mathcal{H})(dt) \geq 0 \), as the instantaneous loss rate at time \( t \). This definition satisfies: (C1) \( \mathcal{L}(\mathcal{H})(dt) = 0 \) for \( t \geq D \), ensuring that no loss is incurred after death; (C2) \( \mathcal{L}(\mathcal{H})(dt) \) depends only on the event history up to time \( t \). Condition (C1) aligns with the nature of recurrent event analysis. Condition (C2) is natural and exhibits the dependence of the loss rate on the observed history up to \( t \) instead of the future. We follow \citet{mao2023nonparametric} to define the generalized while-alive loss rate over the interval \([0, t]\) as:
\begin{equation}
    l(\mathcal{H})(t) = \frac{\mathbb{E}\left[ \mathcal{L}(\mathcal{H})(t) \right]}{\mathbb{E}[D \wedge t]}, \label{while-alive}
\end{equation}
where \(\mathcal{L}(\mathcal{H})(t) = \int_{0}^{t} \mathcal{L}(\mathcal{H})du \), and \(\mathcal{L}(\mathcal{H})(du) = \sum_{k=1}^{K}w_kdN_k(u) + w_DdN_D(u) \) . In this definition, \( w_k \) is the weight assigned to the $k$th type of recurrent event, \( w_D \) is the weight assigned to the terminal event, and \( \mathbb{E}[D \wedge t] \) is the restricted mean survival time (RMST). In definition \eqref{while-alive}, the numerator quantifies the expected cumulative weighted events up to time \( t \). The denominator, \( \mathbb{E}[D \wedge t] \), provides a time-at-risk adjustment by capturing the expected duration of survival over \([0, t]\). This definition accounts for both recurrent and terminal events as component endpoints, and adjusts the loss rate by the duration of survival. Particularly, \eqref{while-alive} corresponds to the exposure-weighted while-alive loss rate \citep{wei2023properties}, which quantifies the average burden per unit time alive across the cohort. In this formulation, individuals who survive longer contribute more exposure time and therefore receive greater weight in the overall average. This is to be differentiated from the patient-weighted while-alive loss rate \citep{ragni2024nonparametric}, defined as $l^{*}(\mathcal{H})(t) = \mathbb{E}\{{\mathcal{L}(\mathcal{H})(t)}/(D\wedge t)\}$, which averages each patient's event-to-survival ratio and thus reflects the typical patient's while-alive loss rate. The two quantities are linked by
\[
l(\mathcal{H})(t) = \mathbb{E}\left\{\frac{D\wedge t}{\mathbb{E}[D\wedge t]} \times \frac{\mathcal{L}(\mathcal{H})(t)}{D\wedge t}\right\},
\]
indicating that $l(\mathcal{H})(t)$ equals the survival-time weighted average of the patient-level while-alive measure. Finally, in the special case where $w_k=0$ for all $k$ and $w_D=1$, \eqref{while-alive} becomes the average hazard of the terminal event in \citet{uno2023ratio} and \citet{uno2024regression}.

\subsection{While-alive generalized linear model}
\label{sec: wl_reg_ind}
We denote \( \bm{Z}_i \) as a \( p \)-dimensional vector of bounded covariates for individual \(i \in \{1,\dots, n\} \). By convention, the first element may be set to 1 to include an intercept. The while-alive loss rate at truncation time \( t \), conditional on \( \bm{Z}_i \), is denoted as \( l(\mathcal{H}_i)(t \mid \bm{Z}_i) \), which we model as
\begin{equation}
\eta \{ l(\mathcal{H}_i)(t \mid \bm{Z}_i) \} = \bm{\beta}(t)^\top \bm{Z}_i.\label{while-alive-model}
\end{equation}
Throughout, we use the term truncation time to denote the evaluation horizon \(t\) at which the while-alive loss rate is defined and modeled.Here $\mathcal H_i$ denotes the filtration generated by the event history up to time $t$ (recurrent counts and aliveness/terminal event). In \eqref{while-alive-model}, \( \bm{\beta}(t)=\{\beta_1(t),\ldots,\beta_p(t)\}^\top \) is a \( p \)-dimensional vector of time-varying parameters, and \( \eta(\cdot) \) is a specified, increasing, and differentiable link function. The choice of \( \eta(\cdot) \) determines the interpretation of the regression coefficients. For example, with a binary treatment variable and a log-link function \( \eta(\cdot) = \log(\cdot) \), the exponentiated coefficient represents the ratio of while-alive loss rates between treatment groups.
The vector $\bm\beta(t)$ indexes covariate effects on the exposure-weighted while-alive loss rate at time $t$. Of note, a time-varying $\bm\beta(t)$ does not by itself imply that the underlying treatment effect on the hazard of death or recurrent events is time-varying. Because the while-alive rate conditions on being alive, both its numerator and denominator evolve with the survivor cohort, so that even constant hazard-scale effects can translate into a time-varying contrast on the while-alive scale. We further illustrate this point in Web Appendix A.2. Furthermore, we view $\beta_j(t)$ as a local or instantaneous covariate effect, in the sense that it captures the contribution of the $j$th covariate to the while-alive loss rate at time $t$. Because clinical decision making is rarely based on an infinitesimal moment, one can use the window-averaged summaries of the local effect curve. 
For example, for an interval $[t_a,t_b]$ , we define
\[
\overline{\beta}_j(t_a,t_b)=\frac{\int_{t_a}^{t_b}\beta_j(u)\,\omega(u)\,du}{\int_{t_a}^{t_b}\omega(u)\,du},
\]
as the time-averaged effect of the $j$th covariate, with weight $\omega(u)$ reflecting the effective information at time $u$ (for example, the probability of being alive and uncensored at $u$; in the absence of such adjustment we set $\omega\equiv 1$). Under the log link, $\exp\{\overline{\beta}_j(t_a,t_b)\}$ is the weighted geometric mean of the instantaneous rate ratios $\exp\{\beta_j(u)\}$ over $ u \in [t_a,t_b]$. In practice, the time windows can be selected based on the scientific question or clinical judgment (e.g., 0-3 months immediately after randomization, 3-6 months during early recovery).

We let \( C_i \) denote the censoring time due to either study termination or loss to follow-up. We assume covariate-dependent censoring such that \( C_i \perp \mathcal{H}_i \mid \bm{Z}_i \) and positivity such that $ \inf_{u\le \tau} G(u\mid \bm{Z}_i) := \inf_{u\le \tau} P(C_i \ge u \mid \bm{Z}_i) \ge \epsilon > 0\ $ amost surely. Define \( U_i = D_i \wedge C_i \) as the observed time and \( \Delta_i = \mathbb{I}(D_i \leq C_i) \) as the indicator for whether the terminal event occurred. We define the observed data for subject \( i \) as \( \mathcal{O}_i = \{ \widetilde{\mathcal{H}}_i, U_i, \Delta_i, \bm{Z}_i \} \), where \( \widetilde{\mathcal{H}}_i \) denotes the observable portion of the event history up to time \( U_i \). We further assume that $\bm{\beta}(t)$ is a vector of continuous functions on $[0,t_V]$, parameterized via a finite-dimensional basis in $t$. Similar to \citet{zhong2022restricted,chen2023clustered}, we express the time-varying coefficients using
\[
\beta_j(t)\;=\;\sum_{r=1}^{R_j}\gamma_{jr}\,J_{jr}(t),\qquad 
\bm J_j(t)=\{J_{j1}(t),\ldots,J_{jR_j}(t)\}^\top,
\]
where $\bm J_j(t)$ denotes the basis functions (e.g., piecewise-constant segments or spline functions with a chosen knot set) for covariate $j$. The stacked design vector can then be written as \(\widetilde{\bm Z}_i(t)\;=\;\big(Z_{i1}\bm J_1(t)^\top,\; \ldots,\; Z_{ip}\bm J_p(t)^\top\big)^\top\), and the parameter vector as 
\(\widetilde{\bm\beta}\;=\;\big(\bm\gamma_1^\top,\ldots,\bm\gamma_p^\top\big)^\top, \bm\gamma_j=\big(\gamma_{j1},\ldots,\gamma_{jR_j}\big)^\top\).
For convenience, we adopt a common basis across covariates, i.e., $R_j\equiv R$ and $\bm J_j(t)\equiv \bm J(t)=\{J_1(t),\ldots,J_R(t)\}^\top$ with the same knot locations, in which case the representation simplifies to $\bm{\beta}(t)=\sum_{r=1}^R \bm{\gamma}_r\,J_r(t)$, $\widetilde{\bm Z}_i(t)=\bm Z_i\otimes \bm J(t)$, $\widetilde{\bm\beta}=\big(\bm\gamma_1^\top,\ldots,\bm\gamma_R^\top\big)^\top$.

To estimate \( \widetilde{\bm{\beta}} \), we create \( V \) batches of data at time points \( t_v \), \( v = 1, \dots, V \), where the component \(\{t_1,\dots, t_V\} \) are sorted in ascending order within the interval \( (0, \tau) \), and \( \tau \) denotes the maximum follow-up time. The stacked estimating function is given by
\begin{align}
    \bm{U}_n(\widetilde{\bm{\beta}}) &= n^{-1} \sum_{i=1}^{n} \sum_{v=1}^{V} \frac{\mathbb{I}(U_i \leq t_v) \Delta_i + \mathbb{I}(U_i > t_v)}{G(U_i \wedge t_v \mid \bm{Z}_i)} \widetilde{\bm{Z}}_i(t_v) \Bigg[ \widetilde{\mathcal{L}}(\mathcal{H}_i)(t_v) - \eta^{-1}\{\widetilde{\bm{\beta}}^\top \widetilde{\bm{Z}}_i(t_v)\}(U_i \wedge t_v) \Bigg], \label{estimating-equation}
\end{align}
where \( G(\cdot) \) is the censoring survival function given covariates, and $\widetilde{\mathcal{L}}(\mathcal{H}_i)(t) = \int_{0}^{t} \widetilde{\mathcal{L}}(\mathcal{H}_i)(du)$,  $\widetilde{\mathcal{L}}(\mathcal{H}_i)(du) = \sum_{k=1}^{K} w_k dN_{i,k}(U_i \wedge u) + w_D d\widetilde{N}_{i,D}(u)$, with 
\(d\widetilde{N}_{i,D}(u) = \mathbb{I}(U_i = u,\Delta_i=1)\). 
The weight in \eqref{estimating-equation}, \(\frac{\mathbb{I}(U_i \le t_v)\Delta_i + \mathbb{I}(U_i > t_v)}{G(U_i\wedge t_v \mid \bm{Z}_i)}\) is the inverse probability of remaining uncensored up to the relevant exposure time \(U_i\wedge t_v\), conditional on covariates. It restores the unbiasedness of the observed-data estimating function relative to its full-data counterpart under covariate-dependent censoring. Operationally, the numerator \(\mathbb{I}(U_i\le t_v)\Delta_i+\mathbb{I}(U_i>t_v)\) partitions the sample into those who die before \(t_v\) (contribute at their death time \(D_i\)) and those who live beyond \(t_v\) (contribute at \(t_v\)); those censored before \(t_v\) contribute zero, and the denominator reweights the contribution by the inverse probability of being uncensored at the truncated follow-up time $U_i\wedge t_v$. The locations of the knots defining the basis functions \( \bm{J}(t) \) need not coincide with the stacking points \( \{t_1, \dots, t_V\} \) defining the estimating equations. 
However, for simplicity and convenience, the stacking points may often be chosen to coincide with the knot locations.

Assuming that \(\{\mathcal{O}_i,\ldots,\mathcal{O}_n\}\)
are independent and identically distributed, we can show that, as \( n \to \infty \), \( \bm{U}_n(\widetilde{\bm{\beta}}) \) converges uniformly to a monotone limiting function \( \bm{u}(\widetilde{\bm{\beta}}) = E [ \widetilde{\bm{Z}}_i \sum_{v=1}^{V} ( \mathcal{L}_i(\mathcal{H})(t_v) - \eta^{-1}\{\widetilde{\bm{\beta}}^\top \widetilde{\bm{Z}}_i(t_v)\}(D_i \wedge t_v) ) ]\). 
Let \( \widehat{\bm{\beta}} \) and \( \widetilde{\bm{\beta}}^0 \) denote the solutions to \( \bm{U}_n(\widetilde{\bm{\beta}}) = 0 \) and \( \bm{u}(\widetilde{\bm{\beta}}) = 0 \), respectively. In Web Appendix A.3, we prove that \( \widehat{\bm{\beta}} \stackrel{p}{\to} \widetilde{\bm{\beta}}^0 \) as \( n \to \infty \), under correct specification of the while-alive regression model and \( G(\cdot) \). 
Finally, the censoring survival function \(G(\cdot\mid\bm{Z}_i)   \) can be estimated using familiar survival analysis tools, such as the Cox proportional hazards model:
\[ \lambda^C(t\mid\bm{Z}_i) = \lambda_0^C(t) \exp(\widetilde{\bm{\theta}}^\top \bm{Z}_i),   \]
and thus \(G(t\mid\bm{Z}_i) = \exp \{-\int_{0}^{t} \lambda_0^C(u) \exp(\widetilde{\bm{\theta}}^\top \bm{Z}_i)  \} \). Under completely independent censoring, a Kaplan-Meier estimator can be used to estimate $G(t \mid\bm{Z}_i)=G(t)$ without covariates.

\begin{remark} \label{rmk:specific_t}
\emph{If only the association effect at a specific time \( t \) is of interest, the estimating equation \eqref{estimating-equation} can be simplified by setting \( V = 1 \) without stacking. That is, we can solve the time-specific coefficient $\bm{\beta}$ from the unstacked estimating equation}
\begin{align*}
    n^{-1} \sum_{i=1}^{n} \frac{\mathbb{I}(U_i \leq t) + \mathbb{I}(U_i > t) }{ G(U_i\wedge t \mid \bm{Z}_i)} \bm{Z}_i\left[\widetilde{\mathcal{L}}(\mathcal{H}_i)(t) - \eta^{-1} \left\{\bm{\beta}^{\top} \bm{Z}_i  \right\}(U_i\wedge t)  \right]=0.
\end{align*}
\emph{This approach represents a localized, or ``landmark'' version of the while-alive regression model at the specified time horizon $t$, instead of borrowing information across multiple landmarks \(\{t_1,\ldots,t_V\}\) via spline smoothing.}
\end{remark}

\subsection{Asymptotic properties} \label{sec:iid_asymptotic}
We characterize the asymptotic properties of the estimator for the regression parameters \(\widehat{\bm{\beta}}\) with independent data under covariate-dependent censoring. The properties with completely independent censoring are provided in the Web Appendix A.5 as a special case. We make the following regularity conditions: 
\begin{enumerate} [(i)]
    \vspace{-0.4cm}\item \( P(T_{i,k} \geq t ) > 0  \) and \(P(D_i \geq t) > 0 \) for \(t \in (0,\tau)\). \label{iid_c1}
    \vspace{-0.4cm} \item The covariate \( Z_{ij} \) is bounded 
    for $i=1,\dots, n$, $j=1,\dots, p$. \label{iid_c2}
    \vspace{-0.4cm} \item \(0 < G(t\mid \bm{Z}_i) \leq 1 \) is absolute continuous for \(t\in (0,\tau]\), and the cumulative hazard function for censoring \(\Lambda^C(t|\bm{Z}_i) = \int_{0}^{t} \lambda^C(u\mid\bm{Z}_i)du \) is absolute continuous for \(t\in (0,\tau]\) \label{iid_c3}
    \vspace{-0.4cm} \item \(\bm{\Omega}(\widetilde{\bm{\beta}}) = \mathbb{E}\left\{\widetilde{\bm{Z}}_i^{\otimes 2}\sum_{v=1}^{V} \left[ \dot{\eta}^{-1} ( \bm{\widetilde{\beta}}^\top \bm{\widetilde{Z}}_i) (D \wedge t_v)  \right]   \right\} \) is positive definite for any $t \in (0,\tau]$, where \(\dot{\eta}^{-1}(.) \) is the derivative of the \(\eta^{-1}(.)\), and \( a^{\otimes 0} = 1 \), \( a^{\otimes 1} = a \), \( a^{\otimes 2} = a a^\top \) for vector \(a\). \label{iid_c4}
    \vspace{-0.4cm} \item The matrix  
    \[\bm{I}_\theta = \mathbb{E}\left\{\int_{0}^{t_v} \left[\left\{\frac{\bm{s}^{(2)}(u,\widetilde{\bm{\theta}})}{s^{(0)}(u,\widetilde{\bm{\theta}})} - \left( \frac{\bm{s}^{(1)}(u,\widetilde{\bm{\theta}})}{s^{(0)}(u,\widetilde{\bm{\theta}})} \right)^{\otimes 2} \right\} s^{(0)}(u,\widetilde{\bm{\theta}})  d\Lambda_{0}^C(u) \right] \right\},\] is positive definite for $v=1,\ldots,V, $
    where \( \bm{s}^{(d)}(u; \widetilde{\bm{\theta}}) = \mathbb{E}\left\{ n^{-1} \sum_{i=1}^{n} \mathbb{I}(U_i \geq u) \bm{Z}_i^{\otimes d} \exp(\widetilde{\bm{\theta}}^\top \bm{Z}_i) \right\} \) for \( d = 0, 1, 2 \), and \( s^{(0)}(u; \widetilde{\bm{\theta}}) \) is bounded away from zero, and \( d\Lambda_0^C(t) = \lambda_0^C(t) \) is the baseline hazard for censoring time. \label{iid_c5}
\end{enumerate}
\begin{theorem} \label{thm:assym_iid}
    Under the above regularity conditions, \(\sqrt{n} (\widehat{\bm{\beta}} - \widetilde{\bm{\beta}}^0 ) \) converge weakly to zero-mean Gaussian distribution with variance $\bm{V} = \bm{\Omega}(\widetilde{\bm{\beta}}^0)^{-1} \bm{\Sigma}(\widetilde{\bm{\beta}}^0) \left\{\bm{\Omega}(\widetilde{\bm{\beta}}^0)^{-1} \right\}^{\top}$, where $\bm{\Sigma}(\widetilde{\bm{\beta}}^0) = E\{ \bm{\phi}_i(\widetilde{\bm{\beta}}^0)^{\otimes 2} \}$, and 
\begin{align*}
\phi_i(\widetilde{\bm{\beta}}^0) & =  \sum_{v=1}^{V}
\left[\varpi_i(t_v;\widetilde{\bm\beta}^0) + \kappa_\theta\left(t_v;\widetilde{\bm\beta}^0,\bm\theta,\Lambda_0^C\right)\,\xi_i(\bm\theta) + \int_{0}^{t_v} \zeta_i(u,\bm\theta)\,d\kappa_\Lambda\left(u,t_v;\widetilde{\bm\beta}^0,\bm\theta,\Lambda_0^C\right)
\right],
\end{align*}
\begin{align*}
\varpi_i(t_v;\widetilde{\bm\beta}^0) &= \frac{\mathbb{I} \left(U_i \le t_v\right)\Delta_i + \mathbb{I} \left(U_i > t_v\right)} {G\!\left(U_i \wedge t_v \mid \bm Z_i\right)} \,\widetilde{\bm Z}_i(t_v)\,
\Big[ \widetilde{L}_i(\mathcal H_i)(t_v) - \eta^{-1} \big\{ (\widetilde{\bm\beta}^0)^\top \widetilde{\bm Z}_i(t_v) \big\} \left(U_i \wedge t_v\right)\Big] ,\\
\xi_i(\bm\theta) &= \int_{0}^{\tau} \left\{ \bm Z_i - \overline{\bm z}(r,\bm\theta) \right\} \, dM_i^C(r),\\
\zeta_i(u,\bm\theta) &= \int_{0}^{u} \frac{1}{s^{(0)}(s,\bm\theta)} \, dM_i^C(s),\\
\kappa_\theta\big(t_v;\widetilde{\bm\beta}^0,\bm\theta,\Lambda_0^C\big) &= \lim_{n\to\infty} \frac{1}{n} \sum_{i=1}^{n} \
\frac{\mathbb{I} \left(U_i \le t_v\right)\Delta_i + \mathbb{I} \left(U_i > t_v\right)} {G\!\left(U_i \wedge t_v \mid \bm Z_i\right)} \,
\widetilde{\bm Z}_i(t_v)\,
\Big[ \widetilde{L}_i(\mathcal H_i)(t_v) - \eta^{-1} \big\{ (\widetilde{\bm\beta}^0)^\top \widetilde{\bm Z}_i(t_v) \big\} \left(U_i \wedge t_v\right)\Big]\\
&\qquad \times \exp\!\left(\bm\theta^\top \bm Z_i\right)
\Bigg[ - \left\{ \int_{0}^{t_v} \frac{\bm s^{(1)}(u,\bm\theta)}{s^{(0)}(u,\bm\theta)} \, d\Lambda_0^C(u) \right\}^{\!\top}  + \Lambda_0^C(t_v)\, \bm Z_i^\top \Bigg]\bm I_\theta^{-1},\\
\kappa_\Lambda\big(u,t_v;\widetilde{\bm\beta}^0,\bm\theta,\Lambda_0^C\big) &=\lim_{n\to\infty} \frac{1}{n} \sum_{i=1}^{n}
\frac{\mathbb{I} \left(U_i \le t_v\right)\Delta_i + \mathbb{I} \left(U_i > t_v\right)} {G\!\left(U_i \wedge t_v \mid \bm Z_i\right)} \,
\exp\!\left(\bm\theta^\top \bm Z_i\right)\,
\widetilde{\bm Z}_i(t_v)\\
&\times \Big[ \widetilde{L}_i(\mathcal H_i)(t_v) - \eta^{-1} \big\{ (\widetilde{\bm\beta}^0)^\top \widetilde{\bm Z}_i(t_v) \big\} \left(U_i \wedge t_v\right)\Big]\,
\mathbb{I}(u\le t_v),
\end{align*}
    where \(M_i^C(t) = \mathbb{I}(U_i \leq t, \Delta_i=0) - \int_{0}^{t} \mathbb{I}(U_i \geq u) d\Lambda_i^C(u) \) is the martingale for the censoring process. 
\end{theorem}
The proof of Theorem \ref{thm:assym_iid} is provided in Web Appendix A.4. In particular, the asymptotic variance matrix $\bm{V}$ explicitly accounts for the variability in estimating the censoring survival function. Theorem \ref{thm:assym_iid} motivates a consistent sandwich variance estimator for $\widehat{\bm\beta}$, which can be used to construct 95\% pointwise confidence interval for the regression coefficient, as well as prediction intervals for the while-alive rate at a given time point (Web Appendix A.4). In particular, the while-alive regression model generalizes the average hazard (AH) regression proposed by \citet{uno2024regression} from a single terminal event to composite endpoints, and from a time-fixed association analysis to time-varying association analysis.
Theorem \ref{thm:assym_iid} also motivates a global test for the \(j\)th covariate effect based on the null hypothesis \(\widetilde{\bm{\beta}}_j = \{ \gamma_{j1}, \dots, \gamma_{jR_j} \} = \bm{0}\). The corresponding Wald statistic is $\chi_{R_j-1}^2 = \widehat{\bm{\beta}}_j^\top \widehat{\bm{\Sigma}}(\widehat{\bm{\beta}})_j^{-1} \widehat{\bm{\beta}}_j$, where \(\widehat{\bm{\beta}}_j\) is the estimator of \(\widetilde{\bm{\beta}}_j\), and \(\widehat{\bm{\Sigma}}(\widehat{\bm{\beta}})_j\) denotes the estimated \(R \times R\) submatrix of the full asymptotic covariance matrix \(\bm{\Sigma}(\widetilde{\bm{\beta}})\). Under the null hypothesis, \(\chi_{R-1}^2\) follows a Chi-squared distribution with \(R_j - 1\) degrees of freedom. The asymptotic results under the completely independent censoring scenario with Kaplan-Meier censoring estimator are provided in Web Appendix A.5. 
\begin{remark}
\emph{Our regression framework extends the nonparametric estimator of \citet{mao2023nonparametric}, which focused only on estimating arm-specific exposure-weighted while-alive loss rate under arm-specific independent censoring. In the special case of a binary treatment $A \in \{0,1\}$, the while-alive regression model specified by $\bm Z_i = (1, A_i)^\top$ in \eqref{while-alive-model} is saturated and can be used to predict arm-specific while-alive loss rate; this is asymptotically equivalent to the nonparametric estimator of \citet{mao2023nonparametric}. Additional details are provided in Web Appendix A.6.}
\end{remark}

\section{Extension to cluster-correlated data} \label{sec:clustered_extension}

We next describe a marginal while-alive regression model as an extension to handle cluster-correlated data. Suppose there are \(M\) independent clusters, where each cluster \(i\) have \(N_i\) subjects for \(i=1,\dots, M\). Denote \(T_{ijk}\) as the $k${th} type of recurrent event, \(D_{ij}\) as the fatal event, \(C_{ij}\) as the censoring time for subject \(j\) in cluster \(i\). Then for subject \(j\) in cluster \(i\), we denote \(U_{ij} = \min(D_{ij},C_{ij})\), \(\Delta_{ij} = \mathbb{I}(D_{ij} \leq C_{ij})\). The covariates \(\bm{Z}_{ij}\) can include both cluster-level and individual-level covariates. The complete history of event for subject \(j\) in cluster \(j\) is denoted as \(\mathcal{H}_{ij}\), thus, we assume covariate-dependent censoring such that \(\bm{C}_i \perp \bm{\mathcal{H}}_i | \bm{Z}_i ,\bm{N}_i\), where \(\bm{\mathcal{H}}_i = (\mathcal{H}_{i1},\dots, \mathcal{H}_{iN_i}  )^\top \), \(\bm{C}_i = (C_{i1},\ldots, C_{iN_i})^\top \), and  \(\bm{Z}_i = (\bm{Z}_{i1},\ldots, \bm{Z}_{iN_i})^\top \) be the collection of the observed information within cluster \(i\). 
With clustered data, individuals within the cluster can be correlated. Let the observed data for cluster \( i \) be denoted as \( \mathcal{O}_i = \{ \bm{\widetilde{\mathcal{H}}}_i, \bm{U}_i, \bm{\Delta}_i, \bm{Z}_i \} \), where \( \bm{\widetilde{\mathcal{H}}}_i = (\widetilde{\mathcal{H}}_{i1}, \dots, \widetilde{\mathcal{H}}_{iN_i}) \), and \(\widetilde{\mathcal{H}}_{ij}\) denotes the observed event histories up to \(U_{ij}\) for cluster \(j\) in cluster \( i=1,\dots, m \). We assume the cluster-level observations \( \mathcal{O}_i\)'s 
are identically independent distributed across clusters. The marginal while-alive regression for clustered data is defined as:
\begin{equation}
    \eta\{l(\mathcal{H}_{ij})(t \mid \bm{Z}_{ij})\} = \bm{\beta}(t)^\top \bm{Z}_{ij}, \label{wl_reg_cluster}
\end{equation}
where the while-alive loss rate \(l(\mathcal{H}_{ij}) (t)\) is defined analogously to \eqref{while-alive} based on the individual event history up to \(t\). 
Similar to \eqref{estimating-equation}, we propose the following unbiased estimating equation under the working independence assumption to estimate \(\bm{\widetilde{\beta}}\) as
\begin{align*}
    \bm{U}_n^{*}(\widetilde{\bm{\beta}}) &= M^{-1} \sum_{i=1}^{M} \sum_{j=1}^{N_i} \sum_{v=1}^{V} \frac{\mathbb{I}(U_{ij} \leq t_v) \Delta_{ij} + \mathbb{I}(U_{ij} > t_v)}{G(U_{ij} \wedge t_v |\bm{Z}_{ij})} \widetilde{\bm{Z}}_{ij} \left[ \widetilde{\mathcal{L}}(\mathcal{H}_{ij})(t_v) - \eta^{-1}\{\widetilde{\bm{\beta}}^\top \widetilde{\bm{Z}}_{ij}(t_v)\}(U_{ij} \wedge t_v) \right]. \label{estimating-equation_cluster}
\end{align*}

In the above estimating equation, we can similarly estimate the censoring survival function \(G(t|\bm{Z}_{ij}) = P(C_{ij} \geq t|\bm{Z}_{ij})\) via the marginal proportional hazard model \citep{chen2023clustered}, given by $\lambda_{ij}^C(t) = \lambda_0^C(t) \exp(\bm{\theta}_C^\top \bm{Z}_{ij})$. We then denote \(\widehat{G}(t|\bm{Z}_{ij})\) as the estimator of \(G(t|\bm{Z}_{ij})\). If censoring is considered independent of covariates \(\bm{Z}_{ij}\), then the estimation of \(G(t)\) can simply proceed with the Kaplan-Meier estimator. We then establish the asymptotic results in Web Appendix A.7. 
With clustered-correlated data, besides accounting for the variability introduced by estimating the censoring survival function, the ``meat'' of the sandwich variance is additionally constructed in two steps. First, we compute the individual contributions as in the independent-data setting. Then, these are aggregated at the cluster level by summing over \(j = 1, \dots, N_i\). Since clusters are assumed to be independent, the variance is constructed based on these independent cluster-level sums rather than the individual-level scores, which is a major difference from the asymptotic results in Theorem \ref{thm:assym_iid}. A consistent variance estimator is obtained by replacing \(\widetilde{\bm{\beta}}\) with its estimator \(\bm{\overline{\beta}}\), and \(G(t \mid \bm{Z}_{ij})\) with its estimator \(\widehat{G}(t \mid \bm{Z}_{ij})\) (Web Appendix A.7). 
The asymptotic results under the completely independent censoring scenario with Kaplan-Meier censoring estimator are provided in Web Appendix A.8. Similar to the case with independent data, by setting \( w_k = 0 \) for all \( k = 1, \dots, K \) and focusing on the parameter at a fixed time point \( t \), the marginal while-alive regression reduces to an average hazard regression for cluster-correlated data (Web Appendix A.9). 

\section{Practical considerations}
\label{cv_selection}

In our weighted estimation equation, smoothing parameters, including the degree of the spline basis function \(d\), and the number and location of knots \(R\), need to be chosen to balance flexibility and overfitting. It may often be impractical to optimize all components jointly. To simplify the process, we consider combinations of \(d\) and \(R\), where knots are placed at equally spaced quantiles of the observed time scale. The optimal pair \((d,R)\) is then selected using a data-driven criterion based on predictive performance tailored to while-alive loss rate estimation. To determine the optimal number of knots \((d,R)\), we adopt the B-fold cross-validation by minimizing the prediction error within the held-out validation data. 

With independent data, we divide the data into \(B\) approximately equal-sized subsets, denoted \(\mathcal{D}_1, \dots, \mathcal{D}_B\). For each fold \(b = 1, \dots, B\), we fit the while-alive regression model with \(R\) knots, excluding the subset \(\mathcal{D}_b\). Let \(\widehat{\bm{\beta}}^{(-b)}\) denote the estimator obtained from this training data (excluding fold \(b\)). We then evaluate the prediction error \(PE_b(R)\) using the held-out data \(\mathcal{D}_b\). Repeating this procedure over all folds gives the overall prediction error
\(
PE(R) = \sum_{b=1}^{B} PE_b(R),
\)
and the optimal number of knots \(\widehat{R}\) is selected as the value of \(R\) that minimizes \(PE(R)\). For while-alive regression, we define the fold-specific prediction error as
\[
PE_b(R) = \sum_{i \in \mathcal{D}_b} \int_{0}^{\tau} \left[ \frac{\mathbb{I}(U_i \leq s)\Delta_i + \mathbb{I}(U_i > s)}{\widehat{G}^{(-b)}(U_i \wedge s \mid \bm{Z}_i)} \left\{ \widetilde{\mathcal{L}}(\mathcal{H}_i)(s) - \eta^{-1}(\widehat{\bm{\beta}}^{(-b)}(s)^\top \widetilde{\bm{Z}}_i(s))(U_i \wedge s) \right\} \right]^2 ds,
\]
where \( \widehat{G}^{-b}(t \mid \bm{Z}_i) \) is the estimated censoring survival probability, obtained by excluding \( \mathcal{D}_b \).

With clustered data, we modify the cross-validation approach by partitioning the data at the cluster level. Specifically, we divide the \(M\) clusters into \(B\) non-overlapping subsets \(\mathcal{C}_1, \dots, \mathcal{C}_B\), each containing one or more clusters. For each fold \(b = 1, \dots, B\), we fit the model using all clusters except those in \(\mathcal{C}_b\), obtaining the estimator \(\overline{\bm{\beta}}^{(-b)}\) and compute the prediction error \(PE_b(R)\) using only the data from the held-out clusters. The total prediction error is then
\(
PE(R) = \sum_{b=1}^{B} PE_b(R),
\)
and the optimal number of knots is selected to minimize this sum. Generalizing the definition from the independent data case, the \(PE_b(R)\) for clustered data is defined as
\[
PE_b(R) = \sum_{i \in \mathcal{C}_b} \sum_{j=1}^{N_i} \int_{0}^{\tau} \left[ \frac{\mathbb{I}(U_{ij} \leq s)\Delta_{ij} + \mathbb{I}(U_{ij} > s)}{\widehat{G}^{-b}(U_{ij} \wedge s \mid \bm{Z}_{ij})} \left\{ \widetilde{\mathcal{L}}(\mathcal{H}_{ij})(s) - \eta^{-1}(\overline{\bm{\beta}}^{(-b)}(s)^\top \widetilde{\bm{Z}}_{ij}(s))(U_{ij} \wedge s) \right\} \right]^2 ds.
\]
In practice, the integral over time in \(PE_b(R)\) can be computed numerically. Taking the independent data as an example, we can define
\[f_i(s) = \left[ \frac{\mathbb{I}(U_{i} \leq s)\Delta_{i} + \mathbb{I}(U_{i} > s)}{G^{(-b)}(U_{i} \wedge s \mid \bm{Z}_{i})} \left\{ \widetilde{\mathcal{L}}(\mathcal{H}_{i})(s) - \eta^{-1}(\widehat{\bm{\beta}}^{(-b)}(s)^\top \widetilde{\bm{Z}}_{i}(s))(U_{i} \wedge s) \right\} \right]^2.\]
We discretize the time interval \([0,\tau]\) into a fine grid of points \(0=s_1 < s_2 ,\dots, s_{T-1} < s_T= \tau\), and approximate the integral using trapezoidal rule with $\int_{0}^{\tau} f_i(s)ds  \approx \sum_{t=1}^{T-1} \frac{s_{t+1} - s_{t}}{2} \left[f_i(s_t) + f(s_{t+1})\right]$. 

\section{Simulation studies}
\label{simu_RCT}

We carry out simulations to examine the finite-sample performance of the proposed regression methods. Focusing on independent data, we follow the approach in \citet{wei2023properties} and generate two types of recurrent events ($k=1,2$) and a fatal event ($k=D$) from a joint frailty model with $n=2000$ individuals \citep{toenges2021comparison}. Each individual has two baseline covariates $Z_{i1}\sim \mathrm{Bernoulli}(0.5)$ and $Z_{i2}\sim \mathcal N(0,1)$, together with a multiplicative frailty $W_i\sim \mathrm{Gamma}(\text{shape}=4.5,\text{rate}=4.5)$ shared across all event types. Conditional on $(\bm Z_i,W_i)$ and being at risk, the \(k^{}\)th recurrent event for individual \(i\) follows a nonhomogeneous Poisson distribution with intensity $\mu_k(t \mid \bm Z_i,D_i\ge t) = \mu_{0k}(t)\,W_i \exp(\bm\alpha_k^\top \bm Z_i)$, $k=1,2$, 
where \(\mu_{0k} (t)\) is the baseline intensity for event type \(k\). The fatal event \(D_i\) has a hazard function $\lambda_D(d\mid \bm Z_i) = \lambda_{0D}(d)\,W_i \exp(\bm\alpha_D^\top \bm Z_i)$, 
and \(\lambda_{0D}(d)\) is the baseline hazard function. We specify a Weibull baseline for type 1 recurrences, a three-interval piecewise exponential baseline for type 2 recurrences, and a Gompertz baseline for the terminal hazard. The while-alive estimand under this joint frailty model admits a closed-form expression \citep{wei2023properties}, with full details provided in Web Appendix A.2.  Baseline and coefficient values for all scenarios are summarized in Table \ref{tab:scenarios_iid}. Right censoring times are generated either from a covariate-dependent exponential distribution \( C_i \sim \mathrm{Exp}\!\left\{c_0 \exp(\bm\theta^\top \bm Z_i)\right\} \), with $(c_0,\bm\theta)$ chosen to achieve approximately $25\%$ or $50\%$ censoring rate (for the fatal event), or a completely independent exponential distribution $C_i\sim \mathrm{Exp}(\lambda_c)$ with $\lambda_c$ tuned to achieve approximately $50\%$ censoring rate.
We consider two scenario sets. Scenario Set I corresponds to a higher recurrent event intensity, whereas Scenario Set II corresponds to a lower recurrent event intensity. Within each set, we examine two weight specifications, $(w_1,w_2,w_D)=(1,1,1)$ representing equal weighting and $(1,2,2)$ with greater emphasis on death and severe recurrences. A complete specification of all configurations is provided in Table \ref{tab:scenarios_iid}.

We fit the while-alive regression model described in Section \ref{sec: wl_reg_ind} using a step-function basis with $R=7$ equally spaced knots at $\{1.0,1.5,2.0,2.5,3.0,3.5,4.0\}$. That is, $\beta_j(t)=\sum_{r=1}^{R} \gamma_{j,r}\,J_r(t)$, and $J_r(t)=\mathbf{1}\{t\ge t_{r-1}\}$, $j=1,2$. We further let the stacking points coincide with the knot locations, and consider a log link $\eta(\cdot)=\log(\cdot)$ in the regression model. The true association effects $\bm{\beta}(t)=\{\beta_1(t),\beta_2(t)\}^\top$ are evaluated directly from the joint frailty data-generating model using Monte Carlo approximation with a super population of size $n=10^6$ with censoring removed; see Web Appendix A.10 for further details. We consider $1000$ simulation replicates for each setting, and evaluate the following performance metrics at the knot locations: absolute bias (ABias), Monte Carlo standard deviation (MCSD), average estimated standard error (AESE), and $95\%$ coverage probability (CP) of the confidence interval estimator.

Tables \ref{tab:scenario_iid_ibd} and \ref{tab:scenario_iid_icab} summarize the simulation results under two censoring scenarios: covariate-dependent censoring (Scenarios I(b) and I(d), censoring rate $=50\%$) and completely independent censoring (Scenarios II(a) and II(b), censoring rate $=50\%$). For each selected time point, we report the Abias, MCSD, AESE and CP of 95\% confidence intervals of the regression coefficient estimators. The results show that the proposed regression estimators are approximately unbiased across all time points and under both censoring settings. The mean standard error estimates obtained from the proposed sandwich variance estimator closely match the empirical Monte Carlo standard deviation throughout, leading to close to nominal coverage. 

Comparing the two weight specifications, $(w_1,w_2,w_D)=(1,1,1)$ versus $(1,2,2)$, the bias patterns are essentially unchanged. Under covariate-dependent censoring (Table \ref{tab:scenario_iid_ibd}), MCSD and AESE tend to be larger with $(1,2,2)$ than with $(1,1,1)$, and the association effects are positively shifted under the $(1,2,2)$ weight specification. Under completely independent censoring (Table \ref{tab:scenario_iid_icab}), MCSD and AESE are lower compared to the covariate-dependent censoring case, and these uncertainty metrics appear to be more similar when different weight specifications are considered. The similarity between the uncertainty metrics arises likely because the censoring survival functions are modeled by Kaplan-Meier estimator without covariates.

In Web Appendix A.10, we report additional simulations under the independent data setting, including Scenarios I(a) and I(c) under covariate-dependent censoring with 25\% censoring rate, Scenarios I(b)/I(d) and II(a)/II(b) with 50\% censoring rate, coupled with low event rates. The findings are consistent with the main results. Web Appendix A.11 extends the data-generating mechanism to cluster-correlated data settings. With $M=60$ clusters of varying sizes, our proposed estimators continue to perform well, with negligible bias and close to nominal coverage. Finally, in Web Appendix A.12, we provide an additional sensitivity analysis when the locations of spline knots and stacking times differ, under Scenarios I(b) and I(d) with 50\% censoring rate. Specifically, we fit the while-alive regression with spline knots at $\{1.0,1.5,2.0,2.5,3.0,3.5,4.0\}$ but consider a finer grid of stacking times at $\{1.0,1.3,1.6,1.9,2.2,2.5,2.8,3.1,3.4,3.7,4.0\}$. This alternative implementation of the regression estimator maintains small bias and nominal coverage, but the MCSD and AESE appear slightly smaller compared to our default implementation.

\section{While-alive regression analysis of the HF-ACTION trial}
\label{sec:data_analysis}
HF-ACTION is a randomized controlled clinical trial to evaluate the efficacy and safety of exercise training among patients with heart failure with median follow-up 30 months \citep{o2009efficacy}.  Patients were randomly assigned to usual care alone or usual care plus aerobic exercise training that consists of 36 supervised sessions followed by home-based training. 
The primary endpoint was a composite of all-cause mortality or all-cause hospitalization. Secondary endpoints included all-cause mortality, the composite of cardiovascular mortality or cardiovascular hospitalization, and the composite of cardiovascular mortality or heart failure hospitalization. We analyzed the data using our proposed method, adjusting for exercise duration from the CPX test (in minutes), best available baseline left ventricular ejection fraction (LVEF), and history of depression (binary), using complete cases after standardization. Our analysis focused on a high-risk subgroup consisting of 719 nonischemic patients with a baseline cardiopulmonary test duration of less than 12 minutes (the analysis of the full study cohort was reported Web Appendix A.13). We specified a log link function and assigned weights \( w_H = 1 \), \( w_D = 2 \) for all-cause of hospitalization and all-cause of death, respectively, emphasizing the greater importance of fatal events. To assess potentially covariate-dependent censoring, we fit a Cox proportional hazards model at a significance level of 0.05 and found that the CPX test was significantly associated with censoring time; accordingly, CPX was included in the censoring model. For this analysis, we modeled time-varying covariate effects using a B-spline basis \citep{gordon1974b}. We formed polynomial with order-\(d\) B-spline basis \(\bm{B}_j(t)\) for \(j=1,\dots,p\) with \(R\) interior knots. Thus the coefficient \(\bm{J}_i(t) = t \bm{B}_j(t)\) is smooth on \([0,\tau]\). The number of interior knots  \(R\) and the degree of spline \(d\) were selected by 5-fold cross-validataion over \(R \in \{2,3,4,5,6\}\) and \(d \in \{1,2,3,4\}\), and the optimal choice minimizing out-of-sample prediction error was \(R=4\) and \(d=2\).  
We report both the treatment effect and the covariate effects \( \bm{\beta}(t) \) under this optimal choice, along with their pointwise corresponding confidence intervals. 

\begin{figure}[ht!]
    \centering
    \includegraphics[width=1\linewidth]{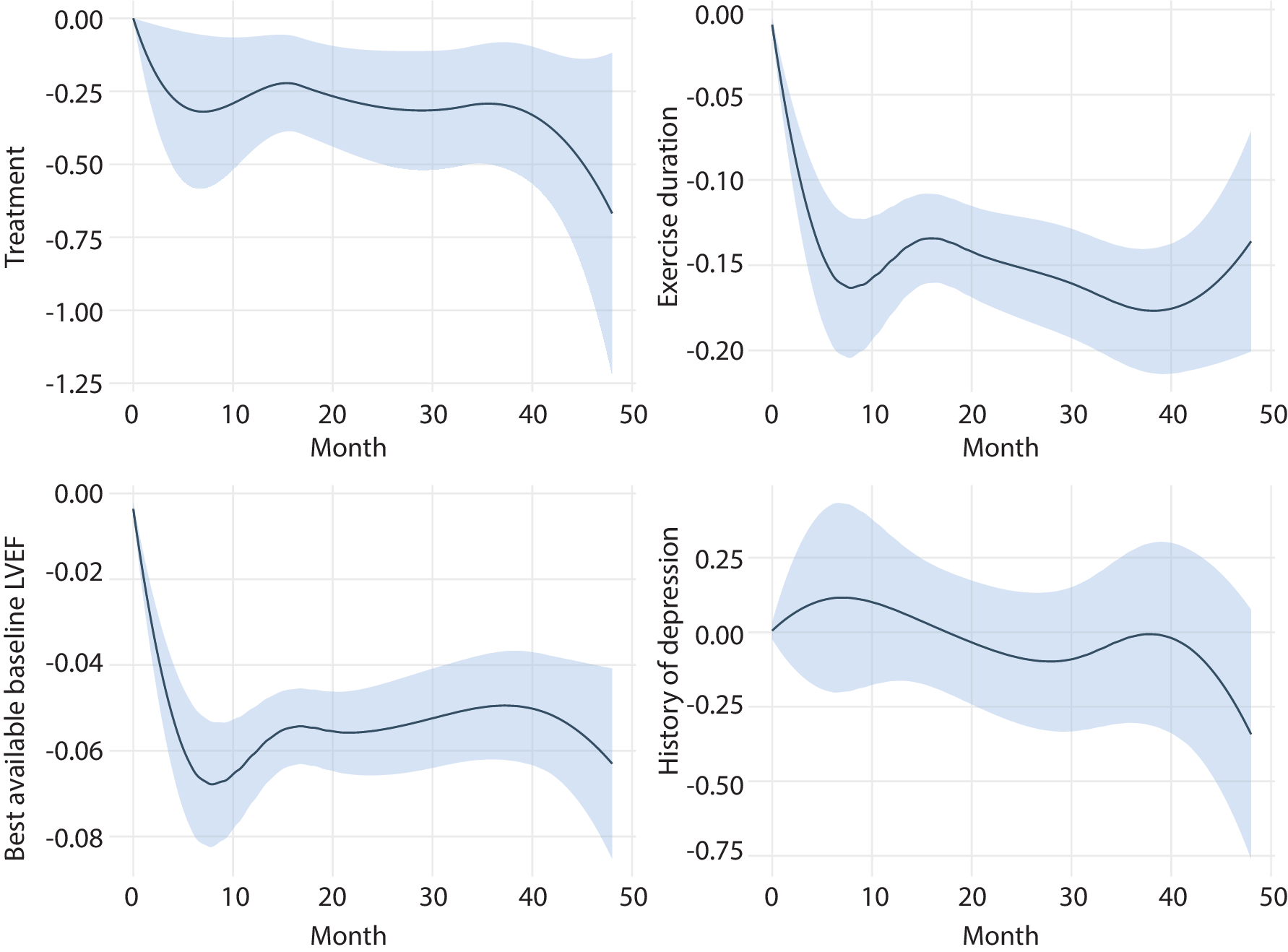}
    \caption{Analysis of the HF-ACTION trial with high-risk subgroup using the while-alive regression model. Each panel presents the estimated covariate effect as a function of time.}
    \label{fig:hf_action_sub}
\end{figure}

Figure \ref{fig:hf_action_sub} presents the time-varying effects of covariates on the while-alive loss rate among this high-risk subgroup. The results reveal substantial variation in the treatment effect over time. Notably, the treatment leads to a significant reduction in the log while-alive loss rate throughout the study period, with the 95\% confidence interval consistently excluding zero. This finding aligns with results from the nonparametric approach proposed by \citet{mao2023nonparametric}. The treatment effect becomes more pronounced after one year, suggesting that the benefits accumulate over time. These results underscore the limitations of traditional models that assume constant effects, as such approaches may fail to detect meaningful time-varying patterns in treatment efficacy, especially during the mid-study period. For instance, at one year, the estimated log-rate difference is \(-0.257\), corresponding to an approximate  \(22.7\%\) (\(\approx 1-e^{-0.257}\)) reduction in the while-alive loss rate for the intervention group relative to control, with a $p$-value of 0.009. By year three, the effect increased to \(-0.293\), corresponding to \(25.4\%\) reduction in while-alive loss rate, with a $p$-value of 0.006. The global test for the treatment effect gives a \(p\)-value of \(<0.001 \).
Additionally, exercise duration and baseline left ventricular ejection fraction (LVEF) show statistically significant and beneficial effects throughout the follow-up period, but with different trajectories over time. In contrast, the effect due to history of depression oscillates around zero, with confidence intervals consistently including zero.



In Web Appendix A.14, we further apply the proposed while-alive regression framework to the pragmatic cluster-randomized trial STRIDE, introduced in Section \ref{sec:data}.

\section{Discussion}
\label{discuss}

Recent regulatory guidance, including the ICH E9(R1) addendum \citep{european2020e9}, has emphasized transparency in defining estimands and handling intercurrent events such as death. As summarized in Web Appendix Table 1, existing regression methods for recurrent and terminal events can be mapped to different strategies to handle death as an intercurrent event. Under that classification, the while-alive regression model adopt a combination of composite and while-alive strategies by directly incorporating death into the outcome and evaluating the burden of events relative to the time patients remain alive \citep{schmidli2023estimands}. This yields interpretable measures of the average event rate among individuals while they remain alive, providing a complementary view to estimands that condition on survival or model death separately. Our proposed regression framework expands this while-alive perspective by modeling the time-varying association with the while-alive loss rate, capturing the instantaneous event burden over time prior to death.

One implementation consideration is the selection of the degree of spline and the number of knots. To balance model fit and prevent overfitting, we select the number of knots via cross-validation. While this data-driven approach is flexible, our sandwich variance treats the selected configuration as fixed, as is standard in the spline regression literature, and it does not address the additional uncertainty arising from cross-validation. As noted by \citet{chatfield1995model}, treating the selected model as fixed may lead to an underestimation of variability. Nevertheless, simulations from the spline regression literature and ours indicate that its practical impact on coverage is typically modest. Recently, \citet{yang2023estimation} proposed a semi-smoothing estimating equation for selecting knots in linear splines, which avoids the need for tuning parameters common in traditional smoothing methods. However, extending this approach to composite endpoints presents additional challenges, particularly in constructing influence functions to estimate the knots, an intriguing avenue for future research. Although cross-validation is used to choose the number of knots, one can still pre-specify knot locations based on the study context. That is, one can choose interior knots at clinical landmarks where effects may plausibly change (e.g., scheduled visits) or calendar-time anchors when secular shifts are expected.

Choosing the weights for the composite endpoint can be controversial and critically depends on the study context and objective. Following \citet{mao2016semiparametric}, our method allows one to set weight reflecting the clinical severity or priority of each event type. This flexibility is useful but introduces subjectivity, and different weight choices may lead to different conclusions. Prior studies have shown that when less frequent but more critical outcomes are combined with more frequent but less severe events, the frequent event can overshadow the composite endpoint unless appropriately weighted \citep[e.g.,][]{baracaldo2023making}.
Regulatory guidance emphasizes that components of a composite should be of comparable clinical importance, and if mortality is included, it should not be down-weighted relative to less severe events \citep{guidance2017multiple, baracaldo2023making}. Specification based on $w_{D}<\max_k \{w_k\}$ may thus be less clinically coherent, whereas dramatic difference between $w_{D}$ and $w_{k}$ may also allow one component to dominate the analysis results \citep{ozga2022weighted}.
In our applications, we used $(w_{H},w_{D})=(1,2)$ to address severity ordering but also to avoid dominance by either component.
In practice, assigning weights to different component outcomes depends on the clinical context and requires a collective effort from an interdisciplinary study team.

Finally, our method requires specifying the stacking time points \(t_v\in\{t_1,\dots,t_V\}\). We allow the differentiation between the stacking times points in the estimating equation from the spline knots and recommend using the knots or finer points for stacking time points. 
A principled consideration is the number of events per time interval generated by the stacking time points. One may choose stacking time points to align with clinical landmarks, but it is often necessary to ensure that each interval contains a sufficient number of events, for numerical stability in the estimation procedure. Although there is no consensus, a reasonable rule of thumb is aiming for at least one or a few events of each type per interval, and sensitivity analysis can be considered based on alternative specifications of the stacking time points. 

\section{Supplementary Material}
\label{sec:supp}

Supplementary material is available online at \url{http://biostatistics.oxfordjournals.org}. Simulation codes are available on Github at \url{https://github.com/fancy575/WA_reg_code}.

\section*{Acknowledgments}
Research in this article was supported by the National Heart, Lung, and Blood Institute (NHLBI, grant number R01-HL168202) and 
National Institute of General Medical Sciences (NIHGM, grant number R01GM152499). The author also thanks the Yale University-Mayo Clinic Center of Excellence in Regulatory Science and Innovation (CERSI) for supporting this study. All statements in this report, including its findings and conclusions, are solely those of the authors and do not necessarily represent the views of the NIH. The data example in Section \ref{sec:data_analysis} was prepared using the HF-ACTION Research Materials obtained from the NHLBI Biologic Specimen and Data Repository Information Coordinating Center (BioLINCC) and does not necessarily reflect the opinions or views of HF-ACTION or NHLBI. The second data example in Web Appendix A.14, is based on de-identified data of the STRIDE study, which was funded primarily by the Patient Centered Outcomes Research Institute (PCORI\textsuperscript{\textregistered}), with additional support from the National Institute on Aging (NIA). Funding for STRIDE is provided and the award managed through a cooperative agreement (5U01AG048270) between the NIA and the Brigham and Women’s Hospital.

\bibliographystyle{apalike}
\bibliography{refs}

\label{LastPage}

\begin{table}[htbp!]
\centering
\caption{Simulation scenarios for the independent data case under the joint frailty model. Type 1 recurrences: Weibull baseline$(\text{scale}=0.50,\text{shape}=1.25)$;  
Type 2 recurrences: piecewise exponential distribution baseline with cuts $(0,1,3)$ and rates $(0.40,0.22,0.10)$;  
Death: Gompertz baseline with parameters $(\kappa,\nu)$.  
Covariate effects are $\mu_1(\bm Z_i)=\bm\alpha_1^\top \bm Z_i$, 
$\mu_2(\bm Z_i)=\bm \alpha_2^\top \bm Z_i$, 
$\mu_D(\bm Z_i)=\bm \alpha_D^\top \bm Z_i$.  
Censoring follows $C_i\sim \text{Exp}(c_0 e^{\bm \theta^\top \bm Z_i})$ or $C_i\sim \text{Exp}(\lambda)$ (independent).  
The Rates column reports realized event rates per person-time for type 1, type 2, and death.}
\label{tab:scenarios_iid}
\resizebox{\textwidth}{!}{
\begin{tabular}{lccccccc}
\toprule
Scenario & $(\kappa,\nu)$ & $\bm \alpha_1$ & $\bm \alpha_2$ & $\bm \alpha_D$ & $(w_1,w_2,w_D)$ & Censoring & Rates \\
\midrule
I(a)   & (0.45, 0.30) & (0.50, -0.80) & (0.30, 0.90) & (0.20, 1.00) & (1,1,1) & $\bm\theta=(0.20,0.2,0.5)$ (25\%) & (1.780,\;0.279,\;0.514) \\
I(b)   & (0.45, 0.30) & (0.50, -0.80) & (0.30, 0.90) & (0.20, 1.00) & (1,1,1) & $\bm\theta=(0.45,0.5,0.5)$ (50\%) & (1.680,\;0.304,\;0.492) \\
I(c)   & (0.45, 0.30) & (0.50, -0.80) & (0.30, 0.90) & (0.20, 1.00) & (1,2,2) & $\bm\theta=(0.20,0.2,0.5)$ (25\%) & (1.780,\;0.279,\;0.514) \\
I(d)   & (0.45, 0.30) & (0.50, -0.80) & (0.30, 0.90) & (0.20, 1.00) & (1,2,2) & $\bm\theta=(0.45,0.5,0.5)$ (50\%) & (1.680,\;0.304,\;0.492) \\
\midrule
II(a)  & (0.10, 0.30) & (0.20, 0.50) & (0.80, 1.00) & (0.20, 1.00) & (1,1,1) & $\bm\theta=(0.17,0.5,0.5)$ (50\%) & (0.599,\;0.387,\;0.188) \\
II(b)  & (0.10, 0.30) & (0.20, 0.50) & (0.80, 1.00) & (0.20, 1.00) & (1,2,2) & $\bm\theta=(0.17,0.5,0.5)$ (50\%) & (0.599,\;0.387,\;0.188) \\
\midrule
IC(a)  & (0.45, 0.30) & (0.50, -0.80) & (0.30, 0.90) & (0.20, 1.00) & (1,1,1) & $\lambda=0.55$ (50\%) & (1.280,\;0.385,\;0.571) \\
IC(b)  & (0.45, 0.30) & (0.50, -0.80) & (0.30, 0.90) & (0.20, 1.00) & (1,2,2) & $\lambda=0.55$ (50\%) & (1.280,\;0.385,\;0.571) \\
\bottomrule
\end{tabular}
}
\end{table}

\begin{table}[htbp!]
\caption{Simulation results under Scenario I(b) and I(d) with $R=7$ equally spaced knots and stacking time at $(1.0,1.5,2.0,2.5,3.0,3.5,4.0)$ and covariate-dependent censoring (50\% censoring rate). Sample size $n=2000$. Metrics: ABias (absolute bias), MCSD (Monte Carlo SD), AESE (asymptotic SE), CP (coverage of the 95\% CI). Column blocks compare weights $(w_1,w_2,w_D)=(1,1,1)$ \textit{vs.} $(1,2,2)$.}
\label{tab:scenario_iid_ibd}
\centering
\small
\begin{tabular}{cccccccccccc}
\toprule
 &  & \multicolumn{5}{c}{$(w_1,w_2,w_D)=(1,1,1)$} & \multicolumn{5}{c}{$(w_1,w_2,w_D)=(1,2,2)$} \\
\cmidrule(lr){3-7} \cmidrule(lr){8-12}
Time & $\bm{\beta}(t)$ & True & ABias & MCSD & AESE & CP & True & ABias & MCSD & AESE & CP \\
\midrule
1.0 & $\beta_1$ &  0.904 & 0.001 & 0.043 & 0.042 & 0.951 &  1.341 & 0.001 & 0.043 & 0.043 & 0.955 \\
    & $\beta_2$ & -0.008 & 0.004 & 0.056 & 0.056 & 0.941 &  0.355 & 0.004 & 0.064 & 0.063 & 0.940 \\
\midrule
1.5 & $\beta_1$ &  0.871 & 0.001 & 0.047 & 0.047 & 0.949 &  1.333 & 0.001 & 0.046 & 0.047 & 0.953 \\
    & $\beta_2$ & -0.146 & 0.004 & 0.055 & 0.055 & 0.938 &  0.195 & 0.004 & 0.066 & 0.066 & 0.937 \\
\midrule
2.0 & $\beta_1$ &  0.839 & 0.002 & 0.052 & 0.053 & 0.952 &  1.312 & 0.001 & 0.053 & 0.054 & 0.963 \\
    & $\beta_2$ & -0.238 & 0.004 & 0.056 & 0.054 & 0.942 &  0.077 & 0.004 & 0.070 & 0.069 & 0.942 \\
\midrule
2.5 & $\beta_1$ &  0.812 & 0.003 & 0.057 & 0.058 & 0.952 &  1.287 & 0.003 & 0.061 & 0.062 & 0.944 \\
    & $\beta_2$ & -0.304 & 0.004 & 0.056 & 0.054 & 0.940 & -0.013 & 0.005 & 0.072 & 0.070 & 0.945 \\
\midrule
3.0 & $\beta_1$ &  0.789 & 0.003 & 0.062 & 0.064 & 0.953 &  1.261 & 0.004 & 0.067 & 0.069 & 0.951 \\
    & $\beta_2$ & -0.354 & 0.005 & 0.056 & 0.055 & 0.934 & -0.084 & 0.006 & 0.072 & 0.072 & 0.940 \\
\midrule
3.5 & $\beta_1$ &  0.768 & 0.005 & 0.069 & 0.069 & 0.946 &  1.235 & 0.007 & 0.076 & 0.076 & 0.941 \\
    & $\beta_2$ & -0.392 & 0.006 & 0.057 & 0.055 & 0.929 & -0.140 & 0.007 & 0.074 & 0.073 & 0.933 \\
\midrule
4.0 & $\beta_1$ &  0.751 & 0.008 & 0.074 & 0.074 & 0.940 &  1.211 & 0.010 & 0.082 & 0.081 & 0.940 \\
    & $\beta_2$ & -0.421 & 0.007 & 0.058 & 0.056 & 0.929 & -0.184 & 0.010 & 0.076 & 0.075 & 0.929 \\
\bottomrule
\end{tabular}
\end{table}

\begin{table}[htbp!]
\caption{Simulation results under Scenario IC(a) and IC(b) with $R=7$ equally spaced knots and stacking time at $(1.0,1.5,2.0,2.5,3.0,3.5,4.0)$ and completely independent censoring. Sample size $n=2000$; censoring rate $50\%$. Metrics: ABias (absolute bias), MCSD (Monte Carlo SD), AESE (asymptotic SE), CP (coverage of the $95\%$ CI). Column blocks compare weights $(1,1,1)$ vs.\ $(1,2,2)$.}
\label{tab:scenario_iid_icab}
\centering
\small
\begin{tabular}{cccccccccccc}
\toprule
 &  & \multicolumn{5}{c}{$(w_1,w_2,w_D)=(1,1,1)$} & \multicolumn{5}{c}{$(w_1,w_2,w_D)=(1,2,2)$} \\
\cmidrule(lr){3-7} \cmidrule(lr){8-12}
Time & $\bm{\beta}(t)$ & True & ABias & MCSD & AESE & CP & True & ABias & MCSD & AESE & CP \\
\midrule
1.0 & $\beta_1$ & 0.904 & 0.002 & 0.037 & 0.038 & 0.958 & 1.341 & 0.002 & 0.039 & 0.039 & 0.961 \\
    & $\beta_2$ & -0.008 & 0.004 & 0.059 & 0.060 & 0.950 & 0.355 & 0.004 & 0.063 & 0.064 & 0.954 \\
\midrule
1.5 & $\beta_1$ & 0.871 & 0.000 & 0.042 & 0.042 & 0.953 & 1.333 & 0.002 & 0.039 & 0.040 & 0.961 \\
    & $\beta_2$ & -0.146 & 0.008 & 0.063 & 0.062 & 0.939 & 0.195 & 0.007 & 0.071 & 0.072 & 0.938 \\
\midrule
2.0 & $\beta_1$ & 0.839 & 0.003 & 0.048 & 0.049 & 0.947 & 1.312 & 0.000 & 0.044 & 0.045 & 0.963 \\
    & $\beta_2$ & -0.238 & 0.010 & 0.070 & 0.071 & 0.930 & 0.077 & 0.009 & 0.083 & 0.084 & 0.928 \\
\midrule
2.5 & $\beta_1$ & 0.812 & 0.005 & 0.055 & 0.056 & 0.953 & 1.287 & 0.002 & 0.050 & 0.052 & 0.956 \\
    & $\beta_2$ & -0.304 & 0.012 & 0.077 & 0.078 & 0.929 & -0.013 & 0.011 & 0.094 & 0.095 & 0.919 \\
\midrule
3.0 & $\beta_1$ & 0.789 & 0.005 & 0.064 & 0.065 & 0.940 & 1.261 & 0.002 & 0.060 & 0.062 & 0.951 \\
    & $\beta_2$ & -0.354 & 0.015 & 0.086 & 0.087 & 0.929 & -0.084 & 0.015 & 0.106 & 0.107 & 0.928 \\
\midrule
3.5 & $\beta_1$ & 0.768 & 0.007 & 0.075 & 0.076 & 0.944 & 1.235 & 0.004 & 0.071 & 0.073 & 0.953 \\
    & $\beta_2$ & -0.392 & 0.020 & 0.095 & 0.096 & 0.932 & -0.140 & 0.021 & 0.118 & 0.119 & 0.930 \\
\midrule
4.0 & $\beta_1$ & 0.751 & 0.010 & 0.086 & 0.087 & 0.935 & 1.211 & 0.007 & 0.084 & 0.085 & 0.935 \\
    & $\beta_2$ & -0.421 & 0.025 & 0.105 & 0.106 & 0.932 & -0.185 & 0.027 & 0.131 & 0.132 & 0.930 \\
\bottomrule
\end{tabular}
\end{table}

\clearpage
\appendix
\title{Supplementary materials for ``While-alive regression analysis of composite survival endpoints'' by Fang, Uno and Li}

\section{Review of regression models for recurrent and terminal events} \label{supp:sec:review}
\allowdisplaybreaks
Table \ref{tab:recurrent_strategies} summarizes common regression modeling frameworks for analyzing recurrent events and a terminal event (death) in survival analysis, highlighting the implied modeling targets, key features, and the corresponding strategies for handling death as an intercurrent event (when treatment is included in the design matrix).

\begin{sidewaystable}[htbp!]
\centering
\caption{An incomplete description of typical regression modeling strategies for recurrent events and a terminal event (death) in survival analysis. Although regression models are more general than estimating treatment effects, when the treatment variable is involved in the regression model in a randomized trial, we also discuss how each method may be mapped to a strategy to handle the terminal event (as an intercurrent event, IE) under the ICH E9 Addendum Estimands Framework \citep{european2020e9}.}
\label{tab:recurrent_strategies}
\renewcommand{\arraystretch}{1.25}
\begin{tabular}{P{4.2cm} P{5.2cm} P{7.0cm} P{2.8cm}}
\toprule
\textbf{Methods} & \textbf{Modeling target} & \textbf{Features} & \textbf{Closest strategies to handle death as an IE} \\
\midrule
Time to first event & All-cause hazard & Death is part of the outcome, but later recurrences are ignored after death & Composite \\[3pt]

Recurrent event model \\[-2pt]\citep{li1997use} & Recurrent event hazard & Death censors the recurrent process & Hypothetical \\[3pt]

Recurrent event model \\[-2pt]\citep{ghosh2002marginal} & Recurrent event rate, acknowledging no recurrence after death & Death truncates the recurrent process & Composite \\[3pt]

Recurrent event model \\[-2pt]\citep{lawless1997analysis} & Recurrent event rate conditional on being alive & Death truncates the recurrent process & While-alive \\[3pt]

Joint model \\[-2pt]\citep{lancaster1998panel} & Recurrent event rate and terminal event hazard & Shared frailty linking recurrent and terminal event processes & Hypothetical \\[3pt]

Multi-state model \\[-2pt]\citep{cook2018multistate} & Transition hazards between states & Death is a terminal event (absorbing state) & Composite \\[3pt]

Composite regression \\[-2pt]\citep{mao2016semiparametric} & Mean of a weighted composite event process & Death is a component of the outcome & Composite \\[3pt]

While-alive regression \\[-2pt](This article)& While-alive loss rate & Death both limits alive-time and can be a component of the outcome definition & Composite while-alive \\
\bottomrule
\end{tabular}
\end{sidewaystable}

\section{Interpreting time-varying effect measures for the while-alive loss rate under different data-generating processes} \label{supp:sec:gen_model}

To focus ideas, we consider the simplest case with only one treatment covariate. Let $A\in\{0,1\}$ denote treatment. Write $S(u\mid A)=\Pr(D>u\mid A)$ as the survival probability, $\lambda_{D}(u)$ as the hazard function for the fatal event, and intensity of the \(k\)th recurrent event as \(\mu_k(u)\) for \(k \in 1,\dots, K\). Thus, we can express the while-alive loss rate as 

\begin{equation*} \label{eq:wa-identity}
l(\mathcal H)(t\mid A)
=\frac{\textstyle \int_0^t S(u\mid A)\Big\{w_D\,\lambda_D(u\mid A)+\sum_{k=1}^K w_k\,\mu_k(u\mid A)\Big\}\,du}{\textstyle \int_0^t S(u\mid A)\,du},
\end{equation*}
with regression model \(\log\{ l(\mathcal H)(t\mid A) \} =\beta_0+ \beta_1(t)A\). We have already know that  $\int_0^t \lambda_D(u\mid A)\,S(u\mid A)\,du = 1 - S(t\mid A)$ and $\int_0^t S(u\mid A)\,du$ is the RMST. We next introduce several different data-generating models (DGMs) and discuss their connections to the while-alive loss rate parameter.
\begin{itemize}
\item[\textbf{DGM 1}] Under a proportional hazard model with death as the only event type, then the log while-alive rate $\beta_1(t)$ does not vary with time. More specifically, set $w_D=1$, and $\lambda_D(u\mid A)=\lambda_0(t)e^{\beta_D A}$ with $S(u\mid A)=\exp\{-\lambda_0(t) e^{\beta_D A} u\}$.  From \eqref{eq:wa-identity}, the while alive loss rate is 
\[
l(\mathcal H)(t\mid A)
=\frac{\int_0^t \lambda_0 e^{\beta_D A}\,S(u\mid A)\,du}{\int_0^t S(u\mid A)\,du}
=\lambda_0 e^{\beta_D A}.
\]
Here, the treatment coefficient is constant with log link, and $\beta_1(t)\equiv \beta_D$ for all $t$. 

\item[\textbf{DGM 2}] Under a shared frailty model with \(W \sim \mathrm{Gamma}(\text{shape}=\alpha,\text{rate}=\alpha)\) and \(\kappa = 1/\alpha\), we assume time-invariant event-specific effects such that
\[
\lambda_D(u\mid A,W)=W\,\lambda_{D0}(u)\,e^{\beta_D A},\qquad 
\mu_k(u\mid A,W)=W\,\mu_{k0}(u)\,e^{\beta_k A}.
\]
Integrating over the frailty gives the marginal survival function \( S(u\mid A)=\bigl(1+\kappa\,\Lambda_A(u)\bigr)^{-1/\kappa} \),  \(\Lambda_A(u)=e^{\beta_D A}\,\Lambda_{D0}(u)\), \(\Lambda_{D0}(u)=\int_0^{u}\lambda_{D0}(s)\,ds\), and the mean frailty among survivors is \(\mathbb{E}(W\mid D>u,A)=\frac{1}{1+\kappa\,\Lambda_A(u)}\). Using \eqref{eq:wa-identity} and conditioning on being alive, the while-alive loss rate is
\begin{align*}
 l(\mathcal H)(t\mid A)
& =\frac{\displaystyle \int_0^t \frac{S(u\mid A)}{1+\kappa\,\Lambda_A(u)}
\Bigl\{w_D\,\lambda_{D0}(u)e^{\beta_D A}+\sum_k w_k\,\mu_{k0}(u)e^{\beta_k A}\Bigr\}\,du}
{\displaystyle \int_0^t S(u\mid A)\,du} \\
& = \frac{w_D\{ 1-S(t\mid A) \} + \int_{0}^{t} S(u \mid A)^{1+\kappa}  \left\{\sum_{k=1}^{K} w_k \mu_{k0}(u) e^{\beta_k A} \right\}\, du  }{\int_{0}^{t} S(u\mid A)\, du } .   
\end{align*}
Although \(\{\beta_D,\beta_k\}\) are constant conditionally on \(W\), the factor \(S(u\mid A)/(1+\kappa\,\Lambda_A(u))\) varies with \(u\) and differs by arm due to frailty depletion among survivors. As a result, the arm-specific time-averaging kernels differ, and the ratio \( \log \left\{l(t\mid 1)/l(t\mid 0) \right\}\) typically varies with \(t\). As a special case, let \(w_D=1\), \(w_k\equiv0\), and \(\lambda_{D0}(u)\equiv \lambda_0\). Then \(S(u\mid A)=\bigl(1+\kappa \lambda_0 e^{\beta_D A} u\bigr)^{-1/\kappa} \), let \( \alpha_A=\kappa\,\lambda_0 e^{\beta_D A}\), the while-alive loss rate is
\[
l(\mathcal H)(t\mid A)=
\begin{cases}
\displaystyle \alpha_A\!\left(1-\frac{1}{\kappa}\right)
\frac{1-(1+\alpha_A t)^{-1/\kappa}}{(1+\alpha_A t)^{\,1-1/\kappa}-1}, & \kappa\neq 1,\\[2ex]
\displaystyle \dfrac{\alpha_A\bigl(1-(1+\alpha_A t)^{-1}\bigr)}{\log(1+\alpha_A t)}, & \kappa=1.
\end{cases}
\]
For any \(\kappa>0\), \(l(\mathcal H)(t\mid 1)/l(\mathcal H)(t\mid 0)\) is a nonconstant function of \(t\). It collapses to the constant \(e^{\beta_D}\) only in the no-frailty limiting case of \(\kappa\to 0\) (in which case DGM 2 reduces to DGM 1).

\item[\textbf{DGM 3}]  We consider the following models in which treatment acts multiplicatively and differently across event types according to 
\[
\lambda_D(u\mid A)=\lambda_{D0}(u)\,e^{\beta_D A}, \qquad \mu_k(u\mid A)=\mu_{k0}(u)\,e^{\beta_k A},
\]
with \(\beta_D\neq \beta_k\) for at least one \(k\). The arm-specific survival function is
\( S(u\mid A)=\exp\!\bigl\{-e^{\beta_D A}\,\Lambda_{D0}(u)\bigr\} \), and   \(\Lambda_{D0}(u)=\int_0^u \lambda_{D0}(s)\,ds\). Then the while-alive loss rate is
\begin{align*}
    l(\mathcal H)(t\mid A) &=\frac{\displaystyle \int_0^t  S(u\mid A)\Bigl\{w_D\,\lambda_{D0}(u)e^{\beta_D A} +\sum_{k=1}^{K} w_k\,\mu_{k0}(u)e^{\beta_k A}\Bigr\}\,du}{\displaystyle \int_0^t S(u\mid A)\,du}\\
    & = \frac{w_D \{1- S(t\mid A)\} + \sum_{k=1}^{K} w_k e^{\beta_k A} \int_{0}^{t} S(u\mid A) \mu_{k0}(u)du }{\displaystyle \int_0^t S(u\mid A)\,du}.
\end{align*}

Because \(S(u\mid A)\) depends on \(\beta_D\), this survival function reweights the event trajectories \(\lambda_{D0}(t\mid A)\), and \(\mu_{k0}(t\mid A)\) differently by arm. Then the mixture of death- and recurrent-driven losses among survivors evolves with time in an arm-specific way. Only under restrictive conditions (e.g., \(\beta_k=\beta_D\) for all \(k\), or \(\mu_{k0} = c_k \lambda_{D0}(u)\), with some constant \(c_k\) ) for \(k=1,\dots, K\),  this survival function reweighting leave the \(l(\mathcal H)(t\mid 1)/l(\mathcal H)(t\mid 0) \) unchanged over time. Therefore, even without frailty, heterogeneous event-specific effects (\(\beta_k \neq \beta_D\), for at least one \(k\)) imply that the while-alive treatment effect \(\beta_1(t)\) typically varies over time.

\item[\textbf{DGM 4}] Consider a joint frailty specification given by \citet{wei2023properties} with a single positive frailty \(W\) (density \(f_W\)) that differentially affects death and recurrent events. Let \(\gamma>0\) and define
\[
\lambda_D(u\mid A,W)=W^{\gamma}\lambda_{D0}(u)e^{\beta_D A}, \qquad \mu_k(u\mid A,W)=W\,\mu_{k0}(u)e^{\beta_k A}.
\]
Then the conditional survival function can be written as \(S(u\mid A,W)=\exp\!\Big\{-W^{\gamma}e^{\beta_D A}\,\Lambda_{D0}(u)\Big\} \), with \(\Lambda_{D0}(u)=\int_{0}^{u}\lambda_{D0}(s)\,ds\).
The exposure-weighted while-alive rate can be written compactly as \citep{wei2023properties}
\begin{align*}
l(\mathcal H)(t\mid A)=\frac{\displaystyle \int_0^{\infty}\!\!\int_0^t \Big\{w_D\,\lambda_{D0}(u)\,W^{\gamma}e^{\beta_D A}+\sum_{k=1}^K w_k\,\mu_{k0}(u)\,W e^{\beta_k A}\Big\}\,S(u\mid A,W)\,du\,f_W(w)\,dw} {\displaystyle \int_0^{\infty}\!\!\int_0^t S(u\mid A,W)\,du\,f_W(w)\,dw}.    
\end{align*}
If \(W \sim \text{Gamma}(\text{shape}=\alpha, \text{rate}=\alpha)\), with \(\kappa = 1/\alpha\), this can be further simplified as
\[
l(\mathcal H)(t\mid A) = \frac{w_D \{ 1- \mathcal{S}(t\mid A) \} + \sum_{k=1}^{K} w_k e^{\beta_k A} \int_{0}^{t} \mu_{k0}(u) \mathcal{S}(u\mid A)^{1+\kappa}\, du  }{ \int_{0}^{t}\mathcal{S}(u \mid A) \, du},
\]
where \(\mathcal{S}(t\mid A)= \left\{1+\kappa \, e^{\beta_D  A} \Lambda_{D0}(t)  \right\}^{-1/\kappa} \). Even though \(\beta_D\) and \(\beta_k\) for \(k=1,\dots, K\) are constant over time, the reweighting factor \(S(u\mid A, W)\) is still time-dependent and changes differently by arm. Consequently, if (\(\gamma\neq 1\)) or event effects are heterogeneous (\(\beta_k\neq\beta_D\) for some \(k\)), the treatment effect \(\beta_1(t)\) typically varies with \(t\). 

\end{itemize}

From the above examples, we can find that the coefficient \(\beta(t)\), even in the simplest binary covariate case, is time-varying on the while-alive scale.  It is only a constant over time in benchmark cases (death-only, exponential proportional hazard model), but can vary over time even when hazard-scale treatment effects are constant in the case with recurrent events.

\section{Proof of the consistency of the proposed method in the independent data setting} \label{supp:sec:proof_iid_consistent}

In this section, we prove the consistency of the estimating equation under the independent data setting. The proof under the cluster-correlated data setting follows almost the same argument and is omitted here. For simplicity, we write the unit of the estimating function for any \(t\) and any individual as 
$$
 U(\widetilde{\bm{\beta}}) =  \frac{\mathbb{I}(U \leq t) \Delta + \mathbb{I}(U > t)}{G(U \wedge t \mid \bm{Z})} \widetilde{\bm{Z}}(t) \left\{ \widetilde{\mathcal{L}}(\mathcal{H})(t) - \eta^{-1}\{\widetilde{\bm{\beta}}^\top \widetilde{\bm{Z}}(t)\}(U \wedge t) \right\}
$$
where  \(\widetilde{\mathcal{L}}(\mathcal{H})(t) = \sum_{k=1}^{K} w_k N_k(U \wedge t) + w_D \widetilde{N}_D(t)\). Then we can show that
\begin{align}
    E[U(\widetilde{\bm{\beta}}) \mid \widetilde{\bm{Z}}(t)] & = E\left[ \frac{\mathbb{I}(U \leq t) \Delta + \mathbb{I}(U > t)}{G(U \wedge t \mid \bm{Z})} \widetilde{\bm{Z}}(t) \widetilde{\mathcal{L}}(\mathcal{H})(t) - \widetilde{\bm{Z}}(t)\,\eta^{-1}\{\widetilde{\bm{\beta}}^\top \widetilde{\bm{Z}}(t)\}(U \wedge t) \mid \widetilde{\bm{Z}}(t) \right] \nonumber \\
    & = \widetilde{\bm{Z}}(t) E\left[ \frac{\mathbb{I}(U \leq t) \Delta + \mathbb{I}(U > t)}{G(U \wedge t \mid \bm{Z})} 
 \left\{ \widetilde{\mathcal{L}}(\mathcal{H})(t) - \eta^{-1}\{\widetilde{\bm{\beta}}^\top \widetilde{\bm{Z}}(t)\}(U \wedge t) \right\} \mid \widetilde{\bm{Z}}(t)\right] \nonumber \\
 & = \widetilde{\bm{Z}}(t) E\left[ \frac{\mathbb{I}(U \leq t) \Delta + \mathbb{I}(U > t)}{G(U \wedge t \mid \bm{Z})}  \times \widetilde{\mathcal{L}}(\mathcal{H})(t) \mid \widetilde{\bm{Z}}(t) \right]  \nonumber \\
 & \quad - \widetilde{\bm{Z}}(t) E\left[ \frac{\mathbb{I}(U \leq t) \Delta + \mathbb{I}(U > t)}{G(U \wedge t \mid \bm{Z})}  \times \eta^{-1}\{\widetilde{\bm{\beta}}^\top \widetilde{\bm{Z}}(t)\}(U \wedge t) \mid \widetilde{\bm{Z}}(t) \right] \nonumber \\
 & = \widetilde{\bm{Z}}(t) E\left[ \frac{\mathbb{I}(U \leq t) \Delta + \mathbb{I}(U > t)}{G(U \wedge t \mid \bm{Z})} \, w_D\,\mathbb{I}(U \leq t) \mid \widetilde{\bm{Z}}(t) \right] \label{ee_survival}\\
 & \quad + \widetilde{\bm{Z}}(t) E\left[ \frac{\mathbb{I}(U \leq t) \Delta + \mathbb{I}(U > t)}{G(U \wedge t \mid \bm{Z})}  \sum_{k=1}^{K} w_k \sum_{q=1}^{\infty} \mathbb{I}(T_{k,q} \leq U \wedge t) \mid \widetilde{\bm{Z}}(t) \right] \label{ee_recurrent}\\
 & \quad -  \widetilde{\bm{Z}}(t) E\left[ \frac{\mathbb{I}(U \leq t) \Delta + \mathbb{I}(U > t)}{G(U \wedge t \mid \bm{Z})} \, \eta^{-1}\{\widetilde{\bm{\beta}}^\top \widetilde{\bm{Z}}(t)\}(U \wedge t)  \mid \widetilde{\bm{Z}}(t)\right].\label{ee_link}
\end{align}
Next, we will focus only on the conditional expectation part. For the \eqref{ee_survival}, we have:
\begin{align*}
    (\ref{ee_survival}) & = E\left[ \frac{\mathbb{I}(D\wedge C \leq t)\mathbb{I}(D \leq C) +\mathbb{I}(D\wedge C > t)  }{P\{C \geq U\wedge t \mid \bm{Z} \}} \, w_D \, \mathbb{I}(D \leq t) \mid \widetilde{\bm{Z}}(t)\right]  \\
    & = w_D  E\left[E\left\{ \frac{\mathbb{I}(D\wedge C \leq t)\mathbb{I}(D \leq C)   }{P\{C \geq U\wedge t\mid \bm{Z} \}} \mid D,\widetilde{\bm{Z}}(t)  \right\}   \mid \widetilde{\bm{Z}}(t)\right]  \\
    & = w_D E\left[E\left\{ \frac{\mathbb{I}(C \geq D\wedge t) \mathbb{I}(D \leq t)}{P\{C\geq D\wedge t \mid \bm{Z}\}}  \mid D, \widetilde{\bm{Z}}(t)   \right\}  \mid \widetilde{\bm{Z}}(t) \right] \\
    & = E\left[w_D\mathbb{I}(D \leq t)\mid \widetilde{\bm{Z}}(t)   \right].
\end{align*}
For the recurrent event part \eqref{ee_recurrent}, we have:
\begin{align}
    (\ref{ee_recurrent}) & = E\left[   \frac{\mathbb{I}(D \wedge C \leq t)\mathbb{I}(D \leq C)  + \mathbb{I}(D\wedge C >  t) }{P\{C \geq U\wedge t \mid \bm{Z}}\}\sum_{k=1}^{K}w_k\sum_{q=1}^{\infty}\mathbb{I}(T_{k,q} \leq D\wedge C \wedge t )  \mid \widetilde{\bm{Z}}(t)   \right] \nonumber \\
     & = E \left[  \frac{\mathbb{I}(D\wedge C \leq t)\mathbb{I}(D\leq C) }{P\{C\geq U\wedge t \mid \bm{Z}\}} \sum_{k=1}^{K}w_k\sum_{q=1}^{\infty}\mathbb{I}(T_{k,q} \leq D\wedge C \wedge  t)  \mid \widetilde{\bm{Z}}(t) \right]  \label{recurrent_p1}\\
      & \quad + E \left[  \frac{\mathbb{I}(D\wedge C > t) }{P\{C\geq U\wedge t \mid \bm{Z}\}} \sum_{k=1}^{K}w_k\sum_{q=1}^{\infty}\mathbb{I}(T_{k,q} \leq D\wedge C \wedge  t )  \mid \widetilde{\bm{Z}}(t)  \right]. \label{recurrent_p2}
\end{align}

The two parts are considered separately. By the total law of conditional expectation, we have
\begin{align*}
    (\ref{recurrent_p1}) & = E\left[ E\left\{\frac{\mathbb{I}(D\wedge C \leq t)\mathbb{I}(D\leq C) }{P\{C\geq U\wedge t \mid \bm{Z}\}} \sum_{k=1}^{K}w_k\sum_{q=1}^{\infty}\mathbb{I}(T_{k,q} \leq D\wedge C \wedge  t ) \mid D,\widetilde{\bm{Z}}(t) \right\} \mid \widetilde{\bm{Z}}(t) \right] \\
    & = E\left[ E\left\{  \frac{\mathbb{I}(D\leq t) \mathbb{I}(C \geq D\wedge t) }{P\{C \geq D\wedge t \mid \bm{Z} \}}        \sum_{k=1}^{K}w_k\sum_{q=1}^{\infty}\mathbb{I}(T_{k,q} \leq D ) \mid D,\widetilde{\bm{Z}}(t)   \right\} \mid \widetilde{\bm{Z}}(t)   \right] \\
    & = E\left[\mathbb{I}(D\leq t) E\left\{  \frac{\mathbb{I}(C\geq D\wedge t)}{P\{C \geq D\wedge t\mid \bm{Z}\}}   \sum_{k=1}^{K}w_k\sum_{q=1}^{\infty}\mathbb{I}(T_{k,q} \leq D )      \mid D,\widetilde{\bm{Z}}(t) \right\}  \mid \widetilde{\bm{Z}}(t)  \right] \\
    & = E\left[ \mathbb{I}(D\leq t) \sum_{k=1}^{K} w_k \sum_{q=1}^{\infty} \mathbb{I}(T_{k,q} \leq D) \mid \widetilde{\bm{Z}}(t) \right]. \\
\end{align*}

Since \(C \perp  \mathcal{H}\mid \bm{Z} \) and \(\sum_{k=1}^{K} w_k \sum_{q=1}^{\infty} \mathbb{I}(T_{k,q} \leq D)\) is measurable with respect to the \(\sigma\)-algebra generated by \(\{D,\widetilde{\bm{Z}}(t) \}\), we have
$$ (\ref{recurrent_p1}) = E\left[\mathbb{I}(D\leq t) \sum_{k=1}^{K} w_k \sum_{q=1}^{\infty} \mathbb{I}(T_{k,q} \leq D) \mid \widetilde{\bm{Z}}(t)  \right]. $$
Similarly, we can show that:
\begin{align*}
    (\ref{recurrent_p2}) & = E\left[ E\left\{ \frac{\mathbb{I}(D\wedge C > t)  }{P\{C \geq U\wedge t \mid \bm{Z} \}}  \sum_{k=1}^{K} w_k \sum_{q=1}^{\infty} \mathbb{I}(T_{k,q} \leq D\wedge C\wedge t) \mid D,\widetilde{\bm{Z}}(t)  \right\} \mid \widetilde{\bm{Z}}(t)  \right]  \\
    & = E\left[E\left\{ \frac{\mathbb{I}(D >t) \mathbb{I}(C >t)}{P(C \geq t \mid \bm{Z})} \sum_{k=1}^{K} w_k \sum_{q=1}^{\infty} \mathbb{I}(T_{k,q} \leq t)          \mid D, \widetilde{\bm{Z}}(t)     \right\}  \mid \widetilde{\bm{Z}}(t)\right] \\
    & = E\left[E\left\{ \mathbb{I}(D> t) \sum_{k=1}^{K} w_k \sum_{q=1}^{\infty} \mathbb{I}(T_{k,q} \leq t)  \mid D,\widetilde{\bm{Z}}(t)\right\} \mid \widetilde{\bm{Z}}(t)            \right] \\
    & =  E\left[ \mathbb{I}(D> t) \sum_{k=1}^{K} w_k \sum_{q=1}^{\infty} \mathbb{I}(T_{k,q} \leq t) \mid \widetilde{\bm{Z}}(t) \right].
\end{align*}

Combine the above, we can obtain 
$$ (\ref{ee_recurrent}) = (\ref{recurrent_p1}) + (\ref{recurrent_p2}) = E\left[ \sum_{k=1}^{K} w_k \sum_{q=1}^{\infty} \mathbb{I}(T_{k,q} \leq D\wedge t) \mid \widetilde{\bm{Z}}(t) \right] $$

Next, we will show the \eqref{ee_link} that
\begin{align*}
    (\ref{ee_link}) & = E \left[  \frac{\mathbb{I}(U \leq t)\Delta  + \mathbb{I}(U > t) }{P\{C \geq U \wedge t \mid \bm{Z}\}} \, \eta^{-1}\{\widetilde{\bm{\beta}}^\top\widetilde{\bm{Z}}(t)\}(U\wedge t)     \widetilde{\bm{Z}}(t)            \right] \nonumber \\
    & = E\left[ \frac{\mathbb{I}(U \leq t)\mathbb{I}(D\leq C) }{P\{C \geq U \wedge t \mid \bm{Z}\}}  \, \eta^{-1}\{\widetilde{\bm{\beta}}^\top \widetilde{\bm{Z}}(t)\} (U\wedge t)  \mid \widetilde{\bm{Z}}(t)\right]  \\
     & \quad + E\left[\frac{\mathbb{I}(U >t)}{P\{C \geq U \wedge t \mid \bm{Z}\}} \, \eta^{-1}\{\widetilde{\bm{\beta}}^\top \widetilde{\bm{Z}}(t) \} (U\wedge t) \mid \widetilde{\bm{Z}}(t)  \right] \\
     & = \eta^{-1}\{\widetilde{\bm{\beta}}^\top \widetilde{\bm{Z}}(t)\} E\left[  E\left\{\frac{\mathbb{I}(D\wedge C \leq t) \mathbb{I}(D\leq C) }{P\{C \geq U \wedge t \mid \bm{Z}\}} (U\wedge t) \mid D,\widetilde{\bm{Z}}(t)  \right\} \mid \widetilde{\bm{Z}}(t) \right] \\
     & \quad + \eta^{-1}\{\widetilde{\bm{\beta}}^\top\widetilde{\bm{Z}}(t)\} E\left[ E\left\{  \frac{\mathbb{I}(D\wedge C > t)}{P\{C \geq U \wedge t \mid \bm{Z}\}} (U\wedge t)    \mid D,\widetilde{\bm{Z}}(t)    \right\}     \mid \widetilde{\bm{Z}}(t) \right] \\
     & =  \eta^{-1}\{\widetilde{\bm{\beta}}^\top\widetilde{\bm{Z}}(t)\} E\left[  \left\{\frac{\mathbb{I}(D\leq t) \mathbb{I}(C \geq D\wedge t)}{P\{C \geq D\wedge t \mid \bm{Z} \}} (D)  \mid D,\widetilde{\bm{Z}}(t)  \right\} \mid \widetilde{\bm{Z}}(t)\right] \\
     & \quad + \eta^{-1}\{\widetilde{\bm{\beta}}^\top\widetilde{\bm{Z}}(t)\} E\left[ E\left\{\frac{\mathbb{I}(D > t) \mathbb{I}(C > t)}{P\{C \geq t \mid \bm{Z} \}} (t) \mid D,\widetilde{\bm{Z}}(t)  \right\} \mid \widetilde{\bm{Z}}(t)\right] \\
     & =  \eta^{-1}\{\widetilde{\bm{\beta}}^\top\widetilde{\bm{Z}}(t)\} E\left[\mathbb{I}(D\leq t) D \mid \widetilde{\bm{Z}}(t) \right] \\
     & \quad + \eta^{-1}\{\widetilde{\bm{\beta}}^\top\widetilde{\bm{Z}}(t)\} E\left[ \mathbb{I}(D >  t)(t) \mid \widetilde{\bm{Z}}(t)  \right]\\
     & = \eta^{-1}\{\widetilde{\bm{\beta}}^\top\widetilde{\bm{Z}}(t)\} E \left[ (D \wedge t)  \mid \widetilde{\bm{Z}}(t) \right].
\end{align*}
Combining all the above results, we can show that when \(P\{C \geq U \wedge t \mid \bm{Z}\}\) is correctly specified, 
\[
E[U(\widetilde{\bm{\beta}}) \mid \widetilde{\bm{Z}}(t)] = \mathbb{E} \left[ \widetilde{\bm{Z}}(t) \Bigg( {\mathcal{L}(\mathcal{H})(t)} - \eta^{-1}\{\widetilde{\bm{\beta}}^\top \widetilde{\bm{Z}}(t)\}(D \wedge t) \Bigg) \right] = 0.
\]
This completes the proof.

\section{Asymptotic properties under the independent data setting, when the censoring process is modeled by Cox proportional hazards regression} \label{supp:sec:proof_iid_asymp_cox}

In this section, we prove the asymptotic properties when  \(G(t \mid \bm Z)\) is estimated through the Cox proportional hazard model.  From \citet{jacobsen1989right,lin2007breslow}, by regularity conditions (i) - (iii) in the main paper
, we know that

\[\widehat{\bm\theta} - \bm\theta = \bm{I}_\theta^{-1} \left[ \frac{1}{n} \sum_{i=1}^{n} \int_{0}^{\tau} \left\{ \bm{Z}_i  - \overline{\bm{z}} (u,\bm\theta)  \right\} dM_i^C(u)   \right] + o_p(n^{-1/2}),   \]
where \(\overline{\bm{z}}(u,\bm\theta) = \frac{\bm{s}^{(1)}(u,\bm\theta)}{s^{(0)}(u,\bm{\theta}) } \), and 
\begin{align*}
    \bm{I}_\theta &= \mathbb{E} \left[ \int_{0}^{\tau} \mathbb{I}(U_i\geq u) \exp(\bm\theta^\top\bm{Z}_i) \left\{ \bm{Z}_i - \overline{\bm{z}}(u,\bm \theta)  \right\}^{\otimes 2}  d\Lambda_0^{C}(u) \right] \\
    & = \mathbb{E} \left( \int_{0}^{\tau} s^{(0)}(u,\bm\theta) \left[ \frac{\bm{s}^{(2)}(u,\bm\theta)}{s^{(0)} (u,\bm{\theta})} -  \left\{ \frac{\bm s^{(1)} (u,\bm \theta)}{ s^{(0)}(u,\bm\theta) }   \right\}^{\otimes 2}   \right] d\Lambda_0^C(u)  \right). 
\end{align*}
Since \(d\Lambda_0^{C}(u)\) is estimated through the Breslow estimator, by condition (v) in the main paper, we know that
\begin{align*}
\widehat{\Lambda}_0^C(u) - \Lambda_0^C(u) & = \frac{1}{n} \sum_{j=1}^{n} \Bigg[
  \int_{0}^{u} \frac{1}{s^{(0)}(s,\bm\theta)}  dM_j^C(s) \\
  & - \int_{0}^{u} \frac{\bm s^{(1)}(s,\bm\theta)}{s^{(0)}(s,\bm\theta)} \, d\Lambda_0^C(s)\;
    \bm I_\theta^{-1} \int_{0}^{\tau} \left\{ \bm Z_j - \overline{\bm z}(r,\bm\theta) \right\} dM_j^C(r) \Bigg] + o_p(n^{-1/2}).
\end{align*}
Combining the above, we know that

\begin{align}
\widehat G(u \mid \bm Z_i) - G(u \mid \bm Z_i)
&= - G(u \mid \bm Z_i) \exp(\bm\theta^\top \bm Z_i)
   \frac{1}{n} \sum_{j=1}^n
   \Bigg[
     \int_{0}^{u} \frac{1}{s^{(0)}(s,\bm\theta)} \, dM_j^C(s) \nonumber\\
& - \left\{ \int_{0}^{u} \frac{\bm s^{(1)}(s,\bm\theta)}{s^{(0)}(s,\bm\theta)} \, d\Lambda_0^C(s) \right\}^{\!\top}
       \bm I_\theta^{-1} \int_{0}^{\tau} \left\{ \bm Z_j - \overline{\bm z}(r,\bm\theta) \right\} dM_j^C(r) \nonumber \\
& + \Lambda_0^C(u)\, \bm Z_i^\top \bm I_\theta^{-1}
       \int_{0}^{\tau} \left\{ \bm Z_j - \overline{\bm z}(r,\bm\theta) \right\} dM_j^C(r)
   \Bigg] + o_p(n^{-1/2}). \label{cox_weigth}
\end{align}

We can express the estimating equation \(\bm{U}_n(\widetilde{\bm{\beta}})\) as

\begin{align}
    \bm{U}_n(\widetilde{\bm{\beta}}) = & \frac{1}{n} \sum_{i=1}^{n} \sum_{v=1}^{V} \frac{\mathbb{I} (U_i\leq t_v) \Delta_i + \mathbb{I}(U > t_v) }{ G(U_i\wedge t_v \mid \bm{Z}_i) } \widetilde{\bm{Z}}_i(t_v) \left[ \widetilde{L}_i(\mathcal{H}_i)(t_v) - \eta^{-1} \left\{ \widetilde{\bm{\beta}}^\top \widetilde{\bm{Z}}_i(t_v) \right\}  (U_i\wedge t_v)   \right]  \nonumber\\
    & + \left\{ \frac{\mathbb{I} (U_i\leq t_v) \Delta_i + \mathbb{I}(U > t_v) }{ \widehat{G}(U_i\wedge t_v \mid \bm{Z}_i) } - \frac{\mathbb{I} (U_i\leq t_v) \Delta_i + \mathbb{I}(U > t_v) }{ G(U_i\wedge t_v \mid \bm{Z}_i) } \right\}\nonumber \\
    &\times \widetilde{\bm{Z}}_i(t_v) \left[ \widetilde{L}_i(\mathcal{H}_i)(t_v) - \eta^{-1} \left\{ \widetilde{\bm{\beta}}^\top \widetilde{\bm{Z}}_i(t_v) \right\}  (U_i\wedge t_v)   \right] .\label{iid_extra}
\end{align}

Replacing \eqref{cox_weigth} into  \eqref{iid_extra} we can show that

\begin{align*}
    \eqref{iid_extra}
    &= \frac{1}{n} \sum_{i=1}^{n} \sum_{v=1}^{V} \frac{\mathbb{I}\!\left(U_i \le t_v\right)\Delta_i + \mathbb{I}\!\left(U_i > t_v\right)} {\widehat{G}\!\left(U_i \wedge t_v \mid \bm Z_i\right)}
    \left\{ \frac{1}{n} \sum_{j=1}^{n} \exp\!\left(\bm\theta^\top \bm Z_i\right)
    \left[ \int_{0}^{U_i \wedge t_v} \frac{1}{s^{(0)}(u,\bm\theta)} \, dM_j^C(u) \right. \right. \\
    &\left. {} - \left( \int_{0}^{U_i \wedge t_v} \frac{\bm s^{(1)}(u,\bm\theta)}{s^{(0)}(u,\bm\theta)} \, d\Lambda_0^C(u) \right)^{\!\top} \bm I_\theta^{-1} \int_{0}^{\tau} \left\{ \bm Z_j - \overline{\bm z}(r,\bm\theta) \right\} dM_j^C(r) \right. \\
    &\left. {} + \Lambda_0^C\!\left(U_i \wedge t_v\right) \bm Z_i^\top \bm I_\theta^{-1} \int_{0}^{\tau} \left\{ \bm Z_j - \overline{\bm z}(r,\bm\theta) \right\} dM_j^C(r) \right] \Bigg\} \\
    &\times \widetilde{\bm Z}_i(t_v)
    \left[\widetilde{L}_i(\mathcal{H}_i)(t_v) - \eta^{-1} \left\{ \widetilde{\bm\beta}^\top \widetilde{\bm Z}_i(t_v) \right\} \left(U_i \wedge t_v\right)
    \right]
    + o_p\left(n^{-1/2}\right).
\end{align*}
Thus,
\begin{align*}
n^{1/2}\,\bm U_n(\widetilde{\bm\beta})
&=n^{-1/2} \sum_{i=1}^{n} \sum_{v=1}^{V}
\frac{\mathbb{I} \left(U_i \le t_v\right)\Delta_i + \mathbb{I} \left(U_i > t_v\right)}
     {G\!\left(U_i \wedge t_v \mid \bm Z_i\right)} \widetilde{\bm Z}_i(t_v) \left[
  \widetilde{L}_i(\mathcal H_i)(t_v) - \eta^{-1} \left\{ \widetilde{\bm\beta}^\top \widetilde{\bm Z}_i(t_v) \right\} \left(U_i \wedge t_v\right)\right]  \\
&+ n^{-1/2}  \sum_{i=1}^{n} \sum_{v=1}^{V} \frac{\mathbb{I}\!\left(U_i \le t_v\right)\Delta_i + \mathbb{I} \left(U_i > t_v\right)} {\widehat G\!\left(U_i \wedge t_v \mid \bm Z_i\right)} \Bigg[\frac{1}{n} \sum_{j=1}^{n} \exp\left(\bm\theta^\top \bm Z_i\right) \\
&\times\Bigg\{\int_{0}^{U_i \wedge t_v} \frac{1}{s^{(0)}(u,\bm\theta)} \, dM_j^C(u)- \left( \int_{0}^{U_i \wedge t_v} \frac{\bm s^{(1)}(u,\bm\theta)}{s^{(0)}(u,\bm\theta)} \, d\Lambda_0^C(u) \right)^{\!\top}\bm I_\theta^{-1}\int_{0}^{\tau} \left\{ \bm Z_j - \overline{\bm z}(r,\bm\theta) \right\} dM_j^C(r) \\
& + \Lambda_0^C\!\left(U_i \wedge t_v\right) \bm Z_i^\top \bm I_\theta^{-1} \int_{0}^{\tau} \left\{ \bm Z_j - \overline{\bm z}(r,\bm\theta) \right\} dM_j^C(r) \Bigg\} \Bigg] \\
& \times \widetilde{\bm Z}_i(t_v) \left[ \widetilde{L}_i(\mathcal H_i)(t_v) - \eta^{-1} \left\{ \widetilde{\bm\beta}^\top \widetilde{\bm Z}_i(t_v) \right\} \left(U_i \wedge t_v\right)
\right] + o_p(1) \\
& = n^{-1/2} \sum_{i=1}^{n} \sum_{v=1}^{V}
\left[ \varpi_i(t_v;\widetilde{\bm\beta}) + \kappa_\theta\left(t_v;\widetilde{\bm\beta},\bm\theta,\Lambda_0^C\right)\,\xi_i(\bm\theta) + \int_{0}^{t_v} \zeta_i(u,\bm\theta)\,d\kappa_\Lambda\left(u,t_v;\widetilde{\bm\beta},\bm\theta,\Lambda_0^C\right) \right]
+ o_p(1) \\
& = n^{-1/2} \sum_{i=1}^{n} \phi_i (\widetilde{\bm{\beta}}) + o_p(1),
\end{align*}
where 
\begin{align*}
\varpi_i(t_v;\widetilde{\bm\beta}) &= \frac{\mathbb{I} \left(U_i \le t_v\right)\Delta_i + \mathbb{I} \left(U_i > t_v\right)} {G\!\left(U_i \wedge t_v \mid \bm Z_i\right)} \,\widetilde{\bm Z}_i(t_v)\,
\Big[ \widetilde{L}_i(\mathcal H_i)(t_v) - \eta^{-1} \big\{ \widetilde{\bm\beta}^\top \widetilde{\bm Z}_i(t_v) \big\} \left(U_i \wedge t_v\right)\Big] ,\\
\xi_i(\bm\theta) &= \int_{0}^{\tau} \left\{ \bm Z_i - \overline{\bm z}(r,\bm\theta) \right\} \, dM_i^C(r),\\
\zeta_i(u,\bm\theta) &= \int_{0}^{u} \frac{1}{s^{(0)}(s,\bm\theta)} \, dM_i^C(s),\\
\kappa_\theta\big(t_v;\widetilde{\bm\beta},\bm\theta,\Lambda_0^C\big) &= \lim_{n\to\infty} \frac{1}{n} \sum_{i=1}^{n}
\frac{\mathbb{I} \left(U_i \le t_v\right)\Delta_i + \mathbb{I} \left(U_i > t_v\right)} {G\!\left(U_i \wedge t_v \mid \bm Z_i\right)} \,
\widetilde{\bm Z}_i(t_v)\,
\Big[ \widetilde{L}_i(\mathcal H_i)(t_v) - \eta^{-1} \big\{ \widetilde{\bm\beta}^\top \widetilde{\bm Z}_i(t_v) \big\} \left(U_i \wedge t_v\right)\Big]\\
&\qquad \times \exp\!\left(\bm\theta^\top \bm Z_i\right)
\Bigg[ - \left\{ \int_{0}^{t_v} \frac{\bm s^{(1)}(u,\bm\theta)}{s^{(0)}(u,\bm\theta)} \, d\Lambda_0^C(u) \right\}^{\!\top}  + \Lambda_0^C(t_v)\, \bm Z_i^\top \Bigg]\bm I_\theta^{-1},\\
\kappa_\Lambda\big(u,t_v;\widetilde{\bm\beta},\bm\theta,\Lambda_0^C\big) &= \lim_{n\to\infty} \frac{1}{n} \sum_{i=1}^{n}
\frac{\mathbb{I} \left(U_i \le t_v\right)\Delta_i + \mathbb{I} \left(U_i > t_v\right)} {G\!\left(U_i \wedge t_v \mid \bm Z_i\right)} \,
\exp\!\left(\bm\theta^\top \bm Z_i\right)\,
\widetilde{\bm Z}_i(t_v)\\
& \times \Big[ \widetilde{L}_i(\mathcal H_i)(t_v) - \eta^{-1} \big\{ \widetilde{\bm\beta}^\top \widetilde{\bm Z}_i(t_v) \big\} \left(U_i \wedge t_v\right)\Big]\,
\mathbb{I}(u\le t_v),\\
\phi_i(\widetilde{\bm{\beta}}) & =  \sum_{v=1}^{V}
\left[\varpi_i(t_v;\widetilde{\bm\beta}) + \kappa_\theta\left(t_v;\widetilde{\bm\beta},\bm\theta,\Lambda_0^C\right)\,\xi_i(\bm\theta) + \int_{0}^{t_v} \zeta_i(u,\bm\theta)\,d\kappa_\Lambda\left(u,t_v;\widetilde{\bm\beta},\bm\theta,\Lambda_0^C\right)
\right].
\end{align*}

By the first-order Taylor expansion around the true value \(\widetilde{\bm{\beta}}^{0}\) and based on assumption (iv) in the main paper gives
\[
n^{1/2} \left\{\bm U_n(\widehat{\bm\beta}) -  \bm U_n(\widetilde{\bm\beta}^0) \right\} =  n^{1/2}  \left\{ \frac{1}{n} \sum_{i=1}^n \frac{\partial \phi_i(\bm\beta)}{\partial \bm\beta^\top} \Bigg|_{\bm\beta=\widetilde{\bm\beta}^0} \right\} (\widehat{\bm\beta} - \widetilde{\bm\beta}^{0}) + o_p(1) =  n^{1/2}\bm{\Omega}(\widetilde{\bm \beta}^0)(\widehat{\bm\beta} - \widetilde{\bm\beta}^0) + o_p(1). 
\]
Recall that \(n^{1/2} U_n(\widetilde{\bm{\beta}}) \) converges weakly to a zero-mean Gaussian distribution with covariance \(\bm{\Sigma}(\widetilde{\bm{\beta}})  = E\{ \bm{\phi}_i (\widetilde{\bm{\beta}}) ^{\otimes 2} \} \), by the Slutsky's theorem, we can show that
\[n^{1/2} (\widehat{\bm{\beta}} - \widetilde{\bm{\beta}}^{0}) \stackrel{p}{\to} \mathcal{N} (\bm{0}, \bm{V}),  \]
where \( \bm{V} = \Omega(\widetilde{\bm{\beta}}^0)^{-1} \bm{\Sigma}(\widetilde{\bm{\beta}}^0) \left\{\Omega(\widetilde{\bm{\beta}}^0)^{-1} \right\}^\top \). 

With the basis function for spline with \(\bm J(t)=(J_1(t),\ldots,J_R(t))^\top\) and \(\bm A(t)=\bm I_p\otimes \bm J(t)^\top\), we write \(\bm\beta(t)=\bm A(t)\widetilde{\bm\beta}\), and \(\widehat{\bm\beta}\) is the estimator of \(\widetilde{\bm{\beta}}\) with sandwich covariance \(\widehat{\bm V}=\widehat{\bm\Omega}^{-1}\widehat{\bm\Sigma} \left\{\widehat{\bm\Omega}^{-1} \right\}^{\top}\). Then, \(\widehat{\bm\beta}(t)=\bm A(t)\widehat{\bm\beta}\) and \(\widehat{\mathrm{Var}}\{\widehat{\bm\beta}(t)\}=\bm A(t)\widehat{\bm V}\bm A(t)^\top\), providing the \(95\%\) pointwise confidence interval for the \(j\)-th coefficient as \(\widehat{\beta}_j(t)\pm 1.96\,\sqrt{\bm e_j^\top \bm A(t)\widehat{\bm V} \bm A(t)^\top \bm e_j}\), where \(\bm e_j\) is the standard orthonormal basis that selects the \(j\)-th component. For the while-alive model \(\eta\{l(\mathcal H_i)(t\mid \bm Z_i)\}=\bm\beta(t)^\top\bm Z_i\), the conditional mean while-alive rate is \(m_i(t)=\eta^{-1}\{\bm\beta(t)^\top\bm Z_i\}\) with estimator \(\widehat m_i(t)=\eta^{-1}\{\widehat{\bm\beta}(t)^\top\bm Z_i\}\). By the delta method, its asymptotic variance at time $t$ is 
\[\widehat{\mathrm{Var}}\{\widehat m_i(t)\}=\big[\dot{\eta}^{-1}\{\widehat{\bm\beta}(t)^\top\bm Z_i\}\big]^2\,\bm Z_i^\top \bm A(t)\widehat{\bm V}_\theta \bm A(t)^\top \bm Z_i,\]
yielding the \(95\%\) pointwise confidence interval \(\widehat m_i(t)\pm 1.96\,\sqrt{\widehat{\mathrm{Var}}\{\widehat m_i(t)\}}\).

\section{Asymptotic properties under the independent data setting, when the censoring process is modeled by the Kaplan-Meier estimator} \label{supp:sec:proof_iid_asymp_km}

In this section, we provide the asymptotic properties for the proposed method for the independent data setting, when the censoring model is estimated by the Kaplan-Meier method.  By \cite{gill1980censor} and regularity conditions in Section 3.3 of the main paper, the Kaplan-Meier estimator can be represented by the linearization process as:
\begin{align} \label{km:martingale}
    n^{1/2}\{\widehat{G}(u) - G(u) \} = - n^{-1/2}G(u) \sum_{i=1}^{n} \int_{0}^{u} \frac{dM_i^C(r)}{ \lim_{n\to\infty} n^{-1} I(U_i \geq r) }   + o_p(1).
\end{align}

Using \eqref{km:martingale}, similar to Section \ref{supp:sec:proof_iid_asymp_cox}, we can express \(n^{1/2} \bm{U}_n(\widetilde{\bm{\beta}})\) as
\begin{align*}
    n^{1/2} \bm{U}_n(\widetilde{\bm{\beta}}) & = n^{-1/2} \sum_{i=1}^{n} \sum_{v=1}^{V} \Bigg(
    \frac{\mathbb{I} \left(U_i \le t_v\right)\Delta_i + \mathbb{I} \left(U_i > t_v\right)}
     {G\!\left(U_i \wedge t_v\right)} \widetilde{\bm Z}_i(t_v) \left[
  \widetilde{L}_i(\mathcal H_i)(t_v) - \eta^{-1} \left\{ \widetilde{\bm\beta}^\top \widetilde{\bm Z}_i(t_v) \right\} \left(U_i \wedge t_v\right)\right]  \\
  &  + \int_{0}^{t_v} \zeta_i^{KM}(u) d\kappa_\Lambda^{KM}(u,t_v;\widetilde{\bm{\beta}}) \Bigg) + o_p(1) \\
  & = n^{-1/2} \sum_{i=1}^{n} \sum_{v=1}^{V} \varpi_i^{KM} (t_v,\widetilde{\bm{\beta}}) + \int_{0}^{t_v} \zeta_i^{KM} (u)d\kappa_\Lambda^{KM}(u,t_v;\widetilde{\bm{\beta}})  + o_p(1) \\
  & = n{^-1} \sum_{i=1}^{n} \phi_i^{KM}(\widetilde{\bm{\beta}}) + o_p(1)
\end{align*}
where 
\begin{align*}
    \varpi_i^{KM} (t_v,\widetilde{\bm{\beta}}) &=  \frac{\mathbb{I} \left(U_i \le t_v\right)\Delta_i + \mathbb{I} \left(U_i > t_v\right)}
     {G\!\left(U_i \wedge t_v\right)} \widetilde{\bm Z}_i(t_v) \left[
  \widetilde{L}_i(\mathcal H_i)(t_v) - \eta^{-1} \left\{ \widetilde{\bm\beta}^\top \widetilde{\bm Z}_i(t_v) \right\} \left(U_i \wedge t_v\right)\right] \\
  \zeta_i^{KM}(u)& = \int_{0}^{u} \frac{dM_i^C(r)}{\lim_{n\to\infty} n^{-1} I(U_i \geq r) } \\
  \kappa_\Lambda^{KM}(u,t_v;\widetilde{\bm{\beta}}) &= \lim_{n\to\infty} n^{-1} \sum_{j=1}^{n} \frac{\mathbb{I} \left(U_j \le t_v\right)\Delta_i + \mathbb{I} \left(U_j > t_v\right)}
     {G\!\left(U_j \wedge t_v\right)} \widetilde{\bm Z}_j(t_v) \\
     & \times \left[
  \widetilde{L}_j(\mathcal H_j)(t_v) - \eta^{-1} \left\{ \widetilde{\bm\beta}^\top \widetilde{\bm Z}_j(t_v) \right\} \left(U_j \wedge t_v\right)\right] \mathbb{I}(U_j\wedge t_v \leq u)\\
  \phi_i^{KM}(\widetilde{\bm{\beta}}) & = \sum_{v=1}^{V} \varpi_i^{KM} (t_v,\widetilde{\bm{\beta}}) + \int_{0}^{t_v} \zeta_i^{KM} (u)d\kappa_\Lambda^{KM}(u,t_v;\widetilde{\bm{\beta}}).
\end{align*}
Then, by the first-order Taylor expansion and Slutsky's theorem, we can show that 
\[n^{1/2} (\widehat{\bm{\beta}} - \widetilde{\bm{\beta}}^{0}) \stackrel{p}{\to} \mathcal{N} (\bm{0}, \bm{V}^{KM}),  \]
where \( \bm{V}^{KM} = \Omega(\widetilde{\bm{\beta}}^0)^{-1} \bm{\Sigma}^{KM}(\widetilde{\bm{\beta}}^0) \left\{\Omega(\widetilde{\bm{\beta}}^0)^{-1} \right\}^\top \), and \(\bm\Sigma^{KM}(\widetilde{\bm{\beta}}^0) = \mathbb{E} \left[ \left\{ \phi_i^{KM}(\widetilde{\bm{\beta}}) \right\}^{\otimes 2} \right]\).

\section{Connection between the proposed regression model and while-alive loss rate estimator in \citet{mao2023nonparametric}} \label{supp:sec:connect_mao}

In this section, we will point out the connection between our proposed regression estimator for the while-alive loss rate and the non-parametric estimator proposed by \citet{mao2023nonparametric}. Specifically, when the regression model only includes a treatment variable such that $\bm Z_i=(1,A_i)^\top$ and $\eta(\cdot)$ is the link function, the regression model implies
\[
l(\mathcal H)(t\mid A=a)=\eta^{-1}\!\left\{\beta_0(t)+\beta_1(t)a\right\},\quad a\in\{0,1\},
\]
so that
\[
l(\mathcal H)(t\mid A=1)=\eta^{-1}\!\left\{\beta_0(t)+\beta_1(t)\right\},\qquad
l(\mathcal H)(t\mid A=0)=\eta^{-1}\!\left\{\beta_0(t)\right\}.
\]
From the estimating equation (3.3), the arm-specific estimator is 
\[
\widehat l(t^\ast\mid A=a)
=\frac{\sum_{i=1}^n I(A_i=a)\,W_i^{(a)}(t^\ast)\,\mathcal{L}_i(t^\ast)}
       {\sum_{i=1}^n I(A_i=a)\,W_i^{(a)}(t^\ast)\,(D_i\wedge t^\ast)}\,,
\qquad a\in\{0,1\},
\]
with IPCW weights $\widehat{W}_i(t^\ast \mid A=a) = \frac{\mathbb{I}(U_i \leq t^{\ast})\Delta_i + \mathbb{I}(U_i > t^{\ast})}{G(U_i\wedge t^{\ast} \mid A=a)}$ based on a consistent estimator $\widehat G(t^\ast\mid A=a)$ as the censoring probability from either the Cox proportional hazard model, or the arm-specific Kaplan-Meier estimator. \citet{mao2023nonparametric} proposed an arm-specific plug-in estimator at $t^\ast$ for while-alive loss rate which is defined as
\begin{equation} \label{mao_est}
    \widehat l^{(a)}(t^\ast) =\frac{\int_0^{t^\ast} \widehat S^{(a)}(u-)\,d\widehat R^{(a)}(u)}
       {\int_0^{t^\ast} \widehat S^{(a)}(u)\,du}\,,
\end{equation}
where \(d\widehat R^{(a)}(u)  = \frac{\sum_{i:A_i=a} Y_i(u) d\mathcal{L}_i(u) }{ \sum_{i:A_i=a}Y_i(u) } \), and  \(\widehat{S}^{(a)}(u)\) is the arm-specific Kaplan-Meier estimator of \(\mathbb{P}(D_i^{(a)} \geq t)\) for \(A=a\). For a finite process \(\mathcal{L}_i(u)\),  we can show that
\begin{align*}
    & \int_{0}^{t^{*}} \widehat{S}^{(a)}(u-) d\widehat{R}^{(a)}(u) = \frac{1}{n_a} \sum_{i:A_i=a} \int_{0}^{t^{*}} \frac{Y_i(u)}{\widehat{G}^{(a)}(u^{-})} d\mathcal{L}_i(u) + o_p(n^{-1/2}),\\
    & \int_{0}^{t^{*}} \widehat{S}^{(a)}(u)du = \frac{1}{n_a} \sum_{i:A_i=a} \int_{0}^{t^{*}} \frac{Y_i(u)}{\widehat{G}^{(a)}(u)} \, du + o_p(n^{-1/2}),
\end{align*}
where \(Y_i(u) = \mathbb{I}(U_i \geq u)\) is risk set.
Because 
\[ \int_{0}^{t^{*}} Y_i(u) d\mathcal{L}_i(u) = \mathcal{L}_i(t^{*}), \quad \int_{0}^{t^{*}} Y_i(u) du = D_i\wedge t^{*}, \]
we can show that, 
\begin{align*}
    &\int_{0}^{t^{*}} \frac{Y_i(u)}{\widehat{G}^{(a)}(u-)} d\mathcal{L}_i(u) = \widetilde{W}_i(t^{*} \mid A = a) \mathcal{L}_i(t^{*}) + o_p(n_a^{1/2}),\\
    &\int_{0}^{t^{*}} \frac{Y_i(u)}{\widehat{G}^{(a)}(u)} du = \widetilde{W}_i(t^{*} \mid A = a) (D_i\wedge t^{*})+ o_p(n_a^{1/2}).
\end{align*}

Plug back into \eqref{mao_est}, we can show that
\[ \widehat{l}^{(a)}(t^{*}) = \frac{\sum_{i=1}^n I(A_i=a)\,W_i^{(a)}(t^\ast)\,L_i(t^\ast)}{\sum_{i=1}^n I(A_i=a)\,W_i^{(a)}(t^\ast)\,(D_i\wedge t^\ast)} + o_p(n^{-1/2}) = \widehat{l}(t^{*} \mid A=a) + o_p(n^{-1/2}). \]

Therefore, the estimator for the while-alive loss rate from our proposed estimating equation is asymptotically equivalent to the non-parametric estimator proposed by \citet{mao2023nonparametric}.

\section{Asymptotic properties under the cluster-correlated data setting, when the censoring process is modeled by marginal Cox proportional hazards regression} \label{supp:sec:proof_cluster_asymp_cox}

In this section, we present the asymptotic properties of the proposed estimator when the censoring distribution is covariate-dependent and modeled via a marginal Cox model $\lambda_{ij}^C(t) = \lambda_0^C(t) \exp(\bm{\theta}_C^\top \bm{Z}_{ij})$, under the cluster-correlated data setting. The regularity conditions are given below: 
 \begin{enumerate}[(I)]
     \vspace{-0.4cm}\item The cluster-level observations \( \{\bm{\mathcal{H}}_i ,\bm{U}_i, \bm{\Delta}_i, \bm{Z}_i \}\) are independent across clusters for \(i=1,\dots, M\). \label{crt_c1}
     \vspace{-0.4cm}\item \( P(T_{ij,k} \geq t ) > 0  \) and \(P(D_{ij} \geq t) > 0 \) for \(t \in (0,\tau)\), \(i=1,\dots, M\), and \(j=1,\dots, N_i\). \label{crt_c2}
     \vspace{-0.4cm}\item The covariate \(Z_{ijl}\) is bounded for \(i=1,\dots, M\), \(j=1,\dots, N_i\), and  \(l=1,\dots, p\). \label{crt_c3}
     \vspace{-0.4cm}\item \(0 < G(t\mid\widetilde{\bm{Z}}_{ij}) \leq 1 \) is absolutely continuous for \(t\in (0,\tau]\), and the cumulative hazard function for censoring \(\Lambda^C(t\mid\widetilde{\bm{Z}}_{ij}) = \int_{0}^{t} \lambda^C(u\mid\widetilde{\bm{Z}}_{ij})\,du\) is absolutely continuous for \(t\in (0,\tau]\). \label{crt_c4}
     \vspace{-0.4cm}\item \(\overline{\bm{\Omega}}(\widetilde{\bm{\beta}}) = \mathbb{E}\!\left\{\sum_{j=1}^{N_i}\sum_{v=1}^{V} \widetilde{\bm{Z}}(t_v)^{\otimes 2} \left[ \dot{\eta}^{-1}\!\big( \widetilde{\bm{\beta}}^\top \widetilde{\bm{Z}}(t_v)\big)\, (D \wedge t_v) \right] \right\} \) is positive definite for any \(t_v\in(0,\tau]\)), where \(\dot{\eta}^{-1}(\cdot) \) is the derivative of \(\eta^{-1}(\cdot)\). \label{crt_c5}
     \vspace{-0.4cm}\item The matrix 
     \[
     \overline{\bm{I}}_\theta =E\!\left\{\int_{0}^{t_v} \left[\left\{\frac{\bm{s}^{(2)}(u;\bm{\theta}_C)}{s^{(0)}(u;\bm{\theta}_C)} - \left( \frac{\bm{s}^{(1)}(u;\bm{\theta}_C)}{s^{(0)}(u;\bm{\theta}_C)} \right)^{\otimes 2} \right\} s^{(0)}(u;\bm{\theta}_C)\, d\Lambda_{0}^C(u) \right] \right\}
     \]
     is positive definite, where \(\bm{s}^{(d)}(u;\bm{\theta}_C) = E\!\left[ \mathbb{I}(U_{ij} \geq u)\, \bm{Z}_{ij}^{\otimes d}\exp(\bm{\theta}_C^\top \bm{Z}_{ij})\right]\) for \(d=0,1,2\), and \(s^{(0)}(u;\bm{\theta}_C)\) is bounded away from zero. \label{crt_c6}
     \vspace{-0.4cm}\item The cluster size \(N_i\) is bounded in probability, i.e.,\(\max_i N_i=O_p(1)\). \label{crt_c7}
 \end{enumerate}

\begin{theorem}\label{thm:assym_cluster}
    Under the regularity conditions, \(\sqrt{M} (\overline{\bm{\beta}} -\bm{\widetilde{\beta}}^0 ) \) converge weakly to a zero-mean Gaussian distribution with variance $\bm{\overline{V}} = \overline{\bm{\Omega}}(\widetilde{\bm{\beta}}^0)^{-1} \overline{\bm{\Sigma}}(\widetilde{\bm{\beta}}^0) \overline{\bm{\Omega}}(\widetilde{\bm{\beta}}^0)^{-1}$, where $\overline{\bm{\Sigma}}(\widetilde{\bm{\beta}}^0) = E\left[ \psi_i ^{\otimes 2} \right]$, and 
\begin{align*}
\phi_{ij}(\widetilde{\bm{\beta}}^0) & =  \sum_{v=1}^{V}
\left[\varpi_{ij}(t_v;\widetilde{\bm\beta}^0) + \kappa_\theta\left(t_v;\widetilde{\bm\beta}^0,\bm\theta_C,\Lambda_0^C\right)\,\xi_{ij}(\bm\theta_C) + \int_{0}^{t_v} \zeta_{ij}(u,\bm\theta_C)\,d\kappa_\Lambda\left(u,t_v;\widetilde{\bm\beta}^0,\bm\theta_C,\Lambda_0^C\right)
\right]\\
\phi_i(\widetilde{\bm{\beta}}^0) & = \sum_{j=1}^{N_i} \phi_{ij}(\widetilde{\bm{\beta}}^0)\\
\end{align*}
where   
\begin{align*}
\varpi_{ij}(t_v;\widetilde{\bm\beta}^0) &= \frac{\mathbb{I} \left(U_{ij} \le t_v\right)\Delta_{ij} + \mathbb{I} \left(U_{ij} > t_v\right)} {G\!\left(U_{ij} \wedge t_v \mid \bm Z_{ij}\right)} \,\widetilde{\bm Z}_{ij}(t_v)\,
\Big[ \widetilde{L}_{ij}(\mathcal H_{ij})(t_v) - \eta^{-1} \big\{(\widetilde{\bm\beta}^0)^\top \widetilde{\bm Z}_{ij}(t_v) \big\} \left(U_{ij} \wedge t_v\right)\Big] ,\\
\xi_{ij}(  \bm\theta_C) &= \int_{0}^{\tau} \left\{ \bm Z_{ij} - \overline{\bm z}(r,  \bm\theta_C) \right\} \, dM_{ij}^C(r),\\
\zeta_{ij}(u,  \bm\theta_C) &= \int_{0}^{u} \frac{1}{s^{(0)}(s,  \bm\theta_C)} \, dM_{ij}^C(s),\\
\kappa_\theta\big(t_v;\widetilde{\bm\beta}^0,  \bm\theta_C,\Lambda_0^C\big) &= \lim_{M\to\infty} \frac{1}{M} \sum_{i=1}^{M} \sum_{j=1}^{N_i}
\frac{\mathbb{I} \left(U_{ij} \le t_v\right)\Delta_{ij} + \mathbb{I} \left(U_{ij} > t_v\right)} {G\!\left(U_{ij} \wedge t_v \mid \bm Z_{ij}\right)} \,
\widetilde{\bm Z}_{ij}(t_v)\,\\
&\qquad \times
\Big[ \widetilde{L}_{ij}(\mathcal H_{ij})(t_v) - \eta^{-1} \big\{ (\widetilde{\bm\beta}^0)^\top \widetilde{\bm Z}_{ij}(t_v) \big\} \left(U_{ij} \wedge t_v\right)\Big]\\
&\qquad \times \exp\!\left(  \bm\theta_C^\top \bm Z_{ij}\right)
\Bigg[ - \left\{ \int_{0}^{t_v} \frac{\bm s^{(1)}(u,  \bm\theta_C)}{s^{(0)}(u,  \bm\theta_C)} \, d\Lambda_0^C(u) \right\}^{\!\top}  + \Lambda_0^C(t_v)\, \bm Z_{ij}^\top \Bigg]\bm I_\theta^{-1},\\
\kappa_\Lambda\big(u,t_v;\widetilde{\bm\beta}^0,  \bm\theta_C,\Lambda_0^C\big) &= \lim_{M\to\infty} \frac{1}{M} \sum_{i=1}^{M} \sum_{j=1}^{N_i}
\frac{\mathbb{I} \left(U_{ij} \le t_v\right)\Delta_{ij} + \mathbb{I} \left(U_{ij} > t_v\right)} {G\!\left(U_{ij} \wedge t_v \mid \bm Z_{ij}\right)} \,
\exp\!\left(  \bm\theta_C^\top \bm Z_{ij}\right)\,
\widetilde{\bm Z}_{ij}(t_v)\\
& \qquad \times \Big[ \widetilde{L}_{ij}(\mathcal H_{ij})(t_v) - \eta^{-1} \big\{ (\widetilde{\bm\beta}^0)^\top \widetilde{\bm Z}_{ij}(t_v) \big\} \left(U_{ij} \wedge t_v\right)\Big]\,
\mathbb{I}(u\le t_v),\\
\phi_{ij}(\widetilde{\bm{\beta}}) & =  \sum_{v=1}^{V}
\left[\varpi_{ij}(t_v;\widetilde{\bm\beta}^0) + \kappa_\theta\left(t_v;\widetilde{\bm\beta}^0, \bm\theta_C,\Lambda_0^C\right)\,\xi_{ij}( \bm\theta_C) + \int_{0}^{t_v} \zeta_{ij}(u, \bm\theta_C)\,d\kappa_\Lambda\left(u,t_v;\widetilde{\bm\beta}^0, \bm\theta_C,\Lambda_0^C\right)
\right].
\end{align*}
\end{theorem}

The asymptotic property for independent data can be viewed as a special case of cluster-dependent data with a cluster size of one. The proof of Theorem  \ref{thm:assym_cluster} is provided below.

Define the marginal martingale increment \(M_{ij}^C(u)=I(U_{ij}\leq u, \Delta_{ij}=0)-\int_0^u\mathbb{I}(U_{ij} \geq r)\exp(\bm\theta_C^\top\bm Z_{ij})\,d\Lambda_0^C(r)\). Since clusters are independent, with arbitrary within-cluster dependence, we use a marginal Cox model to estimate \(G(u\mid\bm Z_{ij})\). Then, by \cite{zhong2022restricted,zhang2011estimating}, similar to Section \ref{supp:sec:proof_iid_asymp_cox}, we can show that for \(u\le\tau\),

\begin{align*}
M^{1/2}\,\bm U_n^{*}(\widetilde{\bm\beta})
&=M^{-1/2} \sum_{i=1}^{M} \sum_{j=1}^{N_{i}} \sum_{v=1}^{V}
\frac{\mathbb{I} \left(U_{ij} \le t_v\right)\Delta_{ij} + \mathbb{I} \left(U_{ij} > t_v\right)}
     {G\!\left(U_{ij} \wedge t_v \mid \bm Z_{ij}\right)} \widetilde{\bm Z}_{ij}(t_v) \left[
  \widetilde{L}_{ij}(\mathcal H_{ij})(t_v) - \eta^{-1} \left\{ \widetilde{\bm\beta}^\top \widetilde{\bm Z}_{ij}(t_v) \right\} \left(U_{ij} \wedge t_v\right)\right]  \\
&+ M^{-1/2}  \sum_{i=1}^{M} \sum_{j=1}^{N_i} \sum_{v=1}^{V} \frac{\mathbb{I}\!\left(U_{ij} \le t_v\right)\Delta_{ij} + \mathbb{I} \left(U_{ij} > t_v\right)} {\widehat G\!\left(U_{ij} \wedge t_v \mid \bm Z_{ij}\right)} \Bigg[\frac{1}{M} \sum_{i'=1}^{M}\sum_{j'=1}^{N_i} \exp\left(\bm\theta_C^\top \bm Z_{ij}\right) \\
&\times\Bigg\{\int_{0}^{U_{ij} \wedge t_v} \frac{1}{s^{(0)}(u,\bm\theta_C)} \, dM_{i'j'}^C(u)- \left( \int_{0}^{U_{ij} \wedge t_v} \frac{\bm s^{(1)}(u,\bm\theta_C)}{s^{(0)}(u,\bm\theta_C)} \, d\Lambda_0^C(u) \right)^{\!\top}\overline{\bm I}_\theta^{-1}\int_{0}^{\tau} \left\{ \bm Z_j - \overline{\bm z}(r,\bm\theta_C) \right\} dM_{i'j'}^C(r) \\
& + \Lambda_0^C\!\left(U_{ij} \wedge t_v\right) \bm Z_{ij}^\top \overline{\bm I}_\theta^{-1} \int_{0}^{\tau} \left\{ \bm Z_{i'j'} - \overline{\bm z}(r,\bm\theta_C) \right\} dM_{i'j'}^C(r) \Bigg\} \Bigg] \\
& \times \widetilde{\bm Z}_{ij}(t_v) \left[ \widetilde{L}_{ij}(\mathcal H_{ij})(t_v) - \eta^{-1} \left\{ \widetilde{\bm\beta}^\top \widetilde{\bm Z}_{ij}(t_v) \right\} \left(U_{ij} \wedge t_v\right)
\right] + o_p(1) \\
& = n^{-1/2} \sum_{i=1}^{n} \sum_{v=1}^{V}
\left[ \varpi_{ij}(t_v;\widetilde{\bm\beta}) + \kappa_\theta\left(t_v;\widetilde{\bm\beta},\bm\theta_C,\Lambda_0^C\right)\,\xi_{ij}(\bm\theta_C) + \int_{0}^{t_v} \zeta_{ij}(u,\bm\theta_C)\,d\kappa_\Lambda\left(u,t_v;\widetilde{\bm\beta},\bm\theta_C,\Lambda_0^C\right) \right]
+ o_p(1) \\
& = n^{-1/2} \sum_{i=1}^{n} \phi_{ij} (\widetilde{\bm{\beta}}) + o_p(1).
\end{align*}
For cluster-correlated data, \(\psi_i(\bm\beta)=\sum_{j=1}^{N_i}\phi_{ij}(\bm\beta)\) is independent for cluster \(i=1,\dots, M\). Then based on regularity condition (\ref{crt_c1}) and (\ref{crt_c7}), by first-order Taylor expansion around \(\widetilde{\bm\beta}^0\), it gives
\[
M^{1/2}(\widehat{\bm\beta}-\widetilde{\bm\beta}^{0})
\ \Rightarrow\ \mathcal N\!\big(\bm 0,\ \overline{\bm V}\big),\qquad
\overline{\bm V}=\overline{\bm\Omega}(\widetilde{\bm \beta}^0)^{-1}\,\overline{\bm\Sigma}\ \left\{\overline{\bm\Omega}(\widetilde{\bm \beta}^0)^{-1}\right\}^\top.
\]

\section{Asymptotic properties under the cluster-correlated data setting, when the censoring process is modeled by the Kaplan-Meier estimator}  \label{supp:sec:proof_cluster_asymp_km}
Similar to Section \ref{supp:sec:proof_iid_asymp_km}, we present below the asymptotic properties of the proposed method for cluster-correlated data when the censoring distribution is estimated using the Kaplan-Meier estimator.
\begin{theorem}
        Under the regularity conditions in Section \ref{supp:sec:proof_cluster_asymp_cox} (except condition (\ref{crt_c6}) ), \(\sqrt{M} (\overline{\bm{\beta}} -\bm{\widetilde{\beta}}^0 ) \) converge weakly to a zero-mean Gaussian distribution with variance $\bm{\overline{V}}^{KM} = \overline{\bm{\Omega}}(\widetilde{\bm{\beta}}^0)^{-1} \overline{\bm{\Sigma}}^{KM}(\widetilde{\bm{\beta}}^0) \overline{\bm{\Omega}}(\widetilde{\bm{\beta}}^0)^{-1}$, where $\overline{\bm{\Sigma}}^{KM}(\widetilde{\bm{\beta}}^0) = E\left[ \{\psi_i^{KM}(\widetilde{\bm{\beta}}^0)\} ^{\otimes 2} \right]$, and 
\begin{align*}
  \phi_{ij}^{KM}(\widetilde{\bm{\beta}}^0) & = \sum_{v=1}^{V} \varpi_{ij}^{KM} (t_v,\widetilde{\bm{\beta}}^0) + \int_{0}^{t_v} \zeta_{ij}^{KM} (u)d\kappa_\Lambda^{KM}(u,t_v;\widetilde{\bm{\beta}})\\
  \phi_i^{KM}(\widetilde{\bm{\beta}}^0) & = \sum_{j=1}^{N_i} \phi_{ij}^{KM}(\widetilde{\bm{\beta}}^0)\\
    \varpi_{ij}^{KM} (t_v,\widetilde{\bm{\beta}}^0) &=  \frac{\mathbb{I} \left(U_{ij} \le t_v\right)\Delta_{ij} + \mathbb{I} \left(U_{ij} > t_v\right)}
     {G\!\left(U_{ij} \wedge t_v\right)} \widetilde{\bm Z}_{ij}(t_v) \left[
  \widetilde{L}_{ij}(\mathcal H_{ij})(t_v) - \eta^{-1} \left\{ (\widetilde{\bm\beta}^0)^\top \widetilde{\bm Z}_{ij}(t_v) \right\} \left(U_{ij} \wedge t_v\right)\right] \\
  \zeta_{ij}^{KM}(u)& = \int_{0}^{u} \frac{dM_{ij}^C(r)}{\lim_{M\to\infty} M^{-1} \sum_{i=1}^{M} \sum_{j=1}^{N_i} I(U_{ij} \geq r) } \\
  \kappa_\Lambda^{KM}(u,t_v;\widetilde{\bm{\beta}}^0) &= \lim_{M\to\infty} M^{-1} \sum_{i'=1}^{M} \sum_{j'=1}^{N_i} \frac{\mathbb{I} \left(U_{i'j'} \le t_v\right)\Delta_{ij} + \mathbb{I} \left(U_{i'j'} > t_v\right)}
     {G\!\left(U_{i'j'} \wedge t_v\right)} \widetilde{\bm Z}_{i'j'}(t_v) \\
     & \times \left[
  \widetilde{L}_{i'j'}(\mathcal H_{i'j'})(t_v) - \eta^{-1} \left\{( \widetilde{\bm\beta}^0)^\top \widetilde{\bm Z}_{i'j'}(t_v) \right\} \left(U_{i'j'} \wedge t_v\right)\right] \mathbb{I}(U_{i'j'}\wedge t_v \leq u).
\end{align*}
        
\end{theorem}

\section{Extension of the average Hazard regression of \citet{uno2024regression} to the cluster-correlated data setting}\label{supp:sec:AH_clustered}

For cluster-correlated data, setting \( w_k = 0 \) for all \( k = 1, \dots, K \) reduces the while-alive regression in equation (3.2) of the main paper to the average hazard model \citep{uno2024regression} at landmark time \( t \), with estimating equation
\begin{align*}
    \bm{U}_n^{*}\{\bm{\beta}(t)\} = M^{-1} \sum_{i=1}^{M} \sum_{j=1}^{N_i}  \frac{\mathbb{I}(U_{ij} \leq t) \Delta_{ij} + \mathbb{I}(U_{ij} > t)}{G(U_{ij} \wedge t \mid \bm{Z}_{ij})} \bm{Z}_{ij} \Big[\mathbb{I}(U_{ij}\leq t,\Delta_{ij}=1) - \eta^{-1}\{\bm{\beta}(t)^\top \bm{Z}_{ij}\}(U_{ij} \wedge t)\Big].
\end{align*}

When the censoring distribution \( G(U_{ij} \wedge t \mid \bm{Z}_{ij}) \) is modeled by a marginal Cox model,  Theorem \ref{thm:assym_cluster} provides the corresponding asymptotic properties for the average hazard model under cluster-dependent data. This can be viewed as an extension of the asymptotic results in \citet{uno2024regression} to address clustering of observations.

\begin{theorem}\label{thm:asymp_ah_cluster}
Under the regularity conditions in Section \ref{supp:sec:proof_cluster_asymp_cox}, 
\(\sqrt{M}\,\{\widehat{\bm{\beta}}(t) - \bm{\beta}(t)^0 \}\) converges weakly to a zero-mean Gaussian distribution with variance 
\(\bm{\overline{V}} = \overline{\bm{\Omega}}\!\big\{\bm{\beta}(t)^0\big\}^{-1} \overline{\bm{\Sigma}}\!\big\{\bm{\beta}(t)^0\big\} \overline{\bm{\Omega}}\!\big\{\bm{\beta}(t)^0\big\}^{-1}\), where \(\overline{\bm{\Sigma}}\!\big\{\bm{\beta}(t)^0\big\} = E\!\left[ \psi_i ^{\otimes 2} \right]\), and
\begin{align*}
\phi_{ij}\!\big\{\bm{\beta}(t)^0\big\}  &=  \varpi_{ij}\!\big\{t;\bm{\beta}(t)^0\big\} 
+ \kappa_\theta\!\big\{t;\bm{\beta}(t)^0,\bm\theta_C,\Lambda_0^C\big\}\,\xi_{ij}(\bm\theta_C) 
+ \int_{0}^{t} \zeta_{ij}(u,\bm\theta_C)\,d\kappa_\Lambda\!\big\{u,t;\bm{\beta}(t)^0,\bm\theta_C,\Lambda_0^C\big\},\\
\phi_i\!\big\{\bm{\beta}(t)^0\big\} 
&= \sum_{j=1}^{N_i} \phi_{ij}\!\big\{\bm{\beta}(t)^0\big\},\\
\varpi_{ij}\{t;\bm{\beta}(t)^0\} &= \frac{\mathbb{I} \left(U_{ij} \le t\right)\Delta_{ij} + \mathbb{I} \left(U_{ij} > t\right)} {G\!\left(U_{ij} \wedge t \mid \bm Z_{ij}\right)} \,\bm{Z}_{ij}\,
\Big[ \big\{ \mathbb{I}(U_{ij}\leq t,\Delta_{ij}=1) - \\
& \qquad \eta^{-1} \big\{ (\bm{\beta}(t)^0\}^\top \bm{Z}_{ij} \big)\,(U_{ij} \wedge t) \big\} \Big] ,\\
\xi_{ij}(\bm\theta) &= \int_{0}^{\tau} \left\{ \bm Z_{ij} - \overline{\bm z}(r,\bm\theta) \right\} \, dM_{ij}^C(r),\\
\zeta_{ij}(u,\bm\theta) &= \int_{0}^{u} \frac{1}{s^{(0)}(s,\bm\theta)} \, dM_{ij}^C(s),\\
\kappa_\theta\big\{t;\bm{\beta}(t)^0,\bm\theta,\Lambda_0^C\big\} &= \lim_{M\to\infty} \frac{1}{M} \sum_{i=1}^{M} \sum_{j=1}^{N_i}
\frac{\mathbb{I} \left(U_{ij} \le t\right)\Delta_{ij} + \mathbb{I} \left(U_{ij} > t\right)} {G\!\left(U_{ij} \wedge t \mid \bm Z_{ij}\right)} \,
\bm{Z}_{ij}\,\\
&\qquad \times \Big[ \big\{ \mathbb{I}(U_{ij}\leq t,\Delta_{ij}=1) - \eta^{-1} \big\{ (\bm{\beta}(t)^0\}^\top \bm{Z}_{ij} \big)\,(U_{ij} \wedge t) \big\} \Big]\\
&\qquad \times \exp\!\left(\bm\theta^\top \bm Z_{ij}\right)
\Bigg[ - \left\{ \int_{0}^{t} \frac{\bm s^{(1)}(u,\bm\theta)}{s^{(0)}(u,\bm\theta)} \, d\Lambda_0^C(u) \right\}^{\!\top}  + \Lambda_0^C(t)\, \bm Z_{ij}^\top \Bigg]\bm I_\theta^{-1},\\
\kappa_\Lambda\big\{u,t;\bm{\beta}(t)^0,\bm\theta,\Lambda_0^C\big\} &= \lim_{M\to\infty} \frac{1}{M} \sum_{i=1}^{M} \sum_{j=1}^{N_i}
\frac{\mathbb{I} \left(U_{ij} \le t\right)\Delta_{ij} + \mathbb{I} \left(U_{ij} > t\right)} {G\!\left(U_{ij} \wedge t \mid \bm Z_{ij}\right)} \,
\exp\!\left(\bm\theta^\top \bm Z_{ij}\right)\,
\bm{Z}_{ij}\\
& \qquad \times \Big[ \big\{ \mathbb{I}(U_{ij}\leq t,\Delta_{ij}=1) - \eta^{-1} \big\{ (\bm{\beta}(t)^0\}^\top \bm{Z}_{ij} \big)\,(U_{ij} \wedge t) \big\} \Big]\,
\mathbb{I}(u\le t).
\end{align*}
\end{theorem}

\section{Simulation results for Scenario I(a), I(c) and II(a), II(b) under the independent data setting} \label{supp::sec:iid_sim_add}

\begin{table}[htbp!]
\caption{Simulation results under Scenario I(a) and I(c) with $R=7$ equally spaced knots and stacking time at $(1.0,1.5,2.0,2.5,3.0,3.5,4.0)$ and covariate-dependent censoring. Sample size $n=2000$; censoring rate = $25\%$. Metrics: ABias (absolute bias), MCSD (Monte Carlo SD), AESE (asymptotic SE), CP (coverage of the $95\%$ CI). Column blocks compare weights $(1,1,1)$ vs.\ $(1,2,2)$.}
\label{tab:scenario_iid_iac}
\centering
\small
\begin{tabular}{cccccccccccc}
\toprule
 &  & \multicolumn{5}{c}{$(w_1,w_2,w_D)=(1,1,1)$} & \multicolumn{5}{c}{$(w_1,w_2,w_D)=(1,2,2)$} \\
\cmidrule(lr){3-7} \cmidrule(lr){8-12}
Time & $\bm{\beta}(t)$ & True & ABias & MCSD & AESE & CP & True & ABias & MCSD & AESE & CP \\
\midrule
1.0 & $\beta_1$ &  0.904 & 0.001 & 0.033 & 0.034 & 0.953 & 1.341 & 0.001 & 0.034 & 0.035 & 0.957 \\
    & $\beta_2$ & -0.008 & 0.000 & 0.050 & 0.050 & 0.952 & 0.355 & 0.000 & 0.055 & 0.055 & 0.963 \\
\midrule
1.5 & $\beta_1$ &  0.871 & 0.001 & 0.034 & 0.035 & 0.959 & 1.333 & 0.001 & 0.032 & 0.033 & 0.958 \\
    & $\beta_2$ & -0.146 & 0.003 & 0.049 & 0.048 & 0.939 & 0.195 & 0.003 & 0.056 & 0.056 & 0.949 \\
\midrule
2.0 & $\beta_1$ &  0.839 & 0.001 & 0.035 & 0.037 & 0.957 & 1.312 & 0.000 & 0.034 & 0.035 & 0.958 \\
    & $\beta_2$ & -0.238 & 0.004 & 0.047 & 0.046 & 0.939 & 0.077 & 0.004 & 0.056 & 0.056 & 0.950 \\
\midrule
2.5 & $\beta_1$ &  0.812 & 0.001 & 0.038 & 0.039 & 0.959 & 1.287 & 0.001 & 0.036 & 0.039 & 0.968 \\
    & $\beta_2$ & -0.304 & 0.004 & 0.046 & 0.045 & 0.940 & -0.013 & 0.004 & 0.056 & 0.057 & 0.945 \\
\midrule
3.0 & $\beta_1$ &  0.789 & 0.002 & 0.040 & 0.041 & 0.957 & 1.261 & 0.002 & 0.040 & 0.042 & 0.960 \\
    & $\beta_2$ & -0.354 & 0.004 & 0.045 & 0.044 & 0.942 & -0.084 & 0.004 & 0.057 & 0.057 & 0.952 \\
\midrule
3.5 & $\beta_1$ &  0.768 & 0.002 & 0.042 & 0.042 & 0.957 & 1.235 & 0.002 & 0.043 & 0.045 & 0.962 \\
    & $\beta_2$ & -0.392 & 0.004 & 0.045 & 0.044 & 0.930 & -0.140 & 0.004 & 0.058 & 0.057 & 0.945 \\
\midrule
4.0 & $\beta_1$ &  0.751 & 0.002 & 0.044 & 0.044 & 0.952 & 1.211 & 0.002 & 0.046 & 0.048 & 0.965 \\
    & $\beta_2$ & -0.421 & 0.004 & 0.045 & 0.044 & 0.931 & -0.185 & 0.005 & 0.058 & 0.058 & 0.942 \\
\bottomrule
\end{tabular}
\end{table}
\begin{table}[htbp!]
\caption{Simulation results under Scenario II(a) and II(b) with $R=7$ equally spaced knots and stacking time at $(1.0,1.5,2.0,2.5,3.0,3.5,4.0)$ and covariate dependent censoring. Sample size $n=2000$; censoring rate = $50\%$. Metrics: ABias (absolute bias), MCSD (Monte Carlo SD), AESE (asymptotic SE), CP (coverage of the $95\%$ CI). Column blocks compare weights $(1,1,1)$ vs.\ $(1,2,2)$.}
\label{tab:scenario_iid_iiab}
\centering
\small
\begin{tabular}{cccccccccccc}
\toprule
 &  & \multicolumn{5}{c}{$(w_1,w_2,w_D)=(1,1,1)$} & \multicolumn{5}{c}{$(w_1,w_2,w_D)=(1,2,2)$} \\
\cmidrule(lr){3-7} \cmidrule(lr){8-12}
Time & $\bm{\beta}(t)$ & True & ABias & MCSD & AESE & CP & True & ABias & MCSD & AESE & CP \\
\midrule
1.0 & $\beta_1$ & 0.506 & 0.003 & 0.036 & 0.036 & 0.955 & 0.917 & 0.001 & 0.017 & 0.018 & 0.957 \\
    & $\beta_2$ & 0.789 & 0.001 & 0.030 & 0.030 & 0.942 & 0.955 & 0.001 & 0.017 & 0.018 & 0.946 \\
\midrule
1.5 & $\beta_1$ & 0.469 & 0.002 & 0.034 & 0.035 & 0.955 & 0.864 & 0.001 & 0.017 & 0.017 & 0.961 \\
    & $\beta_2$ & 0.765 & 0.000 & 0.031 & 0.032 & 0.935 & 0.932 & 0.001 & 0.018 & 0.018 & 0.938 \\
\midrule
2.0 & $\beta_1$ & 0.458 & 0.000 & 0.035 & 0.035 & 0.955 & 0.844 & 0.000 & 0.016 & 0.017 & 0.958 \\
    & $\beta_2$ & 0.748 & 0.000 & 0.032 & 0.033 & 0.940 & 0.915 & 0.001 & 0.020 & 0.020 & 0.938 \\
\midrule
2.5 & $\beta_1$ & 0.455 & 0.000 & 0.036 & 0.036 & 0.956 & 0.838 & 0.000 & 0.017 & 0.017 & 0.957 \\
    & $\beta_2$ & 0.735 & 0.001 & 0.033 & 0.034 & 0.939 & 0.901 & 0.001 & 0.021 & 0.021 & 0.934 \\
\midrule
3.0 & $\beta_1$ & 0.457 & 0.000 & 0.038 & 0.038 & 0.944 & 0.838 & 0.000 & 0.018 & 0.018 & 0.949 \\
    & $\beta_2$ & 0.724 & 0.002 & 0.034 & 0.035 & 0.944 & 0.887 & 0.001 & 0.023 & 0.023 & 0.927 \\
\midrule
3.5 & $\beta_1$ & 0.448 & 0.001 & 0.040 & 0.040 & 0.938 & 0.825 & 0.000 & 0.019 & 0.019 & 0.943 \\
    & $\beta_2$ & 0.715 & 0.001 & 0.036 & 0.037 & 0.931 & 0.879 & 0.001 & 0.024 & 0.024 & 0.924 \\
\midrule
4.0 & $\beta_1$ & 0.444 & 0.000 & 0.042 & 0.042 & 0.939 & 0.819 & 0.000 & 0.020 & 0.020 & 0.951 \\
    & $\beta_2$ & 0.707 & 0.001 & 0.037 & 0.037 & 0.930 & 0.870 & 0.001 & 0.026 & 0.026 & 0.921 \\
\bottomrule
\end{tabular}
\end{table}


In the simulation study, the true values of \(\bm{\beta}(t)^0\) are calculated from the Monte Carlo simulation with \(n=10^{6}\) of individuals (a simulated super-population). For regression model $\eta\{l(\mathcal H)(t\mid \bm Z)\} = \bm\beta(t)^\top \bm Z$, we fix a time grid $\mathcal T$. For each $t \in \mathcal T$, we simulate a large dataset (\(n=10^{6}\) under the data-generating model in Section \ref{supp:sec:gen_model}). The true value $\bm\beta(t)^0$ is calculated by solving the pointwise equation when censoring is removed in the simulated super-population:
\[
\sum_{i=1}^n \bm Z_i \big\{ L_i(t) - (D_i \wedge t)\eta^{-1}\big(\bm\beta(t)^\top \bm Z_i\big) \big\} = \bm 0
\]
Table \ref{tab:scenario_iid_iac} reports the simulation results for Scenarios I(a) and I(c) under covariate-dependent censoring with a censoring rate of 25\%. The covariate effects \(\bm{\beta}(t)^0\) are identical to those in Table 2 of the main paper. 
Our estimators show negligible bias, with absolute bias (Abias) remaining small across all time points. The average estimated standard errors (AESEs) are close to the empirical Monte Carlo standard deviations (MCSDs), and the empirical coverage probabilities (CPs) are near the nominal 95\% level. Compared with the higher censoring rate scenario (50\%) in Table 2, both the MCSD and AESE are consistently smaller, indicating improved efficiency with fewer censored events (hence larger effective sample size). When comparing the weighting schemes, we find that \((w_1, w_2, w_D) = (1, 2, 2)\) gives slightly larger MCSD and AESE than \((w_1, w_2, w_D) = (1, 1, 1)\). 

Table \ref{tab:scenario_iid_iiab} presents the simulation results for Scenarios II(a) and II(b) under covariate-dependent censoring with a censoring rate of 50\%. In these scenarios, the type 1 recurrent event rate per person-time is reduced by approximately 30\%, and the fatal event rate per person-time is reduced by approximately 25\% compared to Scenario I(b) and Scenario I(d). Across all time points and under both weighting schemes, \((w_1, w_2, w_D) = (1, 1, 1)\) and \((w_1, w_2, w_D) = (1, 2, 2)\), our proposed method still provides unbiased estimates of the covariate effects \(\bm{\beta}(t)\), with small ABias and comparable AESE and MCSD values. The empirical 95\% coverage probabilities remain close to the nominal level across all time points.

\section{Simulation study under the cluster-correlated data setting} \label{supp:sec:simu_CRT}

We extend the data-generating mechanism with a joint frailty model from the independent data setting to the cluster-correlated data setting by introducing an additional cluster-level frailty term. We generate \(M=60\) independent clusters with varying sizes \(N_i\) drawn from a discrete uniform distribution on \(\{20,\dots,80\}\). For individual \(j\) in cluster \(i\), baseline covariates are generated from \(Z_{ij1}\sim \mathrm{Bernoulli}(0.5)\) and \(Z_{ij2}\sim \mathcal{N}(0,1)\). Within-cluster dependence is induced through two frailty components: 1) a cluster-level frailty \(B_i\sim \mathrm{Gamma}(\text{shape}=2,\text{rate}=2)\) shared by all members of cluster \(i\), and 2) an individual-level frailty \(W_{ij}\sim \mathrm{Gamma}(\text{shape}=4.5,\text{rate}=4.5)\) shared across event types for individual \(j\). Let \(\mu_{0k}(\cdot)\) and \(\lambda_{0D}(\cdot)\) denote the baseline hazard functions for recurrent event type \(k\in\{1,2\}\) and for the terminal event, respectively. Conditional on \((\bm Z_{ij}, B_i, W_{ij})\), event times are generated from the intensity model for recurrent events and hazard model for terminal event, defined as
\[
\mu_k\!\big(t\mid \bm Z_{ij}, B_i, W_{ij}, D_{ij}\ge t_{ijk}\big)
= \mu_{0k}(t)\, B_i\, W_{ij}\, \exp(\bm\alpha_k^\top \bm Z_{ij}),\qquad k=1,2,
\]
\[
\lambda_D\!\big(d\mid \bm Z_{ij}, B_i, W_{ij}, D_{ij}\ge d_{ij}\big)
= \mu_{0D}(d)\, B_i\, W_{ij}^{\gamma}\, \exp(\bm\alpha_D^\top \bm Z_{ij}),
\]
with \(\gamma=1\) in all settings. 
We specify the baseline rate function for type 1 recurrent event as a Weibull distribution, given by 
\[
\mu_{01}(t) = \frac{1.25}{0.50} \left( \frac{t}{0.50} \right)^{0.25}
\exp\!\left\{ -\left( \frac{t}{0.50} \right)^{1.25} \right\}.
\]
For type 2 recurrent events, we adopt a piecewise-exponential baseline rate function of the form 
\[
\mu_{02}(t) =
\begin{cases}
    0.40\,\exp(-0.40\,t), & 0 \le t < 1, \\[6pt]
    0.22\,\exp\{-0.40 - 0.22\,(t - 1)\}, & 1 \le t < 3, \\[6pt]
    0.10\,\exp\{-0.84 - 0.10\,(t - 3)\}, & t \ge 3.
\end{cases}
\]
For terminal events, we specify a Gompertz baseline hazard function as  \(\lambda_{0D}(d) = 0.30\,\exp(0.30\,d)\). Censoring times are generated from a covariate-dependent exponential distribution, \(C_{ij} \sim \mathrm{Exp}\big\{c_0 \exp(\bm\theta^\top \bm Z_{ij})\big\}\), with \(c_0=0.27\) and \(\bm\theta=(0.50,0.20)\), calibrated to achieve approximately \(50\%\) of censoring rate marginally. We consider two weighting schemes with equal weights \((w_1,w_2,w_D)=(1,1,1)\) and unequal weights \((w_1,w_2,w_D)=(1,2,2)\) assigning greater emphasis to death and severe recurrences. The while-alive regression model in Section 4 is implemented using a spline basis function for \(\bm{\beta}(t)\) with \(R=7\) equally spaced knots at \(\{1.0\ 1.5\ 2.0\ 2.5\ 3.0\ 3.5\ 4.0\}\), specifically
\[
\beta_j(t)=\sum_{r=1}^{R} \gamma_{j,r}\,J_r(t) \qquad
J_r(t)=\mathbb{I}\{t \ge t_{r-1}\} \quad j=1,2.
\]
Stacking times are aligned with the knot locations, and the log link \(\eta(.)=\log (.)\) is used. The true covariate effects \(\bm\beta(t)^0=\left\{\beta_1(t)^0,\beta_2(t)^0\right\}^\top\) are calculated at each landmark time via Monte Carlo simulation with \(M=10^5\) clusters generated under the same data-generating process but with censoring removed. The performance of the proposed method is evaluated at the knot times based on ABias, MCSD, AESE, and CP of the \(95\%\) confidence interval across \(1{,}000\) Monte Carlo replicates.

\begin{table}[htbp!]
\caption{Simulation results with $R=7$ equally spaced knots at $(1.0,1.5,2.0,2.5,3.0,3.5,4.0)$ and covariate-dependent censoring for clustered data. Sample size $M=60$, with cluster size from a discrete uniform distribution \(\{20,\dots, 80\}\); censoring rate =$50\%$. Metrics: ABias (absolute bias), MCSD (Monte Carlo SD), AESE (asymptotic SE), CP (coverage of the $95\%$ CI). Column blocks compare weights $(1,1,1)$ vs.\ $(1,2,2)$.}
\label{tab:scenario_cluster_same}
\centering
\small
\begin{tabular}{cccccccccccc}
\toprule
 &  & \multicolumn{5}{c}{$(w_1,w_2,w_D)=(1,1,1)$} & \multicolumn{5}{c}{$(w_1,w_2,w_D)=(1,2,2)$} \\
\cmidrule(lr){3-7} \cmidrule(lr){8-12}
Time & $\bm{\beta}(t)$ & True & ABias & MCSD & AESE & CP & True & ABias & MCSD & AESE & CP \\
\midrule
1.0 & $\beta_1$ & 0.718 & 0.005 & 0.089 & 0.084 & 0.932 & 1.132 & 0.005 & 0.095 & 0.089 & 0.940 \\
    & $\beta_2$ & -0.105 & 0.001 & 0.057 & 0.057 & 0.938 & 0.229 & 0.001 & 0.059 & 0.060 & 0.948 \\
\midrule
1.5 & $\beta_1$ & 0.666 & 0.004 & 0.084 & 0.081 & 0.937 & 1.090 & 0.005 & 0.090 & 0.087 & 0.938 \\
    & $\beta_2$ & -0.228 & 0.002 & 0.060 & 0.058 & 0.926 & 0.081 & 0.001 & 0.065 & 0.065 & 0.938 \\
\midrule
2.0 & $\beta_1$ & 0.625 & 0.003 & 0.082 & 0.080 & 0.950 & 1.051 & 0.004 & 0.088 & 0.085 & 0.940 \\
    & $\beta_2$ & -0.308 & 0.004 & 0.062 & 0.060 & 0.927 & -0.024 & 0.002 & 0.069 & 0.068 & 0.926 \\
\midrule
2.5 & $\beta_1$ & 0.592 & 0.003 & 0.082 & 0.080 & 0.938 & 1.014 & 0.003 & 0.087 & 0.085 & 0.939 \\
    & $\beta_2$ & -0.363 & 0.004 & 0.063 & 0.061 & 0.922 & -0.102 & 0.004 & 0.071 & 0.071 & 0.927 \\
\midrule
3.0 & $\beta_1$ & 0.565 & 0.002 & 0.082 & 0.081 & 0.943 & 0.981 & 0.002 & 0.087 & 0.085 & 0.932 \\
    & $\beta_2$ & -0.404 & 0.004 & 0.065 & 0.063 & 0.926 & -0.162 & 0.004 & 0.074 & 0.074 & 0.925 \\
\midrule
3.5 & $\beta_1$ & 0.539 & 0.002 & 0.084 & 0.083 & 0.941 & 0.948 & 0.002 & 0.089 & 0.087 & 0.942 \\
    & $\beta_2$ & -0.435 & 0.005 & 0.068 & 0.065 & 0.933 & -0.209 & 0.005 & 0.078 & 0.076 & 0.922 \\
\midrule
4.0 & $\beta_1$ & 0.518 & 0.001 & 0.086 & 0.086 & 0.943 & 0.919 & 0.002 & 0.090 & 0.090 & 0.937 \\
    & $\beta_2$ & -0.459 & 0.005 & 0.070 & 0.067 & 0.917 & -0.246 & 0.005 & 0.081 & 0.079 & 0.918 \\
\bottomrule
\end{tabular}
\end{table}


Table \ref{tab:scenario_cluster_same} presents the simulation results for the while-alive regression model under cluster-correlated data with covariate-dependent censoring at a 50\% censoring rate. Across all time points \(t \in \{1.0, 1.5, 2.0, 2.5, 3.0, 3.5, 4.0\}\), the proposed estimators \(\widehat{\bm{\beta}}(t)\) remain approximately unbiased, with ABias consistently small for both weighting schemes. The average estimated standard errors (AESE) obtained from the sandwich variance estimator closely match the empirical Monte Carlo standard deviations (MCSD), showing accurate variance estimation even in the presence of clustering. The empirical coverage probabilities of the 95\% confidence intervals are also near the nominal level across all time points, suggesting that the cluster-robust sandwich variance estimator is roughly unbiased to capture the true variability.

\section{Additional Simulations when stacking times are different from the knot times in the regression model fit.} \label{supp:sec:siid_stack_knot_diff}

Additional simulation studies are conducted to evaluate the proposed while-alive regression method by decoupling the roles of spline knots and stacking times. We consider the independent data setting to focus ideas. Independent observations from \(n=2000\) individuals are generated under the same joint frailty specification as in the main paper, focusing on Scenarios I(a) and I(c). In both scenarios, censoring is covariate-dependent with a censoring rate of 50\%. The weighting scheme is set to \((w_1,w_2,w_D)=(1,1,1)\) for Scenario I(a) and \((w_1,w_2,w_D)=(1,2,2)\) for Scenario I(c). The while-alive regression model described in Section 3.2 is fitted using a spline basis for \(\bm{\beta}(t)\) with \(R=7\) equally spaced knots at \(\{1.0, 1.5, 2.0, 2.5, 3.0, 3.5, 4.0\}\) as
\[
\beta_j(t)=\sum_{r=1}^{R} \gamma_{j,r}\,J_r(t), \qquad J_r(t)=\mathbb{I}\{t \ge t_{r-1}\}, \quad j=1,2.
\]
To decouple the knot locations from the stacking times, a finer grid of stacking points is specified as 
\(\{t_1,\dots,t_V\}=\{1.0, 1.3, 1.6, 1.9, 2.2, 2.5, 2.8, 3.1, 3.4, 3.7, 4.0\}\).

Table \ref{tab:scenario_iid_iac_diff} presents the simulation results for Scenarios I(a) and I(c) with approximately 50\% of censoring rate, spline knots at \(\{1.0, 1.5, 2.0, 2.5, 3.0, 3.5, 4.0\}\), and stacking times at \\\(\{1.0, 1.3, 1.6, 1.9, 2.2, 2.5, 2.8, 3.1, 3.4, 3.7, 4.0\}\). This design separates the spline knot locations from the stacking times, with \((R \neq V)\), to assess the robustness of the proposed method for while-alive regression. The estimators of \(\bm{\beta}(t)\) remain approximately unbiased across all time points and for both weighting schemes, with small ABias. The AESE closely matches the empirical MCSD, and the coverage of the 95\% confidence intervals is near the nominal level. Compared with Table 2 of the main paper, where stacking times coincide with the spline knots, alignment of the knots and stacking times produces slightly larger standard deviations than those obtained under the finer but non-aligned grid.

\begin{table}[htbp!]
\caption{Simulation results under Scenario I(b) and I(d) with $R=7$ equally spaced knots at $(1.0,1.5,2.0,2.5,3.0,3.5,4.0)$, stacking time at \(\{1, 1.3, 1.6, 1.9, 2.2, 2.5, 2.8, 3.1, 3.4, 3.7, 4.0\} \) and covariate-dependent censoring. Sample size $n=2000$; censoring rate $50\%$. Metrics: ABias (absolute bias), MCSD (Monte Carlo SD), AESE (asymptotic SE), CP (coverage of the $95\%$ CI). Column blocks compare weights $(1,1,1)$ vs.\ $(1,2,2)$.}
\label{tab:scenario_iid_iac_diff}
\centering
\small
\begin{tabular}{cccccccccccc}
\toprule
 &  & \multicolumn{5}{c}{$(w_1,w_2,w_D)=(1,1,1)$} & \multicolumn{5}{c}{$(w_1,w_2,w_D)=(1,2,2)$} \\
\cmidrule(lr){3-7} \cmidrule(lr){8-12}
Time & $\bm{\beta}(t)$ & True & ABias & MCSD & AESE & CP & True & ABias & MCSD & AESE & CP \\
\midrule
1.0 & $\beta_1$ & 0.904 & 0.001 & 0.043 & 0.042 & 0.951 & 1.341 & 0.001 & 0.043 & 0.043 & 0.955 \\
    & $\beta_2$ & -0.008 & 0.004 & 0.056 & 0.056 & 0.941 & 0.355 & 0.004 & 0.064 & 0.063 & 0.940 \\
\midrule
1.5 & $\beta_1$ & 0.871 & 0.013 & 0.045 & 0.045 & 0.938 & 1.333 & 0.004 & 0.045 & 0.045 & 0.946 \\
    & $\beta_2$ & -0.146 & 0.052 & 0.056 & 0.055 & 0.923 & 0.195 & 0.062 & 0.066 & 0.065 & 0.840 \\
\midrule
2.0 & $\beta_1$ & 0.839 & 0.018 & 0.048 & 0.049 & 0.928 & 1.312 & 0.013 & 0.048 & 0.049 & 0.938 \\
    & $\beta_2$ & -0.238 & 0.046 & 0.055 & 0.054 & 0.922 & 0.077 & 0.059 & 0.068 & 0.067 & 0.949 \\
\midrule
2.5 & $\beta_1$ & 0.812 & 0.010 & 0.054 & 0.056 & 0.944 & 1.287 & 0.010 & 0.057 & 0.058 & 0.940 \\
    & $\beta_2$ & -0.304 & 0.022 & 0.056 & 0.054 & 0.922 & -0.013 & 0.029 & 0.071 & 0.069 & 0.924 \\
\midrule
3.0 & $\beta_1$ & 0.789 & 0.012 & 0.059 & 0.062 & 0.943 & 1.261 & 0.014 & 0.064 & 0.066 & 0.935 \\
    & $\beta_2$ & -0.354 & 0.022 & 0.055 & 0.054 & 0.935 & -0.084 & 0.031 & 0.071 & 0.071 & 0.933 \\
\midrule
3.5 & $\beta_1$ & 0.768 & 0.014 & 0.064 & 0.066 & 0.937 & 1.235 & 0.018 & 0.070 & 0.071 & 0.926 \\
    & $\beta_2$ & -0.392 & 0.023 & 0.056 & 0.054 & 0.927 & -0.140 & 0.033 & 0.072 & 0.072 & 0.923 \\
\midrule
4.0 & $\beta_1$ & 0.751 & 0.012 & 0.072 & 0.072 & 0.939 & 1.211 & 0.015 & 0.079 & 0.079 & 0.923 \\
    & $\beta_2$ & -0.421 & 0.015 & 0.057 & 0.055 & 0.921 & -0.184 & 0.021 & 0.075 & 0.074 & 0.922 \\
\bottomrule
\end{tabular}
\end{table}


\section{Additional Analysis of HF-ACTION trial based on the full sample} \label{supp:sec:hf_action_all}

We analyzed the HF-ACTION trial using the proposed while-alive regression method applied to the full trial dataset. 
The full data includes 2048 patients, with 1028 assigned to usual care and 1020 to exercise training. In the usual care group, the median follow-up time is 2.92 years, with 181 deaths (17.6\%) and an average of 1.98 hospitalizations per patient. In the exercise training group, the median follow-up time is 2.85 years, with 167 deaths (16.4\%) and an average of 1.96 hospitalizations per patient. The analysis was performed using a log link and a B-spline basis (same as the main paper in Section 7) for time-varying effects, with weights \(w_H=1\) and \(w_D=2\) to place greater emphasis on death events. The degree of spline \(d\) and the number of knots \(R\) were selected via 5-fold cross-validation, giving \((d,R)=(2,6)\). The stacking times were the same as the interior time points for spline knots.

\begin{figure}[ht!]
    \centering
    \includegraphics[width=0.9\linewidth]{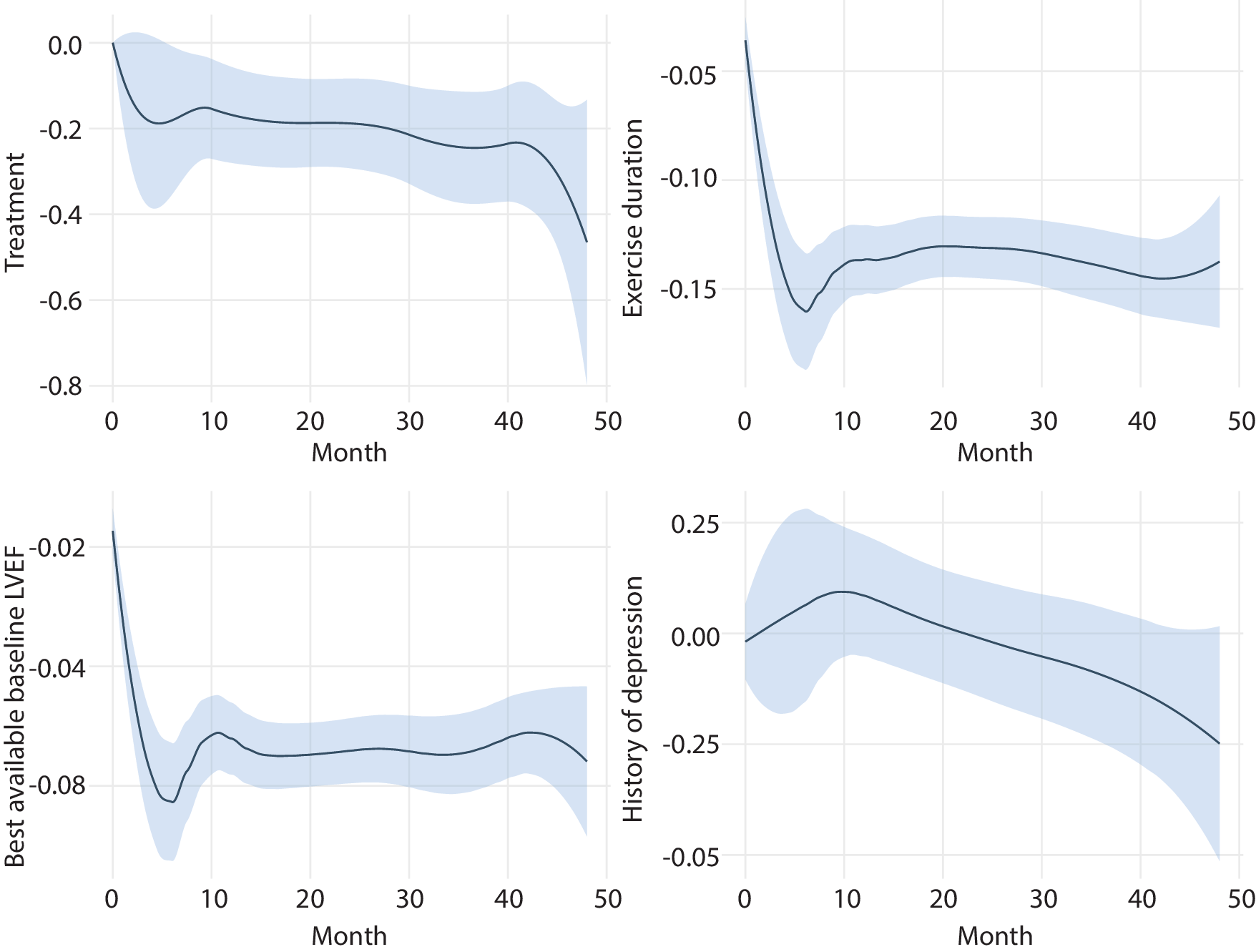}
    \caption{Estimated time-varying covariate effects from the while-alive regression analysis with a log link using the full HF-ACTION trial dataset. Each panel displays the estimated effect trajectory (solid line) and its 95\% pointwise confidence band (shaded area) for covariates including treatment assignment, exercise duration, best available baseline left ventricular ejection fraction (LVEF), and history of depression.}
    \label{fig:hf_action_raw}
\end{figure}

Figure \ref{fig:hf_action_raw} presents the estimated time-varying covariate effects from the proposed method with complete HF-ACTION data. Overall, the treatment effect shows an increasing beneficial trend over the 48-month follow-up, indicating a significant effect in the log while-alive measure between the two intervention arms. Exercise duration and baseline LVEF show statistically significant and increasing effects during the first six months, which then stabilize thereafter, whereas history of depression shows no statistically significant association throughout the study period.

\section{Analysis of STRIDE cluster-randomized trial} \label{supp:sec:stride}

We analyzed the STRIDE cluster-correlated dataset introduced in Section 2 of the main paper using the proposed method described in Section 4.  STRIDE is a pragmatic cluster randomized trial conducted at 86 primary care practice, which were randomized to either multifactorial fall-prevention intervention or usual care. The goal of this trial is to evaluate the effect of the intervention on serious fall injuries among community-dwelling older adults. Across the 86 clusters (average cluster size: 64), about 45\% of participants experienced at least one serious fall injury. Among those with any fall injury, 56\% had a single event, 24\% had two, and 10\% had more than three. In our analysis, we consider both the recurrence of serious fall injuries and death in the while-alive loss rate. We analyze the STRIDE data using our proposed method with log link function by setting the weight \(w_{F}=1\), \(w_D=2\)  for fall injuries and death, respectively, emphasizing the priority of death, and adjust for gender, race (black versus white), and number of chronic diseases. We did not adjust for age because eligibility restricted the cohort to \(\geq 70\) years (mean \(=80\), \(\text{SD}=5.77\)), which provided limited support for spline terms. In exploratory models that included age, we observed numerical instability and variance inflation, and therefore excluded age to maintain stable estimation.
We fit the Cox proportional hazard model to assess potentially covariate-dependent censoring at the significance level of \(0.05\) and found race is significantly associated with the censoring time. Thus, we adjusted for race in the censoring survival function. In this analysis, we used a B-spline basis similar to Section 7 in the main paper, and the optimal combination of degree of spline and number of knots was chosen by 5-fold cross-validation, giving \((d,R) = (1,6)\). The same set of 6 interior time points was used as the stacking time grid.

\begin{figure}[ht!]
    \centering
    \includegraphics[width=1\linewidth]{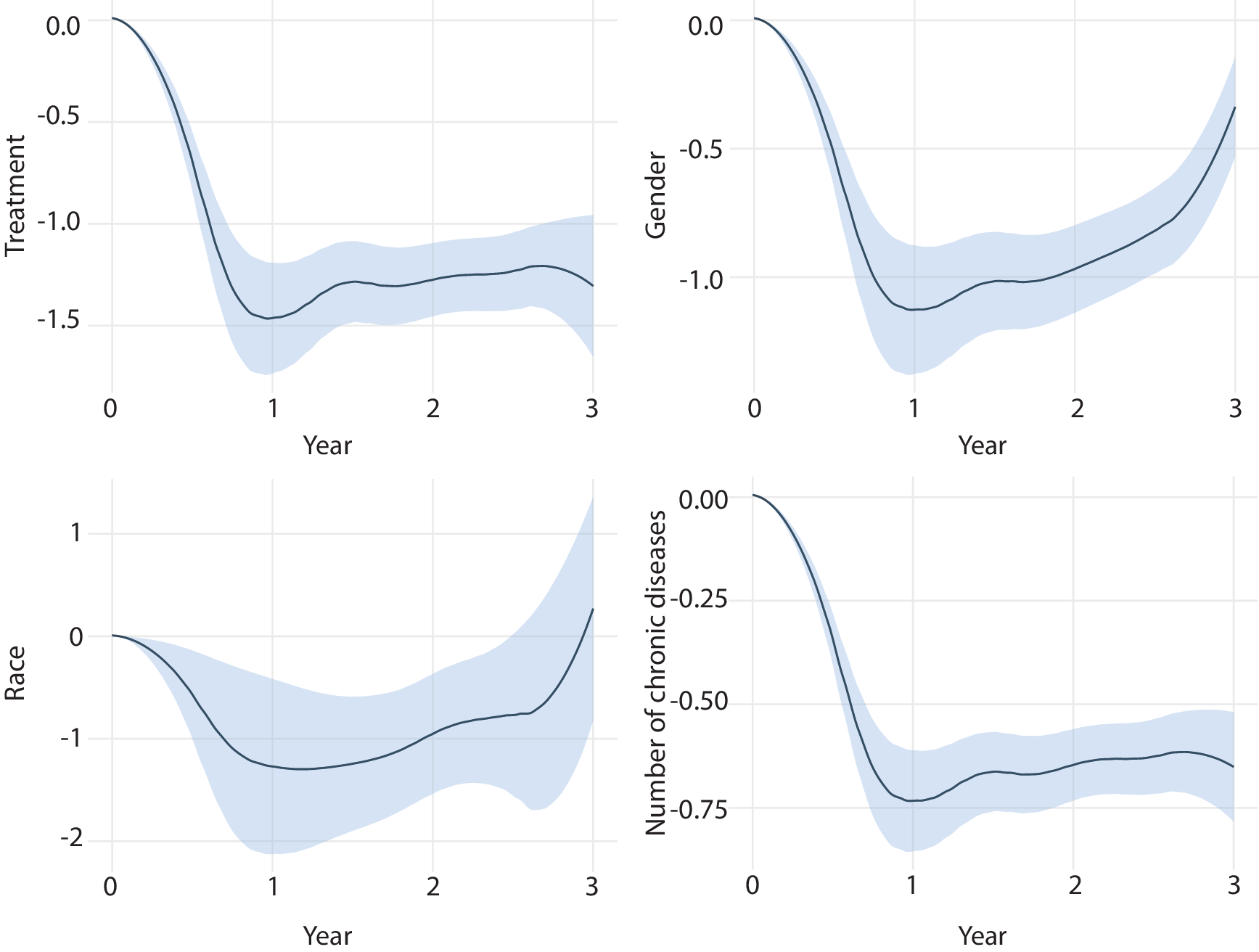}
    \caption{Estimated time-varying covariate effects from the while-alive regression analysis with a log link using the STRIDE trial dataset. Each panel displays the estimated coefficient trajectory (solid line) and its 95\% pointwise confidence band (shaded area) for covariates including treatment assignment, gender, race, and number of chronic diseases. }

    \label{fig:stide_all}
\end{figure}

Figure \ref{fig:stide_all} presents the estimated covariate effects and corresponding 95\% confidence intervals on the log while-alive loss rate over time. The first panel displays a sustained and significant treatment effect, with \(95\%\) confidence intervals consistently excluding zero, indicating a persistent reduction in the cumulative number of adverse events experienced while participants are alive. The treatment effect becomes most pronounced around the first year, with an estimated log-rate difference of \(-1.457\), corresponding to an approximate \(76.7\%\) (\(\approx 1-e^{-1.457}\)) reduction in the while-alive loss rate for intervention group relative to control. After one year, the treatment effect plateaus and maintains a stable beneficial impact. The global Wald test also shows a significant overall effect (\(p<0.001\)). 

To put our analysis in context, we recall that the original primary analysis targeted the time to first self-reported serious fall injury (with death treated as a competing risk), and likewise reported statistically significant difference between groups with hazard ratio 0.90 (see Section 2 in the main paper) \citep{bhasin2020randomized}. In contrast, our while-alive analysis targeted the composite rate of fall injuries and death, emphasizing the cumulative burden of events among survivors rather than the timing of a single initial event. These two analyses aim to offer complementary evidence that the intervention achieved statistically significant benefits under both onset-focused or burden-focused targets. Furthermore, our analysis also reveals how other baseline covariates may exhibit time-varying associations with the while-alive loss rate. For example, the effect of gender increases during one year and gradually declines thereafter, indicating that female participants can accumulate a higher number of recurrent falls or deaths compared to males within the first year, though this difference attenuates over time. Additionally, the number of chronic conditions shows a strong early impact, peaking shortly after baseline and plateauing around one year. This means that participants with a higher comorbidity burden tend to have shorter survival and therefore accumulate fewer recurrent events while alive, reflecting the joint influence of morbidity and mortality in the while-alive framework.

\section{Illustrative Example Using the \texttt{WAreg} Package} \label{supp:sec:package}

In this section, we will show some examples about how to use the \texttt{WAreg} package to fit while-alive regression models for composite survival endpoints under cluster-correlated data. The independent data is a special case of cluster-correlated data with a cluster size of 1; thus, the example is omitted here. This package includes two key functions: \texttt{WA\_fit} and \texttt{WA\_cv}, which are used to fit the proposed while-alive regression model and the K-fold cross-validation to choose the degree of spline and the number of knots.

\subsection{Function Overview}

\paragraph{\texttt{WA\_fit()}} fits the while-alive regression model via estimating equations with inverse probability of censoring weighting (IPCW). Supports inference for both cluster-correlated and independent data, which allows a flexible selection of spline basis function (e.g., B-splines, piecewise polynomials, step functions, time-fixed).

\paragraph{\texttt{WA\_cv()}} performs $K$-fold cross-validation over user-specified grids of basis type, degree of spline, and number of knots. The criterion of selection is to minimize the prediction error.

\subsection{\texttt{WA\_fit}: Arguments and Details}

\begin{description}
\item[\texttt{formula}] A survival formula of the form \verb|Surv(time, status) ~ covariates|, based on a \emph{long-format} data frame. The left-hand side must contain \texttt{time} and \texttt{status}, where \(\text{status}=0\) indicates censoring, \(\text{status} = \{1,\dots, K+1\}\) indicates the recurrent event type, and \(K+1\) is the fatal event.

\item[\texttt{data}] A long-format data frame containing all variables in \texttt{formula} plus identifiers. Each row is an event record for a subject.

\item[\texttt{id}] Character scalar naming the subject identifier column (required).

\item[\texttt{cluster}] (Optional) Character scalar naming the cluster identifier column (for cluster-correlated data). If supplied, standard errors are cluster-robust at the cluster level. If \texttt{NULL}, the method defaults to independent data with relative inference.

\item[\texttt{knots}] Numeric vector specifying the spline basis location (boundaries and, optionally, interior knot position) with length $\ge 2$.

\item[\texttt{tau\_grid}] Numeric vector of $\tau$ values defining the grid over which the estimating equations are stacked (and used for prediction). This is independent of \texttt{knots}.

\item[\texttt{basis}] Types of spline basis option, including
\begin{itemize}
  \item \texttt{"bz"}: B-spline basis (scaled to 0 at $t=0$),
  \item \texttt{"ns"}: natural spline,
  \item \texttt{"ms"}: M-spline (requires \texttt{splines2}),
  \item \texttt{"pl"}: piecewise polynomial of degree \texttt{degree},
  \item \texttt{"st"}: step basis,
  \item \texttt{"tl"}: truncated-linear basis,
  \item \texttt{"il"}: interval-local linear basis,
  \item \texttt{"tf"}: time-fixed with a single value specified in \texttt{tau\_grid}.
\end{itemize}

\item[\texttt{degree}] Non-negative integer controlling polynomial degree of the chosen basis (e.g., \texttt{1} for linear). For \texttt{"tf"} the value is ignored; for \texttt{"bz"} and \texttt{"ns"} it is the spline degree; for \texttt{"pl"} it sets the local polynomial order; for \texttt{"ms"} it is the M-spline degree.

\item[\texttt{link}] Link function for the while-alive loss rate: \texttt{"log"} (default) or \texttt{"identity"}.

\item[\texttt{w\_recur}] Numeric vector of weights for each recurrent-event type (in ascending order of their codes on the \texttt{status} scale). Its length must equal the number of distinct recurrent event codes in the data.

\item[\texttt{w\_term}] Numeric weight for the terminal event.

\item[\texttt{ipcw}] IPCW method: \texttt{"cox"} fits a Cox model for censoring; \texttt{"km"} uses a nonparametric Kaplan--Meier censoring model.

\item[\texttt{ipcw\_formula}] A formula specifying the covariates for the censoring model when \texttt{ipcw = "cox"} (e.g., \verb|~ x1 + x2|). Ignored when \texttt{ipcw = "km"}.

\end{description}

\paragraph{Returns.} An object of class \texttt{"WA"} with elements:
\begin{itemize}
  \item \texttt{est}: coefficient vector estimated from the estimating equation,
  \item \texttt{vcov}, \texttt{variance covariance matrix},
  \item \texttt{basis}, \texttt{degree}, \texttt{link}, \texttt{knots}, \texttt{tau\_grid}, \texttt{Z\_cols},
  \item \texttt{status\_codes}: list of recurrent, terminal, and censor codes,
  \item \texttt{call}, \texttt{formula}.
\end{itemize}
S3 methods includes \verb|print.WA|, \verb|summary.WA|, \verb|predict.WA|, and \verb|plot.WA| for this output.

\subsection{\texttt{WA\_cv}: Arguments and Details}

\begin{description}

\item[\texttt{basis\_set}] Character vector of types of spline basis to evaluate (e.g., \verb|c("bz","pl","il")|).

\item[\texttt{degree\_vec}] Integer vector of degrees to search (e.g., \verb|1:3|).

\item[\texttt{n\_int\_vec}] Integer vector of candidate numbers of \emph{interior} knots (e.g., \verb|c(0,2,4,6)|). 

\item[\texttt{knot\_scheme}] Knot placement scheme: \texttt{"equidist"} (equally spaced on time) or \texttt{"quantile"} (empirical quantiles of observed times).

\item[\texttt{link\_set}] Character vector of link functions to evaluate (subset of \{\texttt{"log"}, \texttt{"identity"}\}).

\item[\texttt{time\_range}] Optional two-vector \verb|c(t_min, t_max)| for the boundary of prediction error. If omitted, \texttt{t\_min} defaults to $0$ and \texttt{t\_max} to the maximum observed time.

\item[\texttt{tau\_grid}] Optional numeric grid for stacking/integration in CV. If \texttt{NULL}, a default dense grid between \texttt{time\_range} bounds is used.

\item[\texttt{K}] Number of folds (default \texttt{5}).

\item[\texttt{seed}] Integer random seed for fold assignment (to ensure replicability).

\end{description}

\paragraph{Returns.} A data frame with configuration and the total cross-validated prediction error \texttt{PE}. Users can select the configuration with minimal \texttt{PE} and refit with \texttt{WA\_fit}.

\subsection{Example with B-spline mode (\texttt{basis = "bz"})}

We use the two simulated datasets \texttt{irt\_dt} and \texttt{crt\_dt} for independent data and cluster-correlated data to show the usage of the function \texttt{WA\_fit}.

The first example is under the independent data.

\begin{verbatim}
library(WAreg)

fit <- WA_fit(
  formula  = Surv(time, status) ~ trt + Z1 + Z2,
  data     = irt_dt,                 # long-format independent data
  id       = "id",
  cluster  = NULL,                   # NULL => for independent data
  knots    = seq(0, 3.5, length.out = 6),   # basis knots position
  tau_grid = seq(0, 3.5, length.out = 6),   # stacking grid
  basis    = "bz", degree = 1, link = "log", w_recur  = c(1, 2), w_term = 2,
  ipcw     = "cox", ipcw_formula = ~ Z1 + Z2
)

summary(fit)
\end{verbatim}

\paragraph{Plot the while-alive loss rate trajectory.}
\begin{verbatim}
plot(
  fit, newdata = irt_dt,
  t_seq = seq(0, 3, length.out = 200),
  id = 1, mode = "wa", smooth = TRUE
)
\end{verbatim}

\begin{figure}[htbp!]
  \centering
  \includegraphics[width=0.7\linewidth]{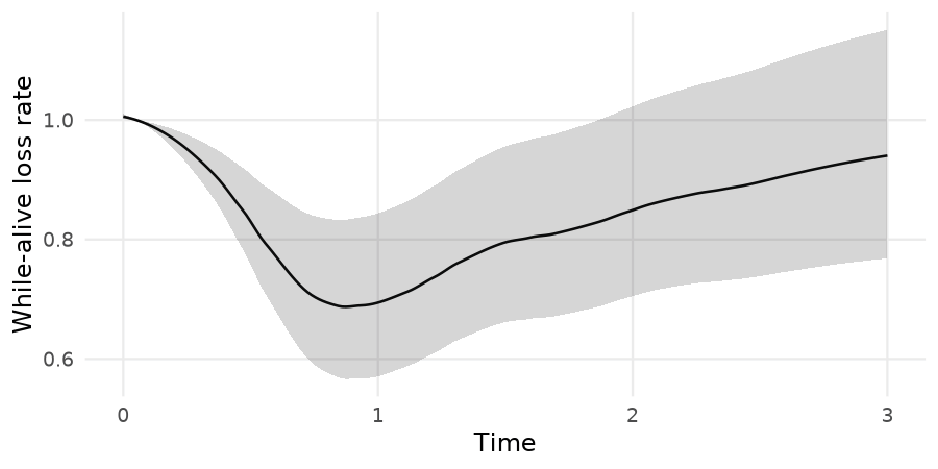}
  \caption{Example plot of the while-alive loss rate from the fitted regression model under independent data}
\end{figure}

\paragraph{Plot the time-varying covariate effect.}
\begin{verbatim}
plot(
  fit, newdata = irt_dt, t_seq = seq(0, 3, length.out = 200),
  id = 1, mode = "cov", covariate = "Z1", ylab_cov = "Covariate effect of Z1", smooth = TRUE
)
\end{verbatim}

\begin{figure}[htbp!]
  \centering
  \includegraphics[width=0.7\linewidth]{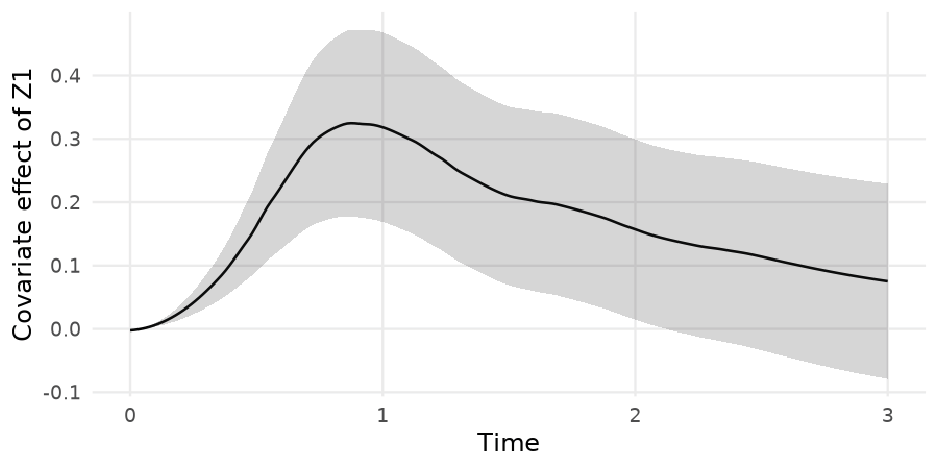}
  \caption{Example plot of the time-varying effect for Z1 from the fitted regression model under independent data}
\end{figure}

\subsection{Example with time-fixed mode (\texttt{basis = "tf"})}
\begin{verbatim}
fit_tf2 <- WA_fit(
  formula  = Surv(time, status) ~ trt + Z1 + Z2, data     = irt_dt, id       = "id",
  cluster  = NULL,
  knots    = c(0, 2),          # two boundaries for "tf"
  tau_grid = 2,                # a single evaluation time
  basis    = "tf", degree = 1, # degree ignored for "tf"
  link     = "log", w_recur  = c(1, 2), w_term   = 2, ipcw     = "cox",
  ipcw_formula = ~ Z1 + Z2
)

pred2 <- predict(
  fit_tf2, newdata = irt_dt[!duplicated(irt_dt$id), ], t_seq = 2
)
head(pred2)

  id t        mu        lb        ub
1  1 2 0.6297689 0.5292576 0.7493684
2  2 2 1.1801194 0.8887261 1.5670540
3  3 2 1.2369475 1.1587066 1.3204716
4  4 2 1.6059896 1.2040297 2.1421420
5  5 2 1.1034374 1.0479120 1.1619050
6  6 2 0.8649552 0.7553303 0.9904905
\end{verbatim}

Next, we illustrate the case of cluster-correlated data through a simulated cluster-randomized trial.

\begin{verbatim}
library(WAreg)

fit <- WA_fit(
  formula  = Surv(time, status) ~ trt + Z1 + Z2,
  data     = crt_dt,                 # long-format cluster-correlated data
  id       = "id",
  cluster  = "cluster",              # cluster identifier
  knots    = seq(0, 3.5, length.out = 6),   # basis shape
  tau_grid = seq(0, 3.5, length.out = 6),   # stacking grid
  basis    = "bz", degree = 1, link = "log",
  w_recur  = c(1, 2), w_term = 2, ipcw     = "cox", ipcw_formula = ~ Z1 + Z2
)

summary(fit)
\end{verbatim}

\paragraph{Plot the while-alive loss rate trajectory.}
\begin{verbatim}
plot(
  fit, newdata = crt_dt, t_seq = seq(0, 3, length.out = 200), id = 1, 
  mode = "wa", smooth = TRUE
)
\end{verbatim}

\begin{figure}[htbp!]
  \centering
  \includegraphics[width=0.7\linewidth]{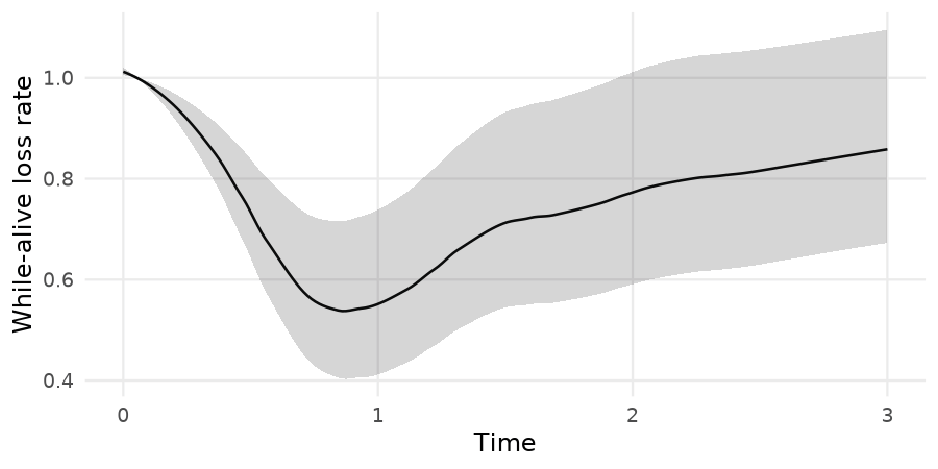}
  \caption{Example plot of the while-alive loss rate from the fitted regression model under cluster-correlated data}
\end{figure}

\paragraph{Plot the time-varying covariate effect.}
\begin{verbatim}
plot(
  fit, newdata = crt_dt, t_seq = seq(0, 3, length.out = 200),
  id = 1, mode = "cov", covariate = "Z1", ylab_cov = "Covariate effect of Z1", smooth = TRUE
)
\end{verbatim}

\begin{figure}[htbp!]
  \centering
  \includegraphics[width=0.7\linewidth]{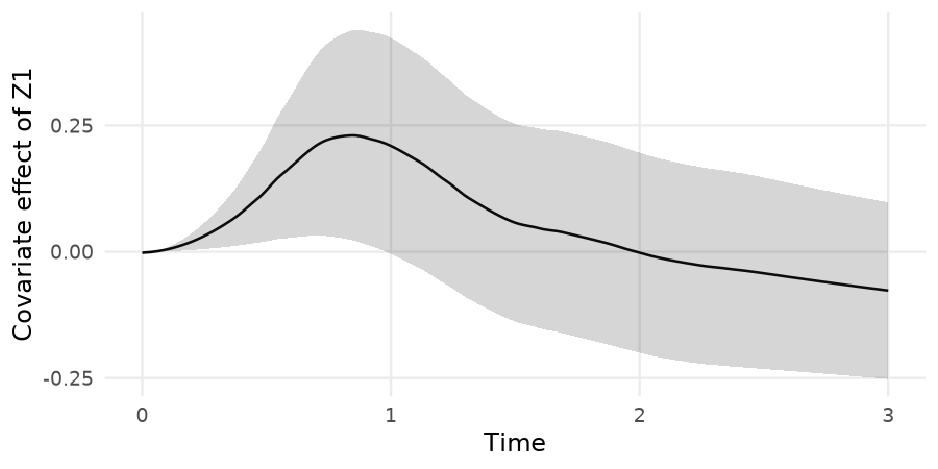}
  \caption{Example plot of the time-varying effect for Z1 from the fitted regression model under cluster-correlated data}
\end{figure}

\subsection{Example with time-fixed mode (\texttt{basis = "tf"})}
\begin{verbatim}
fit_tf2 <- WA_fit(
  formula  = Surv(time, status) ~ trt + Z1 + Z2, data     = crt_dt, id       = "id",
  cluster  = "cluster",
  knots    = c(0, 2),          # two boundaries for "tf"
  tau_grid = 2,                # a single evaluation time
  basis    = "tf", degree = 1, # degree ignored for "tf"
  link     = "log", w_recur  = c(1, 2), w_term   = 2,
  ipcw     = "cox", ipcw_formula = ~ Z1 + Z2
)

summary(fit_tf2)

# Fitted while-alive rate at time = 2 for new data (one row per subject):
pred2 <- predict( fit_tf2, newdata = crt_dt[!duplicated(crt_dt$id), ], t_seq = 2)
head(pred2)

  id t        mu        lb       ub
1  1 2 0.7864806 0.6007492 1.029634
2  2 2 1.5102069 1.0916414 2.089262
3  3 2 1.2300685 1.1817389 1.280375
4  4 2 1.8962564 1.4697644 2.446507
5  5 2 1.0339090 0.9553003 1.118986
6  6 2 1.0146115 0.8339254 1.234447
\end{verbatim}

\label{lastpage}

\end{document}